\documentclass[sigconf,balance=false]{acmart}
\usepackage{popets}

\setcopyright{popets}
\copyrightyear{2024}

\acmYear{2024}
\acmVolume{2024}
\acmNumber{4}
\acmDOI{XXXXXXX.XXXXXXX}
\acmISBN{}
\acmConference{Proceedings on Privacy Enhancing Technologies}
\usepackage{paralist} 
\newcommand{\quotes}[1]{``#1''} 
\usepackage{subfig}
\usepackage{xurl} 
\usepackage{makecell} 
\usepackage{color}
\usepackage{hyperref}
\usepackage{pdflscape} 
\usepackage[edges]{forest}
\usetikzlibrary{arrows.meta,shadows.blur}
\usepackage{pdflscape}
\newcommand{\autorefappendix}[1]{\hyperref[#1]{Appendix~\ref*{#1}}}  
\usepackage[skip=6pt]{caption} 
\usepackage{dblfloatfix} 
\usepackage{colortbl}
\usepackage{changepage} 
\usepackage[para,online,flushleft]{threeparttable}

\usepackage{tcolorbox}
\begin{document}

\title{What Do Privacy Advertisements Communicate to Consumers?}

\author{Xiaoxin Shen}
\orcid{}
\affiliation{%
  \institution{Carnegie Mellon University (CMU)}
  \city{Pittsburgh}
  \state{Pennsylvania}
  \country{USA}}
\email{xiaoxin2@andrew.cmu.edu}

\author{Eman Alashwali}
\authornote{Eman Alashwali was a Collaborating Visitor at CMU while working on this paper.}
\orcid{}
\affiliation{%
  \institution{King Abdulaziz University (KAU) and King Abdullah University of Science and Technology (KAUST)}
  \city{Jeddah}
  \country{Saudi Arabia}}
\email{ealashwali@kau.edu.sa}

\author{Lorrie Faith Cranor}
\orcid{}
\affiliation{%
  \institution{Carnegie Mellon University (CMU)}
  \city{Pittsburgh}
  \state{Pennsylvania}
  \country{USA}}
\email{lorrie@cmu.edu}

\renewcommand{\shortauthors}{Shen et al.}

\begin{abstract}
  When companies release marketing materials aimed at promoting their privacy practices or highlighting specific privacy features, what do they actually communicate to consumers? In this paper, we explore the impact of privacy marketing on: \begin{inparaenum} \item consumers’ attitudes toward the organizations providing the campaigns, \item overall privacy awareness, and \item the actionability of suggested privacy advice\end{inparaenum}. To this end, we investigated the impact of four privacy advertising videos and one privacy game published by five different technology companies. We conducted 24 semi-structured interviews with participants randomly assigned to view one or two of the videos or play the game. Our findings suggest that awareness of privacy features can contribute to positive perceptions of a company or its products. The ads we tested were more successful in communicating the advertised privacy features than the game we tested. We observed that advertising a single privacy feature using a single metaphor in a short ad increased awareness of the advertised feature. The game failed to communicate privacy features or motivate study participants to use the features. Our results also suggest that privacy campaigns can be useful for raising awareness about privacy features and improving brand image, but may not be the most effective way to teach viewers how to use privacy features. 
\end{abstract}


\maketitle
\begin{tcolorbox}
This document is the author’s manuscript for a paper published in Proceedings on Privacy Enhancing Technologies 2024(4) available at: \url{https://petsymposium.org/popets/2024/popets-2024-0126.php} 
\end{tcolorbox}
\section{Introduction} \label{sec:intro}
In recent years, an increasing number of companies have used their privacy practices as a way of differentiating themselves from competitors. Apple launched a video advertisement focusing on the iPhone \quotes{Ask App Not to Track} feature in May 2022. Experts estimate this new iPhone feature caused a 12 billion dollar decline in revenue for Meta~\cite{oflaherty_2022}. The advertisement shows the protagonist, Ellie, stumbling upon an auction where her information is up for sale. As Ellie toggles various privacy protection features on her iPhone, bidders begin to disappear until she is the only one left in the auction room (see~\autoref{fig:apple}, and~\autoref{fig:apple_extended} in~\autorefappendix{app:screen} for illustration)~\cite{apple_2022}. Apple is not alone in showcasing their privacy protection features as a way to gain consumer trust; Samsung, another smartphone giant, launched an advertisement with a similar message in the same year~\cite{samsung_2022}. Google and WhatsApp also have ads highlighting their security and privacy features~\cite{google_2020,whatsapp_2022}, and Twitter released an old-school video game in an attempt to help users gain a better understanding of their privacy policies~\cite{damien24,twitter_2022}. 

While experts have praised some of these privacy ad campaigns~\cite{oflaherty_2022,ulanoff2022} and criticized others~\cite{hurler2022,winkie2022}, it is unclear what consumers take away from these campaigns. In this study, we explore what consumers exposed to corporate privacy ad campaigns are learning from these campaigns, and what impacts these campaigns have on consumers’ privacy awareness and attitudes. 

We conducted 24 semi-structured interviews to evaluate privacy marketing materials released by tech firms (Twitter, Apple, Samsung, WhatsApp, and DuckDuckGo) from a consumer perspective. We explored consumers’ attitudes towards the organizations providing the campaign before and after exposure to the campaigns, and each campaign's impact on consumers' overall privacy awareness. We also evaluated the actionability of the privacy and security advice suggested by the campaigns.

Our exploratory study suggests that awareness of privacy features can contribute to positive perceptions of a company and its products. The ads we tested were more effective than the game in communicating privacy features to our participants. We observed that advertising a single privacy feature using a single metaphor in a short ad increased awareness of the advertised feature among our participants. The game performed poorly in both explaining privacy features and encouraging user engagement with the features. Our results suggest that privacy campaigns might not be the most effective method for educating users on how to use privacy features. In addition, despite their positive impressions, most non-users in our study were reluctant to switch products after seeing ads due to concerns about switching costs, data synchronization, and familiarity with their current operating system. Nonetheless, over time, such campaigns may lead to more interest in a brand's products and might be particularly effective for products and services with low switching costs.

\section{Related Work} \label{sec:related}
In this section, we outline previous research about video advertisements and elements that may impact ad effectiveness. We also introduce research on serious games related to privacy and security. Next, we discuss research on privacy awareness and research on acceptance of privacy and security advice. 

\subsection{Effective Advertisements}
In her book, \textit{Advertising by Design,} Landa described three steps for building any brand image using advertisement: \quotes{get people's attention, keep their attention, call them to action.} She added that video advertisements should also be either entertaining or informative and \quotes{be interesting enough to be viewed again or be shareworthy}~\cite{landa2021}.

Video advertisements contain scenes and imagery carefully crafted by advertisers. Mohanty and Ratneshwar examined the factors that influence a viewer's subjective comprehension of visual metaphors in graphic ads. The researchers found that incongruity, which is the discrepancy between what is actually shown and what it is referring to, could lead to poor comprehension~\cite{mohanty2015}. Mohanty and Ratneshwar also conducted a later study investigating the effects of incongruity on ad effectiveness, and found that a moderate level of incongruity results in the most effective ads, recommending that advertisers be cautious when selecting visual metaphors~\cite{mohanty2016}.

Some video ads are narrative ads that contain a story or plot. Kim et al. found that narrative ads were effective because they generated emotive responses, offered high entertainment value, and created credibility~\cite{kim}. Laurence conducted a study researching the role of storytelling in narrative ads and found that when used to induce positive emotions, participants had more positive brand attitudes~\cite{laurence}. Manyiwa and Ross looked at the impact of negative emotions in advertising and found that participants with higher reported self-efficacy had more favorable views towards ads that used fear~\cite{manyiwa}.

Privacy marketing has the potential to reach a large audience but prior work has not examined its effectiveness. Our research seeks to gain insights into consumer privacy awareness by examining recent privacy marketing campaigns and what consumers are taking away from them. We investigated how the privacy campaigns impacted viewers' impressions of the organizations and how well participants understood privacy metaphors.

\subsection{Privacy and Security Games}
As one of the campaigns in our study was a web-based video game aimed at better acquainting players with the organization's privacy policy, we explored previous research on privacy and security games. Whether tabletop or online, games have long been used as educational tools.

Denning et al. designed \quotes{Control-Alt-Hack} as a classroom tabletop card game aimed to increase security awareness. While participating educators reported that the game increased levels of security awareness, critiques found the game too hard to play and not very fun~\cite{denning2013}. Barnard-Wills and Ashlenden received similar feedback on their card game \quotes{Privacy,} which participants felt was too difficult to play. The researchers noted that educational privacy games should not be designed as a \quotes{graphical skin overlaid on [an] existing game} and recommended that the game mechanisms should actively demonstrate the theoretical model of the \quotes{online privacy ecosystem,} allowing players to better understand the privacy model~\cite{barnard2015}.

Video games can be an effective way of communicating a simplified, simulated model of any concept to a player~\cite{persuasive}. Sheng et al. developed the \quotes{Anti-Phishing Phil} game and found it to be an effective way to educate people to identify phishing URLs. The researchers credited the game's effectiveness to its interactive content~\cite{sheng2007}. Maqsood et al. investigated whether web-based games could help children aged 11 to 13 practice safer online behaviors. They found significant improvements in knowledge and intended behaviors after being exposed to the game and found that the game was \quotes{usable, fun, and relatable}~\cite{maqsood2018}. Google also launched the \quotes{Interland} game aimed at educating children about online safety, along with YouTube videos and other online resources to help parents and teachers raise these issues with children~\cite{moscaritolo}. Thompson and Irvine evaluated \quotes{CyberCiege,} a security game that teaches students basic cybersecurity concepts. They found that students did not treat the game as a lab session, often overlooking the lab manual provided and instead playing straight away. Researchers noted this shift in mentality is something they would keep in mind for future renditions of \quotes{CyberCiege}~\cite{thompson2011}.

\subsection{Privacy Awareness}
Major themes that emerged from the campaigns we selected for our study were social media privacy settings, online privacy and behavioral advertising, and end-to-end encryption in instant messaging. In this section, we explore privacy awareness research related to each theme.

Acquisti and Gross conducted one of the earliest studies about privacy awareness and information-sharing practices on Facebook and found that while most participants claimed to be aware of their profile visibility, there were \quotes{a significant minority of members [who] are unaware of those tools and options}~\cite{acquisti_2006}. Another Facebook privacy study, conducted by Sohoraye et al., focused on users’ knowledge of privacy statements and privacy regulations in their countries~\cite{sohoraye_2015}. The researchers found that while many people are concerned about their personal information, they are relatively unaware of related privacy information~\cite{sohoraye_2015}. Users' lack of awareness of social media privacy settings could lead to misaligned expectations between what they intended to share and reality~\cite{liu2011,madejski2012}. 

Common social media privacy concerns include personal safety, reputation management, social disclosure, and how easy it may be for other users to see their activities~\cite{ahern2007}. Depending on the platform and its intended audience, users have different privacy concerns. Jeong and Kim found that Facebook users have more privacy concerns related to what others may post on their timeline, versus Twitter users who are more concerned about the privacy of tweets they have posted~\cite{jeong2017}. Baruh et al. found that participants' privacy concerns did not significantly correlate with their actual social media usage~\cite{baruh2017}. Building on top of Baruh et al's findings, Barth et al. investigated the privacy paradox and found that users prioritized functionality and usability above possible privacy concerns, regardless of their technical ability or financial situation~\cite{barth_2019}.

Increased privacy awareness may also lead to increased concerns about data used for behavioral advertising. Researchers have found that while users are generally aware of tracking, their understanding of how personalized advertising works may not reflect reality~\cite{rader2014}. Personalized ads and behavioral advertising are considered useful features, but also ones that may bring privacy risks~\cite{ur2012} or become a nuisance~\cite{dehling2019}.

When it comes to instant messaging, security and privacy may have a minor influence on why users choose a particular app. DeLuca et al. found that overall, peer influence had the biggest impact on user choice in messaging apps~\cite{deluca2016}. Abu-Salma et al. interviewed 60 participants to investigate barriers to secure communication tool adoption and found that most participants lacked understanding about end-to-end encryption, potentially limiting their motivation to adopt secure messaging tools~\cite{abu2017}. Other barriers to adoption could include misconceptions that SMS or email may be more secure~\cite{stransky2021l}, or lack of trust in end-to-end encryption~\cite{dechand2019}. Stransky et al. investigated how visualization may impact the perception of messaging security and found that the perception of security depended more on the reputation of the app itself~\cite{stransky2021l}. When attempting to improve user understanding of end-to-end encryption, Bai et al. found that it is most effective to use simple wording with minimal technical detail~\cite{bai2020}.

Privacy advertising campaigns are designed to increase privacy awareness, which can impact consumer choices. Kelley et al. examined privacy as part of a user's decision to download or install an app. The researchers found that users seldom considered privacy information when making a decision to download apps. However, when the authors introduced privacy information into the app store, users were much more likely to consider privacy as a factor in their decision-making process~\cite{kelley}. Tsai et al. found that when privacy information was prominently visible, people were willing to pay a premium for privacy~\cite{tsai2011}. Similarly, Emami-Naeini et al. found that consumers were willing to pay more for Internet-of-Things products with better privacy and security practices, such as de-identified cloud storage~\cite{emami}. 

\subsection{Acceptance of Privacy and Security Advice}
Redmiles et al. developed a set of quality metrics for security and privacy advice: perceived actionability, perceived efficacy, and comprehensibility. The researchers found that generally, while all study participants found the advice actionable, neither the users nor the experts were particularly good at prioritizing advice~\cite{redmiles_2020}. Research on security awareness campaigns also helps to identify key factors that may influence the success or failure of a campaign. Bada et al. evaluated security campaigns across the UK and Africa. The researchers found that security education \quotes{needs to be targeted, actionable, doable and provide feedback,} and that invoking fear is an ineffective way of promoting security awareness~\cite{bada2015}. Subsequently, Das et al. conducted a comprehensive literature review on research related to security and privacy awareness. Based on their review, the researchers developed the Security and Privacy Acceptance Framework (SPAF), identifying three key barriers that could impact a user’s acceptance or rejection of recommended security and privacy advice: awareness, motivation, and ability~\cite{das_2022}.

Technology has also affected the way in which people learn new information. Kross et al. surveyed a nationally representative sample of US adults and found that online safety, security, and privacy are three of the 19 categories of subjects people tried to learn more about through online resources. The researchers identified YouTube as the most commonly used learning channel~\cite{kross_2021}. Akgul et al. analyzed influencer videos on YouTube containing VPN ads. They found that \quotes{VPN ads likely reach billions of viewers, comparable to ambitious industry efforts at influencing users’ understandings of security and privacy tools.} However, their analysis of these ads suggested that the information they contain may not always be correct and may potentially have negative impacts on viewers' understanding of online safety~\cite{akgul_2022}.

\section{Methods} \label{sec:method}
We conducted an exploratory interview study, aimed at investigating the impact of privacy marketing materials on privacy perception. Specifically, we explored how privacy marketing materials, such as privacy ads and games, impact viewers' attitudes towards the organizations providing the campaign, awareness of privacy features and issues, and the actionability of suggested privacy advice. In the subsections below, we first introduce the privacy marketing campaigns that were selected for this study, then outline the recruitment and screening procedure, as well as the interview protocol and data analysis method. 

\subsection{Privacy Marketing Materials}
We searched for privacy marketing materials on YouTube and the web using the keywords \quotes{Privacy Advertisements,} \quotes{Privacy Ads,} and \quotes{Privacy Awareness.} We found four campaigns that were published in 2022 by technology companies. We selected one video ad from each of the four ad campaigns we found. We also selected the Twitter DataDash game, which received some media attention in 2022. Here we provide a short summary of each campaign. Table \ref{priv-mkt} provides more information about each campaign. Additional images from the campaigns are presented in~\autorefappendix{app:screen}.

\subsubsection{Data Auction by Apple.} 
\begin{figure}[t!]
    \centering
    \includegraphics[width=\linewidth]{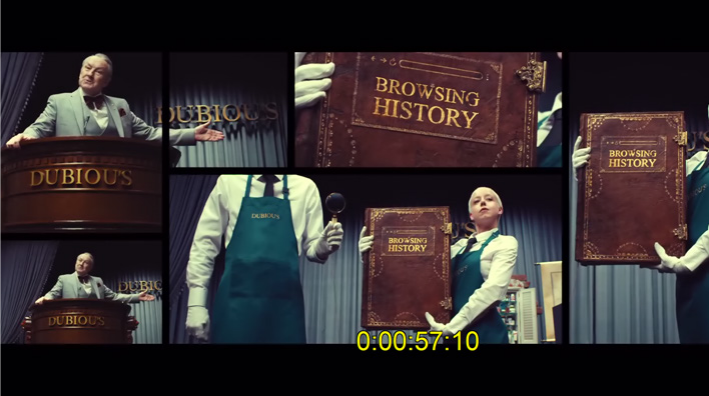}
    \caption{The auctioneer facilitating the auction of the protagonist's data in the Apple ad.}
    \label{fig:apple}
\end{figure}

The protagonist stumbles upon a room packed with people where an auctioneer is selling off her private information to the highest bidder, as shown in~\autoref{fig:apple}. As the information for sale becomes more sensitive, she decides to take action. Using Apple's \quotes{Ask App Not To Track,} she eliminates half of the bidders in the room. Next, by using Apple's \quotes{Protect Mail Activity,} she empties the room, including all remaining bidders, the auction attendants, as well as the auctioneer~\cite{apple_2022}.

\subsubsection{You’re in Control by Samsung.} 
\begin{figure}[t!]
    \centering
    \includegraphics[width=\linewidth]{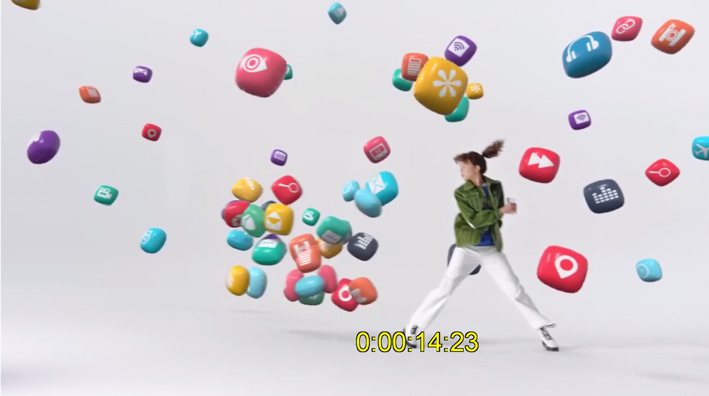}
    \caption{The protagonist struggles to get away from balloons representing apps in the Samsung ad.}
    \label{fig:samsung}
\end{figure}

The protagonist knows that much of her life is on her Samsung phone, where she struggles to get away from floating balloons representing apps, as shown in~\autoref{fig:samsung}. Utilizing Samsung's privacy dashboard, she takes control of her privacy. By using Samsung's permission manager, she turns off the precise location, illustrated with a yellow umbrella from the Weather app, to blend into a sea of yellow umbrellas. As she continues using her phone, she chooses to grant or deny location and microphone permission for apps as needed. Using Samsung’s privacy dashboard, she is able to review all app permission usage in a central location. Samsung’s Knox Security is mentioned at the end of the video~\cite{samsung_2022}.

\subsubsection{A New Era of Personal Privacy with Default End-to-End Encryption by WhatsApp.} 
\begin{figure}[t!]
    \centering\includegraphics[width=\linewidth]{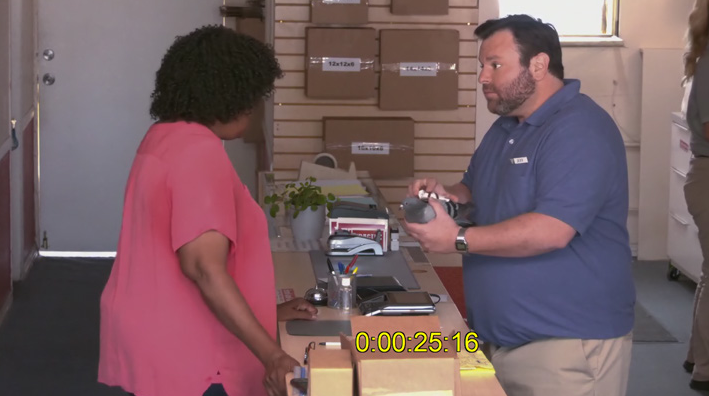}
    \caption{The post office clerk (right) uses a carrier pigeon for sending letters in the WhatsApp ad.}
    \label{fig:whatsapp}
\end{figure}

People arrive at the post office wanting to mail their letters. To their disbelief, the clerk, as shown in~\autoref{fig:whatsapp}, insists on sending those letters using a messenger pigeon, assuring his customers that using pigeons as a delivery method is \quotes{mostly} secure. The ad states that 5.5 billion texts per day are sent without encryption, while with WhatsApp, their conversations would not be one of them~\cite{whatsapp_2022}.

\subsubsection{Watching You by DuckDuckGo.} 
\begin{figure}[t!]
    \centering
    \includegraphics[width=\linewidth]{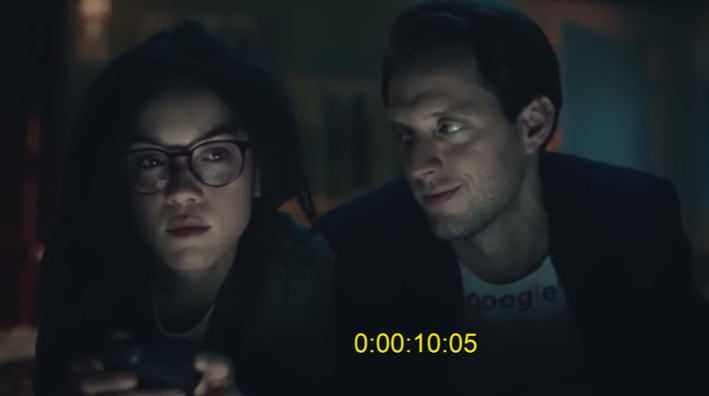}
    \caption{The singer wearing a Google T-shirt (right) shoulder-surfs a person browsing the Internet in the DuckDuckGo ad.}
    \label{fig:ddg}
\end{figure}

A singer in a Google T-shirt singing an altered version of the song \quotes{Every Breath You Take}~\cite{everybreath23}, where lyrics were changed to be relevant to web tracking, e.g., \quotes{Every click you take ... I will be watching you.} As the singer shoulder-surfs people while they browse the Internet, they are visually uncomfortable with the singer’s presence, as shown in~\autoref{fig:ddg}. One person decides to start using DuckDuckGo, causing the singer to be dragged away by an invisible force. An announcer explains that \quotes{The Internet does not have to be so creepy} and introduces DuckDuckGo products~\cite{duckduckgo_2022}.

\subsubsection{DataDash by Twitter.} 
\begin{figure}[t!]
    \centering
    \includegraphics[width=\linewidth]{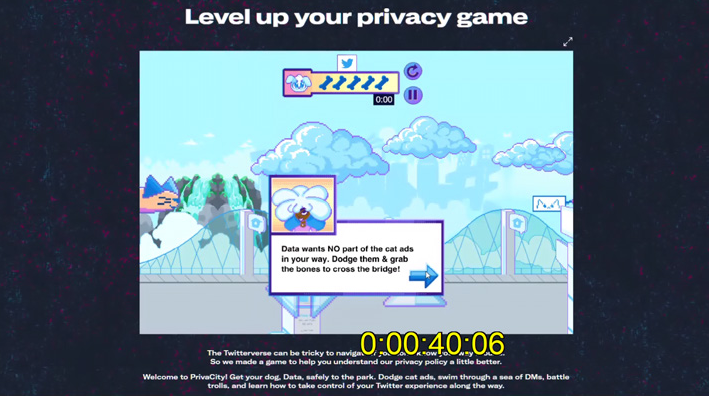}
    \caption{The start of level 1 of the Twitter game.}
    \label{fig:twitter}
\end{figure}

DataDash is an online game simulating the style of 90s old-school platform games, where players journey through PrivaCity with data dog as illustrated in~\autoref{fig:twitter}. Players must collect five artifacts to pass to the next level. The artifacts for levels 1 and 2 look like envelopes containing dog bones. At the end of each level, players are shown a short explanation referring to a particular privacy and safety feature, with a \quotes{Twitter Settings} button that takes them to a specific privacy and safety setting on Twitter. The privacy and safety settings for each level are: \begin{inparaenum}[(1)]\item Ad Preference, \item Direct Messages, \item Location Information, and \item Audience and Tagging\end{inparaenum}. The website where DataDash is hosted explains the purpose of the game: \quotes{The Twitterverse can be tricky to navigate if you don't know your way around. So we made a game to help you understand our privacy policy a little better}~\cite{twitter_2022}. 

\begin{table*}[bt!]
    \centering
    \footnotesize
    \caption{Campaign Details.}
    \begin{threeparttable}
    \begin{tabular}{l|p{3cm}|p{3cm}|l|l}
        \hline
            \textbf{Company} & \textbf{Campaign name} & \textbf{Release Date} & \textbf{Length} & \textbf{Source}\\
            \hline
            Apple & Data Auction & Apple's YouTube channel on May 18, 2022 & 1 min, 35 sec & \url{https://www.youtube.com/watch?v=NOXK4EVFmJY} \\
            \hline
            Samsung & You're in Control & Samsung's YouTube channel on Feb 9, 2022 & 1 min, 35 sec & \url{https://www.youtube.com/watch?v=kZKK80urUZc}\tnote{1}\\
            \hline
            WhatsApp & A New Era of Personal Privacy with Default End-to-End Encryption & WhatsApp's YouTube channel on Oct 17, 2022 & 1 min & \url{https://www.youtube.com/watch?v=zvI4cVGWJhM}\tnote{2} \\
            \hline
            DuckDuckGo & Watching You & DuckDuckGo's YouTube channel on May 17, 2022 & 30 sec & \url{https://www.youtube.com/watch?v=QWpPyYlZXNI} \\
            \hline
            Twitter & DataDash & A blog post by Twitter's Privacy Center~\cite{damien24} on May 11, 2022 & Not Applicable & \url{https://twitterdatadash.com} \\
        \hline
    \end{tabular}
    \begin{tablenotes}
    \item[1] As of June 2024, the original URL is no longer available. An identical video is available at: \url{https://www.youtube.com/watch?v=MayOX9FuvCE}  
    \item[2] As of June 2024, the original URL is no longer available. An identical video is available at: \url{https://www.ispot.tv/ad/25bs/whatsapp-a-new-era-of-privacy}
    \end{tablenotes}
    \end{threeparttable}
    \label{priv-mkt}
\end{table*}

\subsection{Recruitment and Screening}
We recruited participants for our study using convenience snowball sampling and Prolific, a research-oriented online crowdworking platform~\cite{prolific23}. To reduce privacy and security priming, we presented the study as a study on \quotes{Consumer Perspective on Tech Advertising.} We omitted any words related to privacy or security in the study title and purpose, letting the campaigns deliver relevant privacy components naturally. We also did not mention any company name related to the campaigns when we advertised the study. Participants were compensated 1.25 USD for completing a screening survey and those who subsequently participated in the 25-minute interview were compensated 10 USD for their time. 

We designed four treatments to test the five campaigns: Apple, Samsung, Twitter, and WhatsApp-DuckDuckGo. Except for Twitter which was an online game, all other campaigns were short videos ranging from 30 to 95 seconds. WhatsApp and DuckDuckGo ads did not introduce actionable privacy features that we could ask participants about, and were the shortest of all video campaigns, with their combined video length being roughly the same as that of the Apple and Samsung videos. In the interest of maintaining similar interview lengths across all treatments, WhatsApp and DuckDuckGo were grouped together.

We required eligible participants to be aged 18 or above, fluent in English, and located within the United States (the primary target location for the versions of the campaigns we tested). We administered a screening survey that collected demographic information and information about participants’ product or service usage. We used purposive sampling to select from among those who qualified for our study and assign balanced treatment groups. The screening survey is in~\autorefappendix{app:survey}.

For Apple, Samsung, and WhatsApp-DuckDuckGo treatment groups, we screened for 50\% users and 50\% non-users of the relevant product or service since those campaigns appeared to be aimed at both users and non-users of the advertised products and services. The Twitter treatment group was comprised of 100\% Twitter users with some form of online and/or digital gaming experience. We defined users as those who utilize the product or service featured in the campaign at least once every week. Due to the low interview response rate from eligible DuckDuckGo participants, we relaxed the qualifying user condition from using DuckDuckGo \quotes{At least once a week} to \quotes{At least once a month.} We also asked participants about their technical backgrounds and distributed those with technical backgrounds evenly across treatment groups. We defined technical background as a degree or work experience in one or more of the following areas: Computer Science, Information Systems, Information Technology, and Computer Engineering. 

For the Twitter treatment, we screened for Twitter users who participate in some form of digital gaming. We added this requirement after Twitter pilot participants without gaming experience repeatedly mentioned that they would like to stop playing due to frustration. We did not include non-users in this treatment, as the game was designed to inform existing Twitter users rather than interest non-users in using Twitter.

Initially, we screened 181 participants from Prolific using our screening survey and invited 133 eligible participants. Only 24 participants responded to the invitation, including seven who rejected the invitation, and three who missed their appointment. Thus, only 14 participants completed both the survey and the interview. Due to the low response rate from Prolific participants, we also recruited via snowball sampling using email invitations. The first author emailed acquaintances unaware of the research topic using the same information as had been presented to Prolific participants. These acquaintances were also invited to refer others who might be interested. Using this method, we recruited 10 participants. In total, we recruited 24 participants for this study, split evenly into four treatment groups. We ensured that for each treatment group, there were no more than three participants who indicated they had technical expertise. All participants who were recruited through Prolific were paid via the platform, whereas snowball participants were compensated at the same rate, but via Amazon gift cards.

\subsection{Interview Procedure}
We developed a semi-structured interview procedure, with an estimated completion time of 25 minutes. All interviews were conducted over the Zoom video conferencing platform and recorded and transcribed by Zoom. Participants were asked to keep their cameras off throughout the interview. Interviews were conducted between March 31 and April 19, 2023.

First, we asked about participants' general impression of the companies, then screen-shared the advertisements into Zoom for participants to watch. For Twitter DataDash, we sent each participant Twitter login details and the game link. Once  participants had the game open, they were asked to share their screen over Zoom and given 8 minutes of DataDash playtime while the interviewer observed them. In the other treatments, participants spent up to 95 seconds viewing a video advertisement. We asked participants a series of questions aimed at gauging their awareness of privacy features that were shown in the campaign, as well as discussing any privacy and security concepts mentioned by participants. We asked  participants for their impression of the companies again after exposure to the campaigns to see whether the campaign had an impact on this. Then, we asked participants to describe how actionable they thought the privacy advice provided was, and self-rate their motivation and confidence in being able to complete the steps shown in the ad or game. Next, we moved to the actionability exercise where we asked all participants except those in the WhatsApp-DuckDuckGo treatment to find a specific setting that was shown in the ad or game, with a total activity time of 3 minutes. Finally, for ad treatment groups, we re-exposed participants to the ad and asked for concluding remarks and final observations.

Since the WhatsApp and DuckDuckGo videos did not contain other actionable steps aside from switching to the advertised service, we did not include the actionability section of the interview for that treatment, but instead repeated the entire interview procedure for the second ad. The order of the advertisements shown (WhatsApp first, then DuckDuckGo, or vice versa) was randomized. 

The main purpose of the actionability section of the interview was to assess perceived ease of actionability, measured through participants' description of what steps they thought the advertisement had recommended, as well as self-ranked motivation and confidence. We shared an iPhone 12 Pro, a Galaxy A13 smartphone, or the Twitter home page over Zoom. Participants were asked to direct the interviewer on where to go to find the appropriate settings. Before the study, we had restored both smartphones’ desktops to factory default desktop settings.

Video advertising campaigns are generally designed with the assumption that viewers will see ads multiple times~\cite{landa2021}. Therefore, in the three ad treatments, we re-exposed participants to the ads again at the end of the interview to see if this would generate any new insights.

Before recruiting participants for our study, we conducted eight pilot interviews, two per treatment group. We made several revisions to our protocol based on issues discovered during the pilots. Pilot interviews are not included in our results and analysis.

\subsection{Data Analysis}
Interview transcripts were generated automatically using Zoom’s built-in transcription function and manually reviewed with the voice recordings by a member of the research team. Two researchers performed qualitative analysis for each participant interview transcript using Template Analysis, a style of thematic analysis that combines both inductive and deductive coding~\cite{king24, brooks14}. Template Analysis emphasizes hierarchical coding without specific requirements regarding the number of levels and what levels represent. To analyze the data, both researchers first familiarized themselves with the data. Next, the first researcher created an initial codebook (the template) and coded all the data. Then, the second researcher reviewed and refined the codes applied by the first researcher. Both researchers met over multiple sessions to discuss any disagreements in the coding and adjusted the codebook (added, updated, and deleted codes) until they reached a satisfactory version of the codebook, agreeable to both researchers. Our coding started with broader themes, such as \quotes{attitudes,} \quotes{awareness,} and \quotes{actionability,} which encompass narrower themes, such as \quotes{organization impression pre/post-exposure,} to more specific ones, such as \quotes{positive,} \quotes{negative,} etc. For the actionability section (task), after the research team agreed on standard criteria (also provided in~\autorefappendix{app:codebook}) to evaluate task completion, one researcher recorded whether participants completed the activity, and then recorded the participants' self-rated scores regarding motivation and confidence. A diagram summarizing the coding hierarchy is provided in~\autorefappendix{app:diagram}. The codebook along with the specific interview question(s) from which each set of codes were distilled is provided in~\autorefappendix{app:codebook}.

\subsection{Ethical Considerations}
This study protocol was approved by CMU's Institutional Review Board (IRB). Before completing the screening survey, we asked all participants to read and acknowledge their acceptance of the consent form. We asked all interview participants to keep their cameras off. Before participants in the Twitter treatment were asked to share their screen over Zoom we reminded them to avoid accidental information sharing. Since the Twitter game actionability activity required users to log in and navigate Twitter's account settings, participants were provided with an account by the researcher and were not asked to log in using their personal accounts. A member of the research team removed participant names from transcripts along with any personal information participants shared in the interviews. 

Although one of the authors has previously received research funding from two of the companies whose campaign materials were used in this study, we received no funding from any company for this project. The authors, and to the best of our knowledge, the participants, are not associated with any of the companies.

\section{Results} \label{sec:result}
In this section, we discuss the main findings of this study. We use the first letter from each company name to indicate which treatment group the participants were in, followed by the participant number. While our study is qualitative, we provide frequencies of themes to illustrate how common each theme was in our dataset.

\subsection{Participant Demographics}
We interviewed 24 participants, including 13 males, and 11 females. Participants' ages ranged from 18 to 65 years old. Of the 24 participants, 10 indicated they have a technical background. All Apple and Samsung participants who were asked about their smartphone use said they actively use smartphone apps and browse the Internet using their smartphones. All but one DuckDuckGo participant said they actively browse the web and perform search queries using search engines. For more details about participant demographics, refer to~\autoref{tab:demo} in~\autorefappendix{app:p-info}.

\subsection{Attitudes and Impressions of Companies}
We asked participants about their impressions of the companies relevant to their treatment group both before and after viewing the videos or playing the game (Q.5 \& Q.14 in~\autorefappendix{app:interview}). We reviewed the transcripts of these responses and categorized the sentiments they expressed before watching the ads or playing the game into five categories: positive, neutral, negative, mixed-positive (more positive than negative), and mixed-negative (more negative than positive). We classified participants who mentioned an equal number of positive and negative points based on whether the first point they mentioned was positive or negative. We used the same five codes to analyze responses after viewing the ads or playing the game, with two additional codes: positive-plus and negative-plus, to categorize responses that indicated an increased degree of positivity or negativity. We then compared each participant's sentiment before and after viewing the ad or playing the game, and classified the change as: more positive, more negative, or remained the same. 

Our results suggest that most of the campaigns we studied were associated with brands that our participants already viewed fairly positively, except Twitter. Exposure to information about privacy features further improved some participants' impressions of these companies, although, for a few participants, the campaigns worsened their impressions as they viewed claims about privacy in the campaigns with skepticism. The Twitter game also worsened the impression of some participants due to both poor gameplay experience and unclear messaging.

\subsubsection{Attitude and Impression of Companies \textbf{Before} Exposure}
Prior to campaign exposure (Q.5), our results show that most participants had positive or mixed-positive impressions of the companies relevant to their treatments for all treatment groups except Twitter. Among the participants in the Twitter treatment group, only one participant was positive (T6) and one was mixed-positive (T3). 

The top reasons for having a positive attitude towards an organization across all groups were related to having a good product or having a privacy-focused product. Some participants who were not users of a product still remarked that they thought the product was good (A2, S2, S6, WD6 for WhatsApp and DuckDuckGo). S2, a Samsung non-user, remarked: \quotes{I think that they come out with top-notch products. I think I would be almost as happy having a Samsung Android type phone as I would having an iPhone.} DuckDuckGo received the most mentions of privacy-focused aspects as a reason for positive attitudes towards the company, mentioned by both users and non-users. For example, WD5 said they have \quotes{positive feelings about [DuckDuckGo]} due to \quotes{the privacy features.}

Twitter had the most negative impressions, with two negatives (T1, T2) and one mixed-negative impression (T4). Participants mentioned false news (T1, T4) and lack of moderation on the platform (T1, T2). From T2's perspective, \quotes{what Twitter used to be, was not perfect, but tolerable. And now it's just a single rich person's plaything.} It is interesting to note that when discussing their views of Twitter, three participants (T2, T4, T5) either directly mentioned Twitter owner Elon Musk, or implied that the leadership change had some impact on their impression of the organization. One Apple participant (A4) mentioned high prices as the reason for their negative impression of the company. A4 explained: \quotes{an iMac will cost on average, about twice as much ... as a given Windows machine [of] approximately equivalent capability.}

\subsubsection{Attitude and Impression of Companies \textbf{After} Exposure}
After being exposed to the campaigns, 12 of the 30 responses\footnote{WhatsApp responses and DuckDuckGo responses are counted separately, once for each campaign.} reflected the same impression as before while 13, including some participants from every group, expressed more positive attitudes (Q.14). Seven of these participants, including three in the WhatsApp group, cited becoming more informed about privacy features as the reason they became more positive. S3 said that he now felt \quotes{a little bit more comfortable with the privacy features on my phone, and also as soon as we're done here, I’m probably gonna to open up my phone.} S3's response also highlighted another reason that led to improved impression: an increased motivation to use the product. Similarly, WD3 said that the ad made her a \quotes{bit more curious ... I have no idea what it has to offer, and how it would compare to Google.} 

Participants also recognized that organizations may promote privacy as a feature of their products and services. A6 described feeling more positively towards Apple after viewing the ad as \quotes{it shows that they are using data privacy as part of their company brand.} A4, whose impression of Apple stayed the same, explained that the ad didn't \quotes{change my image of Apple so much ... this is the kind of feature [that] ... demonstrate the unique value of its services to users ... This is something you give to users in order to like convince them that their money is well spent by sticking with you.} Participants generally discussed promoting privacy as a feature in a positive or neutral tone.

There were also five participants (from Twitter, DuckDuckGo, and WhatsApp) who, after being exposed to the campaign, had a more negative view of the organization. Skepticism was the top reason for having a worsened impression, mentioned by three participants (WD2, WD4 for DuckDuckGo, WD4 for WhatsApp). Some of those participants questioned the truthfulness of the message and whether the organizations would deliver the messages the ads portrayed. WD2 wanted to do more research on DuckDuckGo, because \quotes{well, how does it make money?} In addition, prior knowledge of negative news about the company such as data breaches could also lead to skepticism. For example, WD4 was skeptical about both WhatsApp and DuckDuckGo due to his exposure to news about those companies. WD4 felt \quotes{Slightly more negative} about DuckDuckGo due to \quotes{what I've read online because they're still reinforcing this ... this silver bullet of search engines,} likely referring to the Microsoft DuckDuckGo tracker incident~\cite{ddgnews_2022}. WD4 admitted that he \quotes{wasn't diligent enough about following up what happened with WhatsApp about a year ago,} likely referring to the WhatsApp data breach~\cite{whatsappnews_2021}. 

For the two Twitter participants who had a more negative view of the organization after being exposed to the campaign, their reasoning stemmed from a poor experience with the DataDash game itself. T1 remarked that \quotes{privacy and safety issues are a very important issue, and and it seems like this, this game didn't really do a good job of taking it like very seriously ... I think there are a lot more better ways to educate than this game ... I think if they did it again have like more real-life implications, and it was more tied to real examples of what can happen with the privacy and data.} T5 had a much more sinister take on what the game represented, concluding that Twitter \quotes{probably [has] no issue selling our data and like exploiting our privacy.} From T5's understanding, \quotes{the bones [represented] the data pieces, or like information ... the dog was probably just someone scooping up data pieces.} 

\subsection{Communicating Privacy}
To assess how well each campaign communicated about privacy, we asked participants to describe the campaign they had just viewed or experienced, how they felt toward the ad or game, and what they perceived as the purpose of the campaigns (Q.7, Q.8, Q.9). From these responses, we explored how well the campaigns increased awareness of privacy features and issues, how well participants understood the privacy metaphors used in the campaigns, and their understanding of the purpose of the campaign.

Overall, our findings suggest that participants were able to grasp broader privacy themes communicated in the campaigns. In general, after being exposed to the campaign, participants noticed at least one main privacy feature showcased, and their perception of the privacy metaphors mostly aligned with that of the research team. The WhatsApp campaign, which appeared to be most successful at communicating a clear message, was short, communicated a single feature (end-to-end encryption), and used a simple metaphor (pigeon to represent unencrypted SMS). On the other hand, Twitter's DataDash game was the only campaign that faced considerable challenges in communicating its privacy feature, with some participants failing to recognize the privacy connection until we explicitly asked about it.

\subsubsection{Privacy Awareness}
Of the 30 responses collected, 25 discussed at least one of the privacy features or functions that were shown in the advertisement or game. The privacy features shown inspired participants to make specific comments related to privacy, with 14 responses explicitly mentioning the word \quotes{privacy} in their replies. 

In the Apple ad, two features were displayed: \quotes{Ask App Not to Track} and \quotes{Protect Mail Activity.} When we asked participants to describe the campaign in their own words, in addition to the general synopsis of the campaign, most Apple participants (A1, A2, A4, A5) included a detailed description of the \quotes{Ask App Not to Track} feature. A5 said: \quotes{the ad is about how ... using an Apple device can help protect your own data, because there are lots of other apps that could track and use your data, and through Apple, you can say that I don't want you to track it.} However, there was only one mention of the second feature shown in the ad \quotes{Protect Mail Activity.} A1 described the protagonist as \quotes{going through and selecting options on her iPhone, asking apps not to track her or protecting her email data.} Participants who did not describe either feature (A3, A6) spoke about the ad in much broader terms, with A3 summarizing the features shown in the ad as \quotes{apps or certain settings on the phone to where you keep all your data private and secure.} Most Apple participants (A2, A3, A4, A5, A6) mentioned issues surrounding personal data, data sharing, or selling.

For Samsung, the campaign included five privacy features: privacy dashboard, location permissions, microphone permissions, camera permissions, and Samsung's Knox Security. The campaign mostly featured the privacy dashboard and location, microphone, and camera sharing permission controls. All participants in this campaign talked about Samsung's permission controls in general. The location sharing control feature captured the most attention, with four participants (S2, S3, S5, S6) mentioning it in their response, while the camera sharing control feature was mentioned by two participants (S5, S6), and the microphone sharing control was mentioned by one (S6). S6 summarized the ad as \quotes{advertising the privacy options that Samsung gives you. Specifically, the option to see which apps are using which features such as camera, microphone, location, and you could fine-grain control how they're using it.}

Each level of Twitter's DataDash game ends with a button that directs the user to the privacy setting relevant to that game level. All participants experienced level 1 with some progressing to latter levels. No participant clicked on the \quotes{Twitter Settings} button at the end of any level, which would have led them to the corresponding privacy setting for that level. When we asked participants to describe the campaign, we had not yet asked any questions containing the words \quotes{privacy} or \quotes{security.} Thus, two Twitter participants (T3, T5), had yet to realize the game was related to privacy. They only made the connection between DataDash and Twitter’s privacy practices later in the interview, after the interviewer asked: \quotes{Were you aware of Twitter’s `Privacy and Safety' features prior to playing DataDash?} Immediately after the interview was over, T5 said that she did not realize there was a privacy connection until halfway through the interview (with her permission we added this comment to her transcript). For the remaining four participants who noticed DataDash's connection to privacy, there were only two participants who mentioned privacy features in their description of the game. T4 said she felt the game was trying \quotes{to show that there are settings to protect ... I guess the purpose of the game was to show that I have more control over my data than I realize.} T6 described the purpose of DataDash as \quotes{to learn more about Twitter’s features through like, like a simple game.} Of all groups, Twitter participants mentioned privacy the least, with T1 commenting that \quotes{I felt like it was a little bit confusing, and it was hard to understand exactly what the game is representing ... It gave me an impression that it can be getting dangerous to be on Twitter with all the different elements of privacy concerns.}

The focal point of the WhatsApp campaign was end-to-end encryption. For example WD3 finished her summary of the ad with this sentence: \quotes{in the end, we find out that 5.5 billion messages are sent in an unsecure way. But if you use Whatsapp, that will not be the case.} When we asked participants to give a summary of the campaign, encryption as a feature was recognized by nearly all participants in this group. WD6, the only participant who did not discuss encryption in WhatsApp specifically, still mentioned unencrypted messages: \quotes{[it] kind of draws that parallel to people sending mail regularly versus just sending unencrypted instant message.} All WhatsApp participants mentioned the security or safety of instant messages when discussing the campaign. For example, W5 mentioned \quotes{having messages encrypted versus unencrypted using a delivery pigeon.}
 
When we asked DuckDuckGo participants to describe the ad, all participants discussed tracking and DuckDuckGo's privacy-friendly search engine and browser. WD4 explained: \quotes{search engines can see everything you type in and all the results that yield from them, and DuckDuckGo is a solution for that.} However, only two participants (WD1, WD2) were specific about the no-tracking feature. WD1 explained: \quotes{Google keeps track and monitors your search history, and DuckDuckGo doesn't.} All participants mentioned DuckDuckGo's privacy-protective features generally. WD3 described being able to \quotes{perform ... similar search queries on [DuckDuckGo] platform, but your privacy won't be impacted.} WD5 was the only one who realized the ad also mentioned the DuckDuckGo mobile app, saying: \quotes{I didn't know before that there was an app for DuckDuckGo.}

We also asked whether participants were aware of the advertised feature (e.g., \quotes{Ask App Not to Track}) or function (e.g., encryption or web tracking) prior to watching or interacting with the campaign, and how they came across it (Q.13, Q.13B). 18 out of 30 responses indicated prior awareness, including 7 who were non-users. Participants commonly reported that they became aware of a privacy feature's existence after being alerted by the app upon first use (A1, A3, S1, S3, S6) or later when using an app (A2, S6, T2). S6 mentioned both types of prompts in her answer: \quotes{it asks you repeatedly when you install it, and then when you use the app.} Others discovered the features through exploring the settings (A2, S2, S4), while some became aware of the feature or concept as it was prominently displayed in the app or website (T5, WD3 for WhatsApp regarding encryption). Participants also reported learning about new features and concepts as a result of updates or policy changes (A5, S5, T4, WD5 regarding the DuckDuckGo web tracking feature). T4 described the types of notifications she received from Twitter: \quotes{when they update the terms. It comes up as like a pop-up on the app. Or sometimes, if they change the terms fully, they send an email.} Interestingly, participants also reported learning about new features and concepts through their usage of other services and apps. WD5 first became aware of encryption through \quotes{us[ing] VPN services in the past, and also PGP.} WD6 described learning about what encryption is through seeing \quotes{a lot of advertising ... on Youtube for VPN like TurboVPN and things along those lines.}

\subsubsection{Participants' Understanding of Privacy Metaphors}
As all of the campaigns use metaphors to explain privacy concepts, we investigated what participants understood about these metaphors and whether they were useful in illustrating these concepts. We selected one prominent symbol from each campaign and asked participants to share what they thought it might represent (Q.10). We wanted to see how the selected symbol was perceived by participants for each campaign, and whether participants of the same campaign felt similarly about what the symbol could represent. Overall, most participants' impressions of these metaphors align with the impressions of the authors of this paper, suggesting that the metaphors are largely effective in communicating to non-experts a similar message as they communicate to experts. However, there were occasions where participants failed to notice the privacy metaphor at all or had a non-privacy-related understanding. Two of the metaphors seemed fairly difficult for our participants to interpret correctly.

When we asked participants in the Apple treatment group about the auctioneer, all of them gave responses related to companies selling personal data. Participants said that the auctioneer either represented data sales (A2, A3, A6) or other parties such as Internet Service Providers (ISP), smartphone companies, or app companies (A1, A4, A5). A1 explained that \quotes{what they would are really hoping is that you associate ... the auctioneer with other cell phone companies ... they're implying that if you are with another cell phone [company] ... that phone company will be able to sell your data.} A5 made similar comments, theorizing that \quotes{the phone company here could be Apple, or it could be Samsung.}

On the other hand, when we asked participants in the Samsung treatment group about the yellow umbrella the protagonist used to hide herself from exposing her exact location, only one participant (S3) said that the umbrella \quotes{was trying to show ... whether you're giving gross location data or precise location.} Others felt that the umbrella represented a warning (S2, S5) or protection of information in general such as location or activity (S1). Two participants (S4, S6) had no idea what the umbrella represented, with S6 saying she \quotes{didn't notice at all.}

We asked participants in the Twitter treatment group about the envelopes containing bones that were collected in level 1 and level 2 of the game and received the most diverse interpretations among the campaigns we studied. Two participants (T1, T5) associated the envelopes with data collection, with T5 saying the envelope represented \quotes{pieces of data.} Others felt the envelope could be a tweet (T2), spam (T4), or opening a direct message (T6). T3 had the most unique take, stating that he \quotes{kind of thought it was a cat, but maybe there was something deeper with that I don't know.} (Indeed the co-authors were not in agreement, and one of our co-authors also thought the envelopes looked like cats.) 

Participants had a mostly uniform understanding of the carrier pigeon in the WhatsApp video, generally themed around instant messaging or the communication channel used. Some thought the pigeon represented SMS message text (WD1, WD2) or unsafe messages (WD3, WD5). Others felt the pigeon referred to the communication platform (WD3) or the network (WD4), as well as outdated technology (WD1, WD6). 

DuckDuckGo's ad followed a singer wearing a Google T-shirt shoulder surfing others while they browsed the Internet. Most participants said the singer represented Google (WD2, WD3, WD5, WD6) or the search engine (WD4). 

\subsubsection{Participants' Perceived Purpose of the Campaign}
Participants' thoughts about the main purpose of the campaigns (Q.9) can be categorized into four groups: advertising privacy services or creating a more privacy-focused corporate image, promoting the company as being different from its competitors, providing information, and entertainment.

Using the campaign to advertise privacy services or create a privacy-focused company image was the most common perceived purpose, with 24 responses across all campaigns, including all participants in the Apple, Samsung, and DuckDuckGo groups, citing this as the purpose of the campaigns. A1 described the main purpose of the Apple ad as \quotes{to sell iPhones, to present them as the most secure option for protecting your identity and your data from unauthorized use.} Participants also felt companies were using these campaigns to advertise themselves as being privacy-focused. S6 explained that \quotes{the main purpose of video is to make the statement that Samsung very much cares about your privacy.}

Others also felt that the campaigns were issued by the companies to promote themselves as being better than their competitors. We saw 15 comments regarding this across four campaigns (Apple, Samsung, WhatsApp, and DuckDuckGo). All participants in the WhatsApp group cited this as the reason behind the campaign, with WD3 stating: \quotes{I think it was purely to get more users to switch on to WhatsApp, instead of using other probably less safe platforms with instant messaging.} Most participants in this group also viewed the purpose of the DuckDuckGo campaign as differentiating themselves from competitors. Five DuckDuckGo participants cited this as the main reason behind the campaign. For example, WD6 said: \quotes{the main purpose of the video was to show consumers that there is another option or a search platform if they feel like Google is invading their privacy.}

Participants also picked up on potential education efforts and the informative nature of the campaigns, with 13 responses across all groups commenting on this intent. Four Twitter participants (T1, T2, T4, T6) said that DataDash was trying to inform them of some sort of privacy feature or information about Twitter. T1 said he felt \quotes{the main purpose of the game was ... intended for education.} T2 echoed T1's response but was worried the message might have gotten lost: \quotes{I think it was to educate people about how Twitter uses our data, but I’m not quite sure it would get the message across correctly because I think people would be so focused on actually [navigating the game] and getting the objects and scoring the points just to beat it that they might miss the entire point of the game.} Indeed, both T3 and T5 failed to recognize the privacy connection between DataDash and Twitter until the interviewer asked a question regarding Twitter's privacy and safety features. Instead, T3 and T5 thought the purpose of the game was entertainment. T3 described the game as \quotes{a diversion ... something people would do to pass time} and T5 was \quotes{not super sure} about the intent of the game \quotes{but it was fun.} The Twitter DataDash website explained that the game was designed \quotes{to help you understand our privacy policy a little better ... and learn how to take control of your Twitter experience along the way.}

\subsubsection{Impact of Re-Exposure}
Since the campaigns we chose were all released in the year before our study, we also asked whether participants had previously seen the ad or played the game before the study (Q.6). Only four participants (A5, WD4 \& WD5 for WhatsApp, WD2 for DuckDuckGo) had previously seen the campaigns. Of those four, three participants came across the campaign while watching TV or using a streaming service, with one participant (WD2) stating he saw the ad after having \quotes{read an article about ... the Super Bowl ads.}

Previously exposed participants were able to pinpoint specific details that might be overlooked by others the first time they viewed the campaign during the study. For example, WD4 commented on a scene in the WhatsApp campaign, stating: \quotes{it's really funny how the user that they zoomed in on while texting was using iMessage, which to my knowledge, is end-to-end encrypted.} In addition, A5 commented: \quotes{it's also very clever that Apple and its other marketing puts the phone over the person's face, which is also another way of signaling that Apple is a barrier between what you are doing on the Internet and what people can see about you.} A5, who was recruited via snowball sampling, also proactively reached out after the interview and provided a screenshot of similar Apple billboard marketing she had seen in her neighborhood, showing the phone-over-face imagery. With her permission, we included this in our study data.

For non-previously exposed participants, upon second viewing, they were also able to pick up on more details. After re-watching the DuckDuckGo ad, WD6 remarked: \quotes{I don't know if I missed something the first time, but I feel like I saw more of the advertisement than I did the last time.} S5, a Samsung non-user, commented on a new detail she observed when watching the ad again regarding how Android features seem \quotes{more granular in terms of they're literally giving you a list of things that are tracking you.} 

\subsection{Actionability}
Three of the campaigns we investigated offered actionable advice related to privacy features in their product or service. We explored participant perceptions of both perceived (\autoref{action_perceived}) and actual (\autoref{action_actual}) actionability of the Apple, Samsung, and Twitter campaigns. We then observed whether participants exposed to all campaigns reported any motivation to switch or continue using the products (\autoref{sec:motiv_switch}). 

Overall we found that the campaigns were not all that effective at teaching viewers how to use privacy features. When we asked participants to direct the interviewer to the privacy features, they seemed to rely on their prior knowledge of the OS and intuition about how to use the interface rather than the information provided in the campaign. Some features, such as Apple's \quotes{Ask App Not to Track,} are presented differently depending on whether the user is opening a newly installed app or searching through settings. Thus, the interface shown in the campaign may not exactly match what a user sees when they try to use the feature. In addition, whether participants could easily find the relevant Twitter privacy settings seemed to have little to do with their experience playing the game.

\subsubsection{\textbf{Perceived} Actionability of Privacy and Security Advice} \label{action_perceived}
Before the activity, participants were asked to rank their confidence in being able to perform the steps or go to the setting shown in the campaign on a 5-point scale, with 1 being \quotes{Not at all confident} and 5 being \quotes{Fully confident} (Q.18). Apple participants rated themselves as a 4 or 5 in confidence, with most Samsung participants rating themselves at 5. Most Twitter participants rated their confidence at 3. Participants were also asked to rate their motivation to perform the steps shown in the ads or game on a 5-point scale, with 1 being \quotes{Not at all motivated} to 5 \quotes{Extremely motivated} (Q.17). Most Apple participants ranked themselves at 3 or 5, and the majority of the Samsung participants ranked themselves at 4 or 5. Half of the Twitter participants ranked their motivation at 1. These results suggest that a negative interaction with the privacy campaign could diminish the participants' confidence and motivation to enact privacy protections as recommended by the campaigns.

\subsubsection{\textbf{Actual} Actionability of Privacy and Security Advice} \label{action_actual}
\begin{figure}[!tbp]
\captionsetup[subfigure]{font=large}
\centering
\resizebox{\columnwidth}{!}{%
\begin{minipage}[b]{\textwidth}
    \begin{minipage}[b]{0.49\textwidth}
        \subfloat[Apple's \quotes{Ask App Not to Track} feature.\label{fig:activity_apple_01}]{%
        \includegraphics[width=\textwidth]{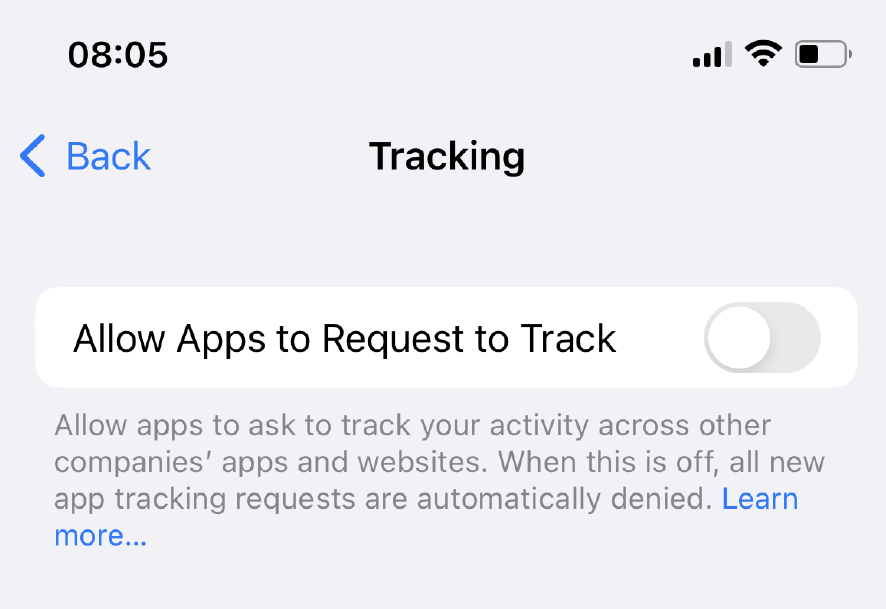}
        \Description{activity-apple-01}
        }
    \end{minipage}
    \vspace{2pt}
    \hfill
    \begin{minipage}[b]{0.49\textwidth}
        \centering
        \subfloat[Apple's \quotes{Protect Mail Activity} feature.\label{fig:activity_apple_02}]{%
        \includegraphics[width=\textwidth]{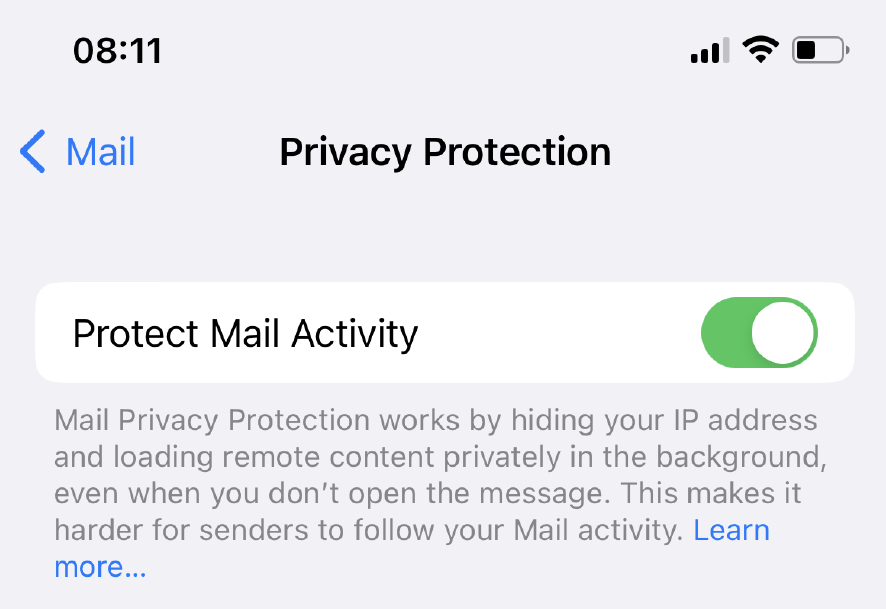}
        \Description{activity-apple-02}
        }
    \end{minipage}
\end{minipage}    
}
\caption{Apple \quotes{Ask App Not to Track} and \quotes{Protect Mail Activity} features.} \label{fig:apple_activity}
\end{figure}

\begin{figure}[!tbp]
\captionsetup[subfigure]{font=large}
\centering
\resizebox{\columnwidth}{!}{%
\begin{minipage}[b]{\textwidth}
    \begin{minipage}[b]{0.49\textwidth}
        \subfloat[Samsung location permission for the weather app.\label{fig:activity_samsung_01}]{%
        \includegraphics[width=\textwidth]{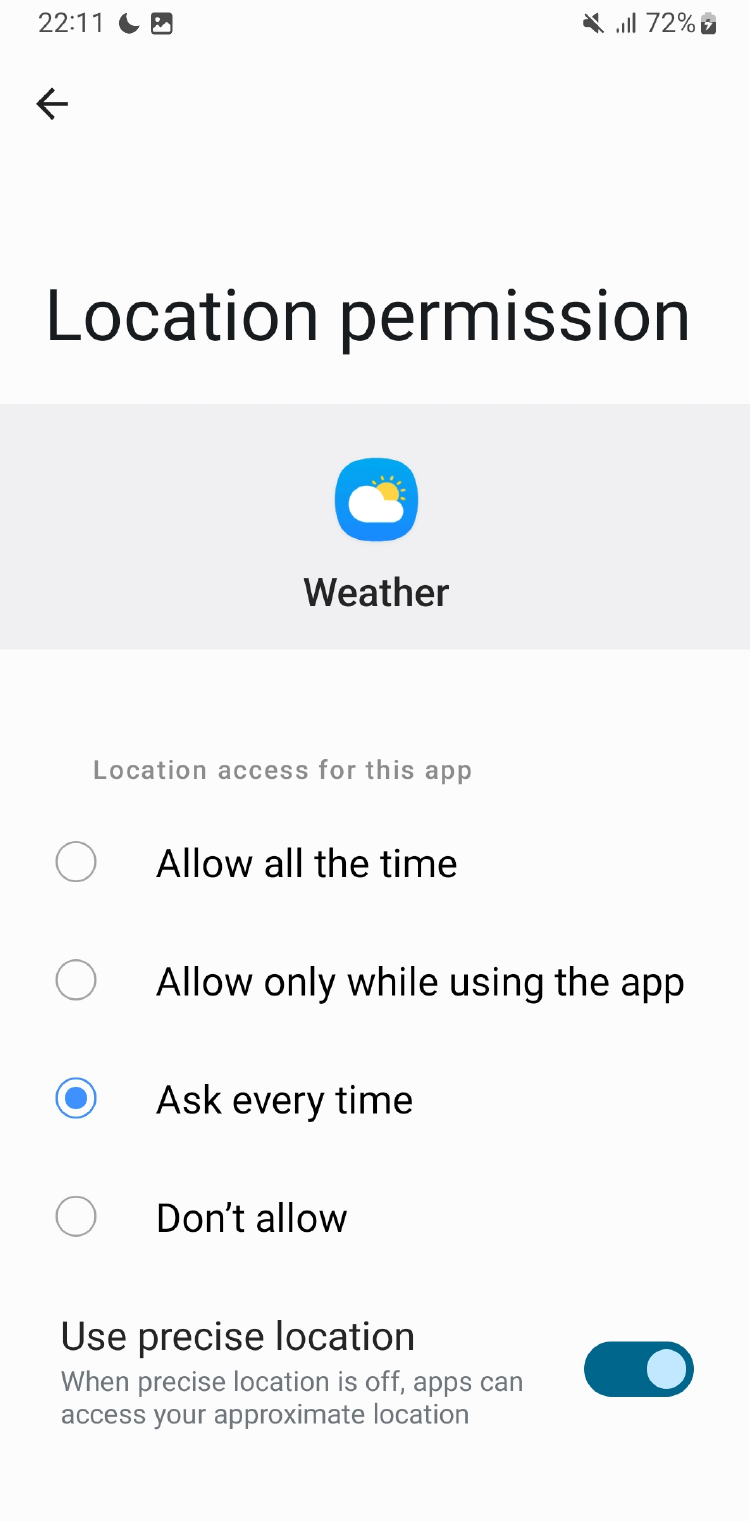}
        \Description{activity-samsung-01}
        }
    \end{minipage}
    \hfill
    \begin{minipage}[b]{0.49\textwidth}
        \subfloat[Samsung microphone permission for the Zoom app.\label{fig:activity_samsung_02}]{%
        \includegraphics[width=\textwidth]{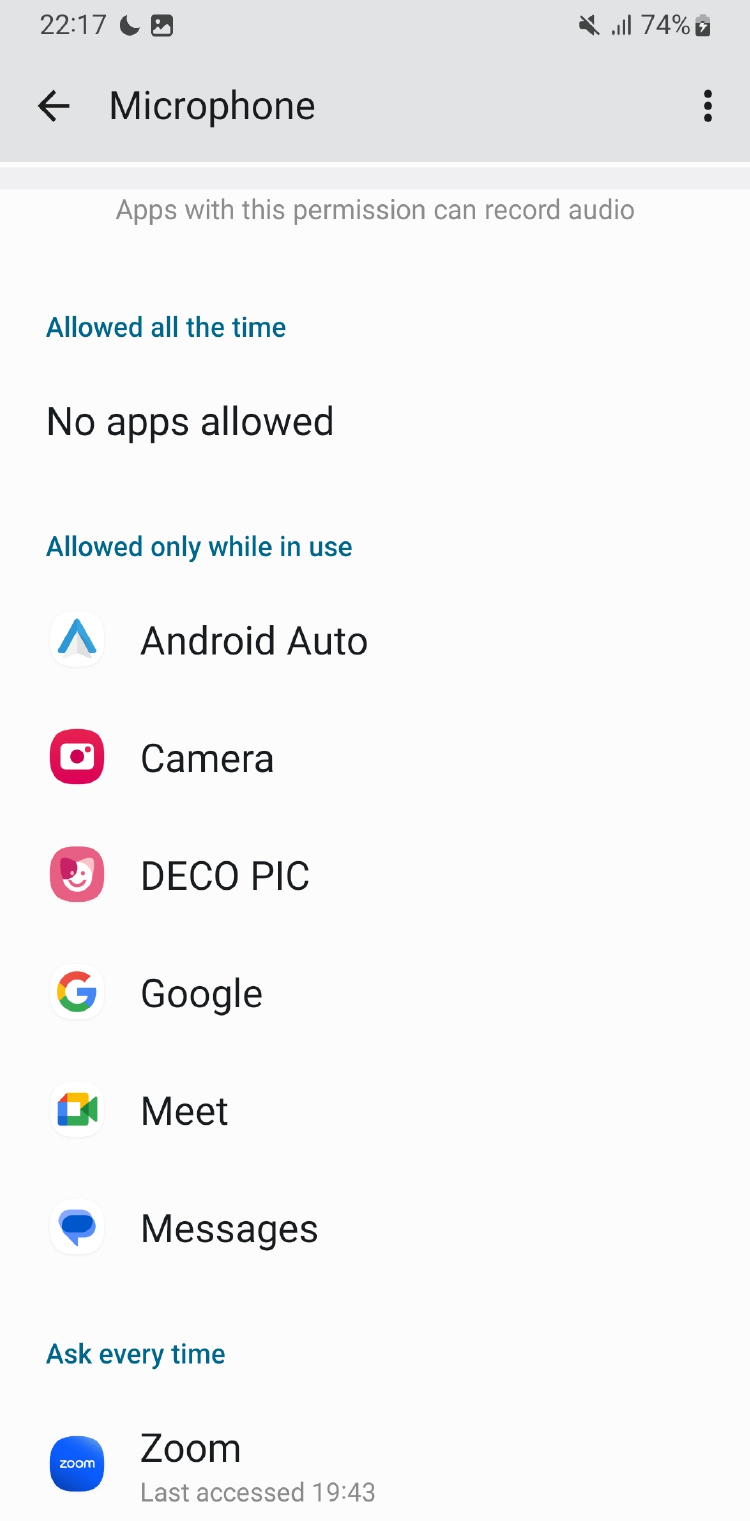}
        \Description{activity-samsung-02}
        }
    \end{minipage}
\end{minipage}    
}
\caption{Samsung app's location and microphone permission control features.} \label{fig:samsung_activity}
\end{figure}

\begin{figure}[!tbp]
\centering
\resizebox{\columnwidth}{!}{%
\includegraphics[width=\columnwidth]{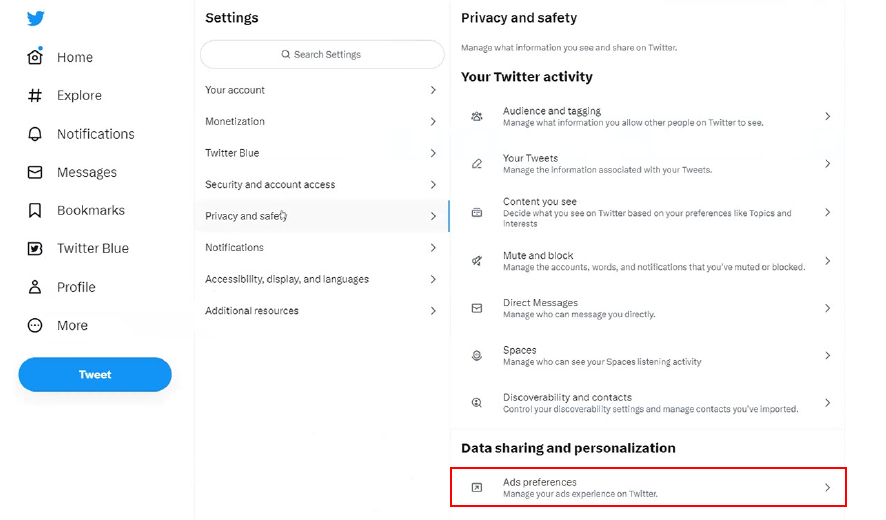}
}
\caption{Twitter's \quotes{Ads preferences} setting.} \label{fig:twitter_activity}
\end{figure}

To test the actual actionability of privacy and security advice in the campaigns, participants were asked to direct the interviewer to a particular privacy setting or feature that was shown in the campaign (Section 6 \& Section 7 of the interview). We asked participants in the Apple treatment group to turn on the \quotes{Ask App Not to Track} and \quotes{Protect Mail Activity} features, shown in \autoref{fig:apple_activity}. We asked participants in the Samsusung group to change location permissions for the Weather App and change microphone permissions for the Zoom app, as shown in~\autoref{fig:samsung_activity}. Finally, we asked Twitter participants to go to the \quotes{Ad Preference} setting, as shown in~\autoref{fig:twitter_activity}.

All Apple participants were able to find the \quotes{Ask App Not to Track} setting eventually. However, only two participants (A5, A6, both users) were able to successfully complete the task without help. During this process, A3 requested help twice to locate the setting, and A4 used Google to find Apple’s online documentation. In addition, both users and non-users (A1, A2, A3) did not realize that they had completed the task. The \quotes{Ask App Not to Track} feature was presented in the ad as a pop-up that is prompted when opening a newly installed app for the first time. However, it did not show how someone who had already used an app would find the setting, which would have been more consistent with the scenario shown in the video since the protagonist had already installed the apps that were tracking her. Within Apple settings, the setting can be found under Settings $\rightarrow$ Privacy and Security $\rightarrow$ Tracking. This discrepancy was noticed by A5 (a user), who knew she had found the correct setting, adding: \quotes{I didn't know that there was a settings page to manage it. But it does make sense.} 

Only one Apple participant (A1) was able to confidently locate the \quotes{Protect Mail Activity} setting without help. One participant (A3) did not continue due to having used up their total allocated activity time (3 minutes for the two activities). All remaining participants were able to perform the activity, but either required help or were unsure they had completed the task. During this activity, two participants, a non-user (A4) and a user (A6), both consulted Google to find Apple’s documentation for this feature, with A6 also bringing out their own devices to navigate in a more familiar setting. One participant (A5, user) was also unsure whether she had reached the correct setting, and was quickly reassured when the feature had the same name as shown in the ad. 

All Samsung participants were able to change the location permission for the Weather App. However, only S5 and S6 were able to complete the activity without help. S1 was unsure that she had completed the activity. The device that was used for the Samsung group was a Galaxy A13, and after the desktop was returned to factory settings, the main setting icon was not immediately available on the home screen. The setting icon becomes visible either when one swipes down on the screen, revealing a small gear on the top right corner, or when one swipes up revealing the master setting icon. This default configuration stumped S2, S3, and S5. S3 was a self-described \quotes{loyal Samsung customer} who had purchased at least three flagship Galaxy phones in the past. S2 requested help, and S3 used Google twice in an attempt to locate Samsung’s master setting icon. While the setting icon was not available on the home screen, the Google search bar was. S5 directed the interviewer to type \quotes{Setting} into the Google search bar and the OS displayed a \quotes{From Your Apps: Setting} icon at the bottom of the search, allowing S5 to get into the main setting and progress from there. After the activity, S3 commented: \quotes{if I had never seen the Samsung app ad before I would have gone to the right place first. Because I saw the Samsung ad, I wanted to go to the Samsung privacy first to see if it was in there and then backed out back to the other one.} Both S2 and S3 were unable to continue through to the next activity due to having used up their total allocated activity time.

Only two Samsung participants (S1, S6) were able to change microphone permissions for Zoom without help. S4 completed the activity with help. S5 used the Google search bar as she did for the prior activity. However, this time when \quotes{Mi} was entered, the icons displayed under \quotes{From Your Apps} did not include \quotes{Microphone,} and S5 was unable to complete the activity.

Finally, for the Twitter group, all participants were able to locate the \quotes{Ad Preference Setting,} with T3 and T4 doing so confidently and without help. However, there were minimal connections between participants' ease of being able to carry out the privacy advice and what they learned playing the game. T3, T5, and T6 all credited Twitter's intuitive design, with T5 stating: \quotes{That was actually pretty easy. I didn't know that that was a setting but I feel like where I found it, or where we found it was pretty straightforward.} 
\subsubsection{Motivation to Switch or Continue Using the Products} \label{sec:motiv_switch}
At the end of the interview, we asked users if the campaigns motivated them to continue using their product (Q.22), and for non-users, we asked if the campaigns motivated them to consider switching to the product shown (Q.21).

Eight of the 18 participants who are already users of the device or service said the campaigns motivated them to continue or increase their current usage. A1 said the campaign had no impact on her choice of smartphones, as her decision is based on \quotes{other factors.} She said she would stay with Apple because: \quotes{I'm not really interested in switching to another brand because it would mess everything up. I've got my iCloud calendars and ... all that would have to be redone. It would be a huge pain, so I would have to really be motivated by something to try to switch from Apple.} Motivations for continued or increased usage include already being a user and familiarity. For example, S4 explained he is motivated to continue using Samsung smartphones because he is familiar with them. Other participants who reported that the campaign motivated them to continue or increase usage referred to feeling reassured by the privacy-focused message the company is conveying, with A6 explaining that \quotes{[by] bank[ing] their brand on being the data privacy brand ... even if [Apple is] not selling my data today, they're making a public commitment to be the data privacy brand.}

All Twitter participants except one (T4) felt that the game did not motivate them to continue using Twitter. Two participants followed up their answers with comments critical of the game. For example, T3 stated that he felt the game \quotes{wouldn't drive me to action to do anything.} T2 was much more critical of the game, going on to say that \quotes{the game doesn't quite have bearing on why I use Twitter to begin with, it might just be something that a few devs made in their spare time over the course of a couple weeks or months, and they decided to put a data protection spin on it to get it in front of more people.} 

The campaigns only motivated two non-user study participants to consider switching (S2, WD6 for WhatsApp). When asked what motivated them to consider switching, S2 stated: \quotes{It seemed like it was a simpler process based on the ad the way the phone was set up.} WD6 said the campaign \quotes{has me thinking about potentially using that to communicate with some of the people I currently talk to on Snapchat.} In addition, the WhatsApp and DuckDuckGo campaigns inspired a few non-users (WD3, WD6 regarding DuckDuckGo, WD5 regarding WhatsApp) to look into what these services have to offer. WD3 explained that the advertisement \quotes{encouraged me to look into what [DuckDuckGo] is, but not to switch, just to explore what it is.}

Non-users who are not interested in switching offered similar reasoning to users, with S5 saying that she would not switch due to \quotes{exit cost for me as an iOS user to switch to an Android operating system, like moving all the contacts, or like photos, or like even memos ... with the current iOS system, I can already turn on all the features that I am aware of, and it's just easier for me as a user to keep using my current operating system.} Again, data synchronization and familiarity with their current device and app were credited as the main reasons not to switch. S5 and S6 highlighted the fact that they have similar features in their current smartphones, and the ad did not motivate them to incur the costs of switching from their current smartphone.

\section{Limitations} \label{sec:limitation}
Our study's small sample size, with 6 participants per group, limits our ability to draw quantitative conclusions or compare across campaigns and participant characteristics. However, our sample size is typical of qualitative interview studies that focus on surfacing themes rather than drawing quantitative conclusions~\cite{distler}.

In the screening survey, we did not ask participants whether they are affiliated with any of the five companies whose campaigns were selected for this study. However, participants' open-ended answers did not indicate any affiliation and we are not aware of any participant affiliated with any of the companies mentioned in the study.

During the activity phase of the interview, we observed some signs of the Hawthorne effect~\cite{hawthorne}, with one Twitter participant stating that they chose not to click on the Twitter setting button as they thought we would like them to focus on the game. Similar observations were also seen in other groups, where non-users often opted not to ask for help even when they were stuck. While we gave reassurance throughout the interview by asking participants to act as they normally would, participants may not have acted so. After the first few participants, we began listing examples of \quotes{normal} actions, such as using Google or looking on one’s own device, which somewhat alleviated this problem.

We also observed some signs that participants in the WhatsApp-DuckDuckGo treatment group were primed by the first video when watching the second video. For example, we observed instances where participants would answer a question about campaign 2, by comparing it to campaign 1. This is mitigated somewhat by the fact that we randomized the order in which the videos were presented to participants.

\section{Discussion} \label{sec:discussion}
Our exploratory study provides insights into the effectiveness of five privacy-focused marketing campaigns. We distilled themes from our interviews that led to six actionable recommendations for future privacy campaigns. Although the campaigns we studied were all for commercial products, most of our takeaways are likely also applicable to privacy education and awareness campaigns from privacy advocacy groups and educational organizations.

\textbf{Increased privacy awareness leads to positive impressions.} Our findings suggest that awareness of privacy features contributes to positive perceptions of a company or its products. Campaigns that increased awareness of privacy features improved some participants' impressions of the associated companies, although they had no impact on the impressions of other participants. In our study, being privacy-focused was credited as a reason for a positive impression of a brand both before and after exposure to the campaigns. This shows that building a privacy-focused brand could help build positive impressions. Being suspicious of a company's privacy practice post-exposure to the campaigns could also lead to a negative impression, such as T5's takeaway from DataDash being that Twitter would sell user data. 

\textbf{Avoid invoking skepticism.} Campaigns should strive to avoid raising skepticism from viewers and avoid giving conflicting messages. Participants who voiced skepticism towards a company usually expressed concerns about the organization not being able to live up to claims advertised in the campaign, such as questioning whether DuckDuckGo could make money without tracking browsing history. Skepticism towards an organization seemed to be amplified if the participant had prior knowledge about the organization, such as exposure to negative news (e.g., data breach), which may conflict with the message that is conveyed in the campaign.

\textbf{Be careful with visual metaphors.} In designing privacy awareness advertisements using visual metaphors, companies need to ensure that viewer comprehension of the metaphor will not arouse negative connotations. One example is Apple's auctioneer being perceived as representing smartphone companies and possibly Apple itself. This could become a double-edge sword that causes viewers to question the company's intent as they can associate the negative connotation as portrayed in the campaign with the organization that issued the campaign. Another more obvious example is when DataDash presented Twitter as a company that the user needed to protect themselves from, which likely goes against any intent Twitter had with launching DataDash. Prior work in the advertising literature cautions advertisers to be careful when selecting visual metaphors, and our results support this~\cite{mohanty2015,mohanty2016}. Creating positive imagery becomes particularly important when viewers are likely to be exposed to the campaign more than once. For those who were previously exposed, continually reinforcing a message can be effective in building a privacy-focused brand image or allowing viewers to pick up on details that may not match the campaign's message.

\textbf{Effective privacy ads are short and to the point.} We found that the most effective privacy awareness ads were those that were short, focusing on one privacy feature, using a metaphor that is easy to comprehend. This is consistent with Bai et al.'s finding that the concept of encryption was best communicated when using simple wording with little technical detail~\cite{bai2020}. In our study, we see that participants often walked away with impressions of one main feature. For example, all Apple participants noticed the \quotes{Ask App Not to Track} feature that was shown first, while only two noticed the \quotes{Protect Mail Activity} feature shown later. Similarly, most participants discussed the location permission control feature in the Samsung ad that was shown first, while only a few noticed the camera and microphone permission control features shown later. WhatsApp and DuckDuckGo campaigns were both short and addressed a single feature. Participants seemed to focus their attention on this feature (Encryption for WhatsApp and Private Browsing for DuckDuckGo) and had a good understanding of the campaign messages. Twitter's DataDash is on the opposite end of the spectrum, where too many abstract elements clouded the campaign's messages.

\textbf{DataDash was not effective in privacy education.} Our results suggest that DataDash was not an effective privacy education game, mainly due to a lack of connection between the game and Twitter's privacy features and settings. To use games to educate players about privacy and security concepts, the game should be fun, relatable, and interactive~\cite{sheng2007,maqsood2018}. While some of our participants found DataDash fun, participants had trouble relating the game content to Twitter's privacy features. Barnard-Wills and Ashenden advised against reducing privacy education games to a \quotes{graphical skin overlaid on [an] existing game,} with little connection between the gaming mechanisms and online privacy systems~\cite{barnard2015}. Our results support their view, suggesting that DataDash likely fell under this category, with little connection between the game mechanism and Twitter's privacy features.

\textbf{Privacy marketing may not be the most effective way to educate users about how to use privacy features.} We found that privacy ads and marketing campaigns may not be the best way to show viewers \textit{how} to use privacy tools and features. The activity segment of the interview highlights the importance of having demonstrations in the ad to match what a user will experience. This may be difficult when user experiences may vary. On the other hand, if the campaign makes users aware of a feature and conveys that it is easy to use, users may be motivated to figure out how to use that feature on their own. 

To make privacy advice more actionable, what is shown in the campaign should remain closely aligned with the actual product or service. From the Apple activity, we see that the difference between the \quotes{Ask App Not to Track} pop-up users receive after installing an app and the \quotes{Allow Apps to Request to Track} setting can lead to confusion should viewers decide to explore the settings as shown. Given participants' ease of finding the settings advertised relied on their familiarity with the OS and whether the settings interface was intuitive, it highlights the importance of alignment between the feature advertised and the actual feature. 
 
\textbf{Privacy campaigns may be most effective for products that don't incur switching costs and for building positive brand associations over time}.
Privacy campaigns may not motivate non-users to switch, but could motivate some non-users to explore the product, and could motivate users to continue usage. Participants said they were not motivated to switch products and pointed out that switching costs can be high, especially for smartphones. However, if people associate a particular brand with privacy, it may sway them to switch brands the next time they need to replace their device. In addition, for products with lower switching costs that can be used on a trial basis without giving up an existing service (e.g., messaging apps), privacy campaigns may have a more immediate impact, encouraging viewers to try a service.

\section{Conclusion} \label{sec:conclusion}
We conducted an interview study to explore what participants took away from five privacy campaigns. Participants gained an increased awareness of privacy features in the products mentioned in the video ad campaigns. However, participants exposed to campaigns that demonstrated features being used did not gain a clear understanding of how to use those features. In addition, the game did not communicate privacy features effectively to our participants, who largely viewed the game as more entertaining than educational. The campaigns that used short, focused videos with a simple metaphor seemed to communicate most effectively about privacy issues and how consumers might protect themselves. 

\section{Acknowledgment} \label{sec:ack}
Eman Alashwali acknowledges the financial support of the Ibn Rushd Program at King Abdullah University of Science and Technology (KAUST). This work was funded in part by the Innovators Network Foundation. The authors thank the participants for their time and valuable insights. 

\bibliographystyle{ACM-Reference-Format}
\bibliography{consumer-ref-2}


\begin{thebibliography}{56}


\ifx \showCODEN    \undefined \def \showCODEN     #1{\unskip}     \fi
\ifx \showDOI      \undefined \def \showDOI       #1{#1}\fi
\ifx \showISBNx    \undefined \def \showISBNx     #1{\unskip}     \fi
\ifx \showISBNxiii \undefined \def \showISBNxiii  #1{\unskip}     \fi
\ifx \showISSN     \undefined \def \showISSN      #1{\unskip}     \fi
\ifx \showLCCN     \undefined \def \showLCCN      #1{\unskip}     \fi
\ifx \shownote     \undefined \def \shownote      #1{#1}          \fi
\ifx \showarticletitle \undefined \def \showarticletitle #1{#1}   \fi
\ifx \showURL      \undefined \def \showURL       {\relax}        \fi
\providecommand\bibfield[2]{#2}
\providecommand\bibinfo[2]{#2}
\providecommand\natexlab[1]{#1}
\providecommand\showeprint[2][]{arXiv:#2}

\bibitem[Abu-Salma et~al\mbox{.}(2017)]%
        {abu2017}
\bibfield{author}{\bibinfo{person}{Ruba Abu-Salma}, \bibinfo{person}{M~Angela
  Sasse}, \bibinfo{person}{Joseph Bonneau}, \bibinfo{person}{Anastasia
  Danilova}, \bibinfo{person}{Alena Naiakshina}, {and} \bibinfo{person}{Matthew
  Smith}.} \bibinfo{year}{2017}\natexlab{}.
\newblock \showarticletitle{Obstacles to the {A}doption of {S}ecure
  {C}ommunication {T}ools}. In \bibinfo{booktitle}{\emph{Proc. Symposium on
  Security and Privacy (SP)}}. \bibinfo{pages}{137--153}.
\newblock


\bibitem[Acquisti and Gross(2006)]%
        {acquisti_2006}
\bibfield{author}{\bibinfo{person}{Alessandro Acquisti} {and}
  \bibinfo{person}{Ralph Gross}.} \bibinfo{year}{2006}\natexlab{}.
\newblock \showarticletitle{Imagined {C}ommunities: {A}wareness, {I}nformation
  {S}haring, and {P}rivacy on the {F}acebook}. In
  \bibinfo{booktitle}{\emph{Proc. Privacy Enhancing Technologies}},
  \bibfield{editor}{\bibinfo{person}{George Danezis} {and}
  \bibinfo{person}{Philippe Golle}} (Eds.). \bibinfo{publisher}{Springer Berlin
  Heidelberg}, \bibinfo{pages}{36--58}.
\newblock


\bibitem[Ahern et~al\mbox{.}(2007)]%
        {ahern2007}
\bibfield{author}{\bibinfo{person}{Shane Ahern}, \bibinfo{person}{Dean Eckles},
  \bibinfo{person}{Nathaniel~S Good}, \bibinfo{person}{Simon King},
  \bibinfo{person}{Mor Naaman}, {and} \bibinfo{person}{Rahul Nair}.}
  \bibinfo{year}{2007}\natexlab{}.
\newblock \showarticletitle{Over-{E}xposed? {P}rivacy {P}atterns and
  {C}onsiderations in {O}nline and {M}obile {P}hoto {S}haring}. In
  \bibinfo{booktitle}{\emph{Proc. Conference on Human Factors in Computing
  Systems (CHI)}}. \bibinfo{pages}{357--366}.
\newblock


\bibitem[Akgul et~al\mbox{.}(2022)]%
        {akgul_2022}
\bibfield{author}{\bibinfo{person}{Omer Akgul}, \bibinfo{person}{Richard
  Roberts}, \bibinfo{person}{Moses Namara}, \bibinfo{person}{Dave Levin}, {and}
  \bibinfo{person}{MIchelle~L. Mazurek}.} \bibinfo{year}{2022}\natexlab{}.
\newblock \showarticletitle{Investigating {I}nfluencer {VPN} {A}ds on
  {Y}ou{T}ube}. In \bibinfo{booktitle}{\emph{Proc. Symposium on Security and
  Privacy (SP)}}. \bibinfo{pages}{876--892}.
\newblock


\bibitem[{Apple}(2022)]%
        {apple_2022}
\bibfield{author}{\bibinfo{person}{{Apple}}.} \bibinfo{year}{2022}\natexlab{}.
\newblock \bibinfo{booktitle}{\emph{Privacy on i{P}hone | {D}ata {A}uction |
  {A}pple}}.
\newblock
\urldef\tempurl%
\url{https://www.youtube.com/watch?v=NOXK4EVFmJY}
\showURL{%
\tempurl}
\newblock
\shownote{Accessed on Nov. 07, 2022}.


\bibitem[Bada et~al\mbox{.}(2015)]%
        {bada2015}
\bibfield{author}{\bibinfo{person}{Maria Bada}, \bibinfo{person}{Angela~M.
  Sasse}, {and} \bibinfo{person}{Jason R.~C. Nurse}.}
  \bibinfo{year}{2015}\natexlab{}.
\newblock \showarticletitle{{C}yber {S}ecurity {A}wareness {C}ampaigns: {W}hy
  {D}o {T}hey {F}ail to {C}hange {B}ehaviour?}. In
  \bibinfo{booktitle}{\emph{Proc. International Conference on Cyber Security
  for Sustainable Society}}. \bibinfo{pages}{118--131}.
\newblock


\bibitem[Bai et~al\mbox{.}(2020)]%
        {bai2020}
\bibfield{author}{\bibinfo{person}{Wei Bai}, \bibinfo{person}{Michael Pearson},
  \bibinfo{person}{Patrick~Gage Kelley}, {and} \bibinfo{person}{Michelle~L
  Mazurek}.} \bibinfo{year}{2020}\natexlab{}.
\newblock \showarticletitle{Improving {N}on-{E}xperts' {U}nderstanding of
  {E}nd-to-{E}nd {E}ncryption: {A}n {E}xploratory {S}tudy}. In
  \bibinfo{booktitle}{\emph{Proc. European Symposium on Security and Privacy
  Workshops (EuroS\&PW)}}. \bibinfo{pages}{210--219}.
\newblock


\bibitem[Barnard-Wills and Ashenden(2015)]%
        {barnard2015}
\bibfield{author}{\bibinfo{person}{David Barnard-Wills} {and}
  \bibinfo{person}{Debi Ashenden}.} \bibinfo{year}{2015}\natexlab{}.
\newblock \showarticletitle{Playing with {P}rivacy: {G}ames for {E}ducation and
  {C}ommunication in the {P}olitics of {O}nline {P}rivacy}.
\newblock \bibinfo{journal}{\emph{Political Studies}} \bibinfo{volume}{63},
  \bibinfo{number}{1} (\bibinfo{year}{2015}), \bibinfo{pages}{142--160}.
\newblock


\bibitem[Barth et~al\mbox{.}(2019)]%
        {barth_2019}
\bibfield{author}{\bibinfo{person}{Susanne Barth}, \bibinfo{person}{Menno D.~T.
  de Jong}, \bibinfo{person}{Marianne Junger}, \bibinfo{person}{Pieter~H.
  Hartel}, {and} \bibinfo{person}{Janina~C. Roppelt}.}
  \bibinfo{year}{2019}\natexlab{}.
\newblock \showarticletitle{Putting the {P}rivacy {P}aradox to the {T}est:
  {O}nline {P}rivacy and {S}ecurity {B}ehaviors {A}mong {U}sers with
  {T}echnical {K}nowledge, {P}rivacy {A}wareness, and {F}inancial {R}esources}.
\newblock \bibinfo{journal}{\emph{Telematics and Informatics}}
  \bibinfo{volume}{41} (\bibinfo{year}{2019}), \bibinfo{pages}{55--69}.
\newblock


\bibitem[Baruh et~al\mbox{.}(2017)]%
        {baruh2017}
\bibfield{author}{\bibinfo{person}{Lemi Baruh}, \bibinfo{person}{Ekin Secinti},
  {and} \bibinfo{person}{Zeynep Cemalcilar}.} \bibinfo{year}{2017}\natexlab{}.
\newblock \showarticletitle{Online {P}rivacy {C}oncerns and {P}rivacy
  {M}anagement: A {M}eta-{A}nalytical {R}eview}.
\newblock \bibinfo{journal}{\emph{Journal of Communication}}
  \bibinfo{volume}{67}, \bibinfo{number}{1} (\bibinfo{year}{2017}),
  \bibinfo{pages}{26--53}.
\newblock


\bibitem[Brooks and King(2014)]%
        {brooks14}
\bibfield{author}{\bibinfo{person}{Joanna Brooks} {and} \bibinfo{person}{Nigel
  King}.} \bibinfo{year}{2014}\natexlab{}.
\newblock \showarticletitle{Doing {T}emplate {A}nalysis: {E}valuating an {E}nd
  of {L}ife {C}are {S}ervice}.
\newblock \bibinfo{journal}{\emph{SAGE Research Methods Cases Part 1}}
  (\bibinfo{year}{2014}).
\newblock


\bibitem[Damien(2022)]%
        {damien24}
\bibfield{author}{\bibinfo{person}{Damien}.} \bibinfo{year}{2022}\natexlab{}.
\newblock \bibinfo{booktitle}{\emph{Our {R}eimagined {P}rivacy {P}olicy}}.
\newblock
\urldef\tempurl%
\url{https://privacy.x.com/en/blog/2022/our-reimagined-privacy-policy}
\showURL{%
\tempurl}
\newblock
\shownote{Accessed on Jun. 12, 2024}.


\bibitem[Das et~al\mbox{.}(2022)]%
        {das_2022}
\bibfield{author}{\bibinfo{person}{Sauvik Das}, \bibinfo{person}{Cori
  Faklaris}, \bibinfo{person}{Jason~I. Hong}, {and} \bibinfo{person}{Laura~A.
  Dabbish}.} \bibinfo{year}{2022}\natexlab{}.
\newblock \showarticletitle{The {S}ecurity \& {P}rivacy {A}cceptance
  {F}ramework ({SPAF})}.
\newblock \bibinfo{journal}{\emph{Foundations and Trends{\textregistered} in
  Privacy and Security}} \bibinfo{volume}{5}, \bibinfo{number}{1-2}
  (\bibinfo{year}{2022}), \bibinfo{pages}{1--143}.
\newblock


\bibitem[De~Luca et~al\mbox{.}(2016)]%
        {deluca2016}
\bibfield{author}{\bibinfo{person}{Alexander De~Luca}, \bibinfo{person}{Sauvik
  Das}, \bibinfo{person}{Martin Ortlieb}, \bibinfo{person}{Iulia Ion}, {and}
  \bibinfo{person}{Ben Laurie}.} \bibinfo{year}{2016}\natexlab{}.
\newblock \showarticletitle{Expert and {N}on-{E}xpert {A}ttitudes {T}owards
  ({S}ecure) {I}nstant {M}essaging}. In \bibinfo{booktitle}{\emph{Proc.
  Symposium on Usable Privacy and Security (SOUPS)}}.
  \bibinfo{pages}{147--157}.
\newblock


\bibitem[Dechand et~al\mbox{.}(2019)]%
        {dechand2019}
\bibfield{author}{\bibinfo{person}{Sergej Dechand}, \bibinfo{person}{Alena
  Naiakshina}, \bibinfo{person}{Anastasia Danilova}, {and}
  \bibinfo{person}{Matthew Smith}.} \bibinfo{year}{2019}\natexlab{}.
\newblock \showarticletitle{In {E}ncryption {W}e {D}on’t {T}rust: The
  {E}ffect of {E}nd-to-{E}nd {E}ncryption to the {M}asses on {U}ser
  {P}erception}. In \bibinfo{booktitle}{\emph{Proc. European Symposium on
  Security and Privacy (EuroS\&P)}}. \bibinfo{pages}{401--415}.
\newblock


\bibitem[Dehling et~al\mbox{.}(2019)]%
        {dehling2019}
\bibfield{author}{\bibinfo{person}{Tobias Dehling}, \bibinfo{person}{Yuchen
  Zhang}, {and} \bibinfo{person}{Ali Sunyaev}.}
  \bibinfo{year}{2019}\natexlab{}.
\newblock \showarticletitle{Consumer {P}erceptions of {O}nline {B}ehavioral
  {A}dvertising}. In \bibinfo{booktitle}{\emph{Proc. Conference on Business
  Informatics (CBI)}}, Vol.~\bibinfo{volume}{1}. \bibinfo{pages}{345--354}.
\newblock


\bibitem[Denning et~al\mbox{.}(2013)]%
        {denning2013}
\bibfield{author}{\bibinfo{person}{Tamara Denning}, \bibinfo{person}{Adam
  Lerner}, \bibinfo{person}{Adam Shostack}, {and} \bibinfo{person}{Tadayoshi
  Kohno}.} \bibinfo{year}{2013}\natexlab{}.
\newblock \showarticletitle{Control-{A}lt-{H}ack: {T}he {D}esign and
  {E}valuation of a {C}ard {G}ame for {C}omputer {S}ecurity {A}wareness and
  {E}ducation}. In \bibinfo{booktitle}{\emph{Proc. Conference on Computer \&
  Communications Security (CCS)}}. \bibinfo{pages}{915--928}.
\newblock


\bibitem[Distler et~al\mbox{.}(2021)]%
        {distler}
\bibfield{author}{\bibinfo{person}{Verena Distler}, \bibinfo{person}{Matthias
  Fassl}, \bibinfo{person}{Hana Habib}, \bibinfo{person}{Katharina Krombholz},
  \bibinfo{person}{Gabriele Lenzini}, \bibinfo{person}{Carine Lallemand},
  \bibinfo{person}{Lorrie~Faith Cranor}, {and} \bibinfo{person}{Vincent
  Koenig}.} \bibinfo{year}{2021}\natexlab{}.
\newblock \showarticletitle{A systematic {L}iterature {R}eview of {E}mpirical
  {M}ethods and {R}isk {R}epresentation in {U}sable {P}rivacy and {S}ecurity
  {R}esearch}.
\newblock \bibinfo{journal}{\emph{ACM Transactions on Computer-Human
  Interaction (TOCHI)}} \bibinfo{volume}{28}, \bibinfo{number}{6}
  (\bibinfo{year}{2021}), \bibinfo{pages}{1--50}.
\newblock


\bibitem[{DuckDuckGo}(2022)]%
        {duckduckgo_2022}
\bibfield{author}{\bibinfo{person}{{DuckDuckGo}}.}
  \bibinfo{year}{2022}\natexlab{}.
\newblock \bibinfo{booktitle}{\emph{Duck{D}uck{G}o: {W}atching {Y}ou}}.
\newblock
\urldef\tempurl%
\url{https://www.youtube.com/watch?v=QWpPyYlZXNI}
\showURL{%
\tempurl}
\newblock
\shownote{Accessed on Jan. 24, 2023}.


\bibitem[Emami-Naeini et~al\mbox{.}(2023)]%
        {emami}
\bibfield{author}{\bibinfo{person}{Pardis Emami-Naeini},
  \bibinfo{person}{Janarth Dheenadhayalan}, \bibinfo{person}{Yuvraj Agarwal},
  {and} \bibinfo{person}{Lorrie~Faith Cranor}.}
  \bibinfo{year}{2023}\natexlab{}.
\newblock \showarticletitle{Are {C}onsumers {W}illing to {P}ay for {S}ecurity
  and {P}rivacy of {IoT} {D}evices?}. In \bibinfo{booktitle}{\emph{Proc. USENIX
  Security Symposium}}.
\newblock


\bibitem[Esp\'osito(2021)]%
        {whatsappnews_2021}
\bibfield{author}{\bibinfo{person}{Filipe Esp\'osito}.}
  \bibinfo{year}{2021}\natexlab{}.
\newblock \bibinfo{booktitle}{\emph{{W}hatsApp {V}ulnerability {C}ould {L}ead
  to {S}ensitive {D}ata {L}eakage}}.
\newblock
\urldef\tempurl%
\url{https://9to5mac.com/2021/09/02/whatsapp-vulnerability-could-lead-to-sensitive-data-leakage}
\showURL{%
\tempurl}
\newblock
\shownote{Accessed on Oct. 09, 2023}.


\bibitem[Fadilpa\u{s}i\'c(2022)]%
        {ddgnews_2022}
\bibfield{author}{\bibinfo{person}{Sead Fadilpa\u{s}i\'c}.}
  \bibinfo{year}{2022}\natexlab{}.
\newblock \bibinfo{booktitle}{\emph{{D}uckDuckGo in {H}ot {W}ater over {H}idden
  {T}racking {A}greement with {M}icrosoft}}.
\newblock
\urldef\tempurl%
\url{https://www.techradar.com/news/duckduckgo-in-hot-water-over-hidden-tracking-agreement-with-microsoft}
\showURL{%
\tempurl}
\newblock
\shownote{Accessed on Oct. 09, 2023}.


\bibitem[{Google Ads}(2020)]%
        {google_2020}
\bibfield{author}{\bibinfo{person}{{Google Ads}}.}
  \bibinfo{year}{2020}\natexlab{}.
\newblock \bibinfo{booktitle}{\emph{A {Y}ear in {P}rivacy {\textbar} {P}rivacy
  {W}eek at {G}oogle}}.
\newblock
\urldef\tempurl%
\url{https://www.youtube.com/watch?v=zmhMA1Xiukk}
\showURL{%
\tempurl}
\newblock
\shownote{Accessed on Nov. 07, 2022}.


\bibitem[Hurlerand and Wodinsky(2022)]%
        {hurler2022}
\bibfield{author}{\bibinfo{person}{Kevin Hurlerand} {and}
  \bibinfo{person}{Shoshana Wodinsky}.} \bibinfo{year}{2022}\natexlab{}.
\newblock \bibinfo{booktitle}{\emph{Twitter’s New Privacy Policy Is a Video
  Game That Sucks}}.
\newblock
\urldef\tempurl%
\url{https://gizmodo.com/twitter-privacy-policy-video-game-data-dash-1848912387}
\showURL{%
\tempurl}
\newblock
\shownote{Accessed on Oct. 28, 2023}.


\bibitem[Jeong and Kim(2017)]%
        {jeong2017}
\bibfield{author}{\bibinfo{person}{Yongick Jeong} {and}
  \bibinfo{person}{Yeuseung Kim}.} \bibinfo{year}{2017}\natexlab{}.
\newblock \showarticletitle{Privacy {C}oncerns on {S}ocial {N}etworking
  {S}ites: {I}nterplay {A}mong {P}osting {T}ypes, {C}ontent, and {A}udiences}.
\newblock \bibinfo{journal}{\emph{Computers in Human Behavior}}
  \bibinfo{volume}{69} (\bibinfo{year}{2017}), \bibinfo{pages}{302--310}.
\newblock


\bibitem[Kelley et~al\mbox{.}(2013)]%
        {kelley}
\bibfield{author}{\bibinfo{person}{Patrick~Gage Kelley},
  \bibinfo{person}{Lorrie~Faith Cranor}, {and} \bibinfo{person}{Norman Sadeh}.}
  \bibinfo{year}{2013}\natexlab{}.
\newblock \showarticletitle{Privacy as {P}art of the {A}pp {D}ecision-{M}aking
  {P}rocess}. In \bibinfo{booktitle}{\emph{Proc. Conference on Human Factors in
  Computing Systems (CHI)}}. \bibinfo{pages}{3393--3402}.
\newblock


\bibitem[Kim et~al\mbox{.}(2017)]%
        {kim}
\bibfield{author}{\bibinfo{person}{Eunjin Kim}, \bibinfo{person}{S Ratneshwar},
  {and} \bibinfo{person}{Esther Thorson}.} \bibinfo{year}{2017}\natexlab{}.
\newblock \showarticletitle{Why {N}arrative {A}ds {W}ork: {A}n {I}ntegrated
  {P}rocess {E}xplanation}.
\newblock \bibinfo{journal}{\emph{Journal of Advertising}}
  \bibinfo{volume}{46}, \bibinfo{number}{2} (\bibinfo{year}{2017}),
  \bibinfo{pages}{283--296}.
\newblock


\bibitem[King(2024)]%
        {king24}
\bibfield{author}{\bibinfo{person}{Nigel King}.}
  \bibinfo{year}{2024}\natexlab{}.
\newblock \bibinfo{booktitle}{\emph{Template {A}nalysis}}.
\newblock
\newblock
\shownote{Accessed on Feb. 13, 2024}.


\bibitem[Kross et~al\mbox{.}(2021)]%
        {kross_2021}
\bibfield{author}{\bibinfo{person}{Sean Kross}, \bibinfo{person}{Eszter
  Hargittai}, {and} \bibinfo{person}{Elissa~M. Redmiles}.}
  \bibinfo{year}{2021}\natexlab{}.
\newblock \showarticletitle{Characterizing the {O}nline {L}earning {L}andscape:
  {W}hat and {H}ow {P}eople {L}earn {O}nline}.
\newblock \bibinfo{journal}{\emph{Proc. ACM Hum.-Comput. Interact.}}
  \bibinfo{volume}{5}, \bibinfo{number}{CSCW1}, Article
  \bibinfo{articleno}{146} (\bibinfo{year}{2021}),
  \bibinfo{numpages}{19}~pages.
\newblock


\bibitem[Landa(2016)]%
        {landa2021}
\bibfield{author}{\bibinfo{person}{Robin Landa}.}
  \bibinfo{year}{2016}\natexlab{}.
\newblock \bibinfo{booktitle}{\emph{Advertising by {D}esign: {G}enerating and
  {D}esigning {C}reative {I}deas {A}cross {M}edia}}.
\newblock \bibinfo{publisher}{John Wiley \& Sons}.
\newblock


\bibitem[Laurence(2018)]%
        {laurence}
\bibfield{author}{\bibinfo{person}{Dessart Laurence}.}
  \bibinfo{year}{2018}\natexlab{}.
\newblock \showarticletitle{Do {A}ds that {T}ell a {S}tory {A}lways {P}erform
  {B}etter? {T}he {R}ole of {C}haracter {I}dentification and {C}haracter {T}ype
  in {S}torytelling {A}ds}.
\newblock \bibinfo{journal}{\emph{International Journal of Research in
  Marketing}} \bibinfo{volume}{35}, \bibinfo{number}{2} (\bibinfo{year}{2018}),
  \bibinfo{pages}{289--304}.
\newblock


\bibitem[Liu et~al\mbox{.}(2011)]%
        {liu2011}
\bibfield{author}{\bibinfo{person}{Yabing Liu}, \bibinfo{person}{Krishna~P
  Gummadi}, \bibinfo{person}{Balachander Krishnamurthy}, {and}
  \bibinfo{person}{Alan Mislove}.} \bibinfo{year}{2011}\natexlab{}.
\newblock \showarticletitle{Analyzing {F}acebook {P}rivacy {S}ettings: {U}ser
  {E}xpectations vs. {R}eality}. In \bibinfo{booktitle}{\emph{Proc. Internet
  Measurement Conference (IMC)}}. \bibinfo{pages}{61--70}.
\newblock


\bibitem[Madejski et~al\mbox{.}(2012)]%
        {madejski2012}
\bibfield{author}{\bibinfo{person}{Michelle Madejski}, \bibinfo{person}{Maritza
  Johnson}, {and} \bibinfo{person}{Steven~M Bellovin}.}
  \bibinfo{year}{2012}\natexlab{}.
\newblock \showarticletitle{A {S}tudy of {P}rivacy {S}ettings {E}rrors in an
  {O}nline {S}ocial {N}etwork}. In \bibinfo{booktitle}{\emph{Proc. Pervasive
  Computing and Communications Workshops}}. \bibinfo{pages}{340--345}.
\newblock


\bibitem[Manyiwa and Brennan(2012)]%
        {manyiwa}
\bibfield{author}{\bibinfo{person}{Simon Manyiwa} {and} \bibinfo{person}{Ross
  Brennan}.} \bibinfo{year}{2012}\natexlab{}.
\newblock \showarticletitle{Fear {A}ppeals in {A}nti-{S}moking {A}dvertising:
  {H}ow {I}mportant is {S}elf-{E}fficacy?}
\newblock \bibinfo{journal}{\emph{Journal of Marketing Management}}
  \bibinfo{volume}{28}, \bibinfo{number}{11-12} (\bibinfo{year}{2012}),
  \bibinfo{pages}{1419--1437}.
\newblock


\bibitem[Maqsood et~al\mbox{.}(2018)]%
        {maqsood2018}
\bibfield{author}{\bibinfo{person}{Sana Maqsood}, \bibinfo{person}{Christine
  Mekhail}, {and} \bibinfo{person}{Sonia Chiasson}.}
  \bibinfo{year}{2018}\natexlab{}.
\newblock \showarticletitle{A {D}ay in the {L}ife of {J}os: {A} {W}eb-based
  {G}ame to {I}ncrease {C}hildren's {D}igital {L}iteracy}. In
  \bibinfo{booktitle}{\emph{Proc. Conference on Interaction Design and
  Children}}. \bibinfo{pages}{241--252}.
\newblock


\bibitem[McCarney et~al\mbox{.}(2007)]%
        {hawthorne}
\bibfield{author}{\bibinfo{person}{Rob McCarney}, \bibinfo{person}{James
  Warner}, \bibinfo{person}{Steve Iliffe}, \bibinfo{person}{Robbert
  Van~Haselen}, \bibinfo{person}{Mark Griffin}, {and} \bibinfo{person}{Peter
  Fisher}.} \bibinfo{year}{2007}\natexlab{}.
\newblock \showarticletitle{The Hawthorne Effect: a randomised, controlled
  trial}.
\newblock \bibinfo{journal}{\emph{BMC medical research methodology}}
  \bibinfo{volume}{7}, \bibinfo{number}{1} (\bibinfo{year}{2007}),
  \bibinfo{pages}{1--8}.
\newblock


\bibitem[Mohanty and Ratneshwar(2015)]%
        {mohanty2015}
\bibfield{author}{\bibinfo{person}{Praggyan~(Pam) Mohanty} {and}
  \bibinfo{person}{S. Ratneshwar}.} \bibinfo{year}{2015}\natexlab{}.
\newblock \showarticletitle{Did {Y}ou {G}et {I}t? {F}actors {I}nfluencing
  {S}ubjective {C}omprehension of {V}isual {M}etaphors in {A}dvertising}.
\newblock \bibinfo{journal}{\emph{Journal of Advertising}}
  \bibinfo{volume}{44}, \bibinfo{number}{3} (\bibinfo{year}{2015}),
  \bibinfo{pages}{232--242}.
\newblock


\bibitem[Mohanty and Ratneshwar(2016)]%
        {mohanty2016}
\bibfield{author}{\bibinfo{person}{Praggyan~(Pam) Mohanty} {and}
  \bibinfo{person}{S. Ratneshwar}.} \bibinfo{year}{2016}\natexlab{}.
\newblock \showarticletitle{Visual {M}etaphors in {A}ds: {T}he {I}nverted-{U}
  {E}ffects of {I}ncongruity on {P}rocessing {P}leasure and {A}d
  {E}ffectiveness}.
\newblock \bibinfo{journal}{\emph{Journal of Promotion Management}}
  \bibinfo{volume}{22}, \bibinfo{number}{3} (\bibinfo{year}{2016}),
  \bibinfo{pages}{443--460}.
\newblock


\bibitem[Moscaritolo(2017)]%
        {moscaritolo}
\bibfield{author}{\bibinfo{person}{Angela Moscaritolo}.}
  \bibinfo{year}{2017}\natexlab{}.
\newblock \bibinfo{booktitle}{\emph{Google's \quotes{{I}nterland} {G}ame
  {M}akes {O}nline {S}afety {E}ducation {F}un}}.
\newblock
\urldef\tempurl%
\url{https://www.pcmag.com/news/googles-interland-game-makes-online-safety-education-fun}
\showURL{%
\tempurl}
\newblock
\shownote{Accessed on Nov. 24, 2023}.


\bibitem[O'Flaherty(2022)]%
        {oflaherty_2022}
\bibfield{author}{\bibinfo{person}{Kate O'Flaherty}.}
  \bibinfo{year}{2022}\natexlab{}.
\newblock \bibinfo{booktitle}{\emph{Apple {S}lams {F}acebook and {G}oogle
  {W}ith {B}old {N}ew {P}rivacy {A}d}}.
\newblock
\urldef\tempurl%
\url{https://www.forbes.com/sites/kateoflahertyuk/2022/05/25/apple-slams-facebook-and-google-with-bold-new-privacy-ad}
\showURL{%
\tempurl}
\newblock
\shownote{Accessed on Nov. 07, 2022}.


\bibitem[Persuasive~Games(2023)]%
        {persuasive}
\bibfield{author}{\bibinfo{person}{LLC Persuasive~Games}.}
  \bibinfo{year}{2023}\natexlab{}.
\newblock \bibinfo{booktitle}{\emph{Persuasive {G}ames}}.
\newblock
\urldef\tempurl%
\url{https://persuasivegames.com/games}
\showURL{%
\tempurl}
\newblock
\shownote{Accessed on Feb. 15, 2024}.


\bibitem[Prolific(2023)]%
        {prolific23}
\bibfield{author}{\bibinfo{person}{Prolific}.} \bibinfo{year}{2023}\natexlab{}.
\newblock \bibinfo{booktitle}{\emph{Prolific}}.
\newblock
\urldef\tempurl%
\url{https://www.prolific.co}
\showURL{%
\tempurl}
\newblock
\shownote{Accessed on Sep. 11, 2023}.


\bibitem[Rader(2014)]%
        {rader2014}
\bibfield{author}{\bibinfo{person}{Emilee Rader}.}
  \bibinfo{year}{2014}\natexlab{}.
\newblock \showarticletitle{Awareness of {B}ehavioral {T}racking and
  {I}nformation {P}rivacy {C}oncern in {F}acebook and {G}oogle}. In
  \bibinfo{booktitle}{\emph{Proc. Symposium On Usable Privacy and Security
  (SOUPS)}}. \bibinfo{pages}{51--67}.
\newblock


\bibitem[Redmiles et~al\mbox{.}(2020)]%
        {redmiles_2020}
\bibfield{author}{\bibinfo{person}{Elissa~M. Redmiles}, \bibinfo{person}{Noel
  Warford}, \bibinfo{person}{Aritha Jayanti}, \bibinfo{person}{Aravind Koneru},
  \bibinfo{person}{Sean Kross}, \bibinfo{person}{Miraida Morales},
  \bibinfo{person}{Rock Stevens}, {and} \bibinfo{person}{Michelle~L. Mazurek}.}
  \bibinfo{year}{2020}\natexlab{}.
\newblock \showarticletitle{A {C}omprehensive {Q}uality {E}valuation of
  {S}ecurity and {P}rivacy {A}dvice on the {W}eb}. In
  \bibinfo{booktitle}{\emph{Proc. USENIX Security Symposium}}.
  \bibinfo{pages}{89--108}.
\newblock


\bibitem[{Samsung}(2022)]%
        {samsung_2022}
\bibfield{author}{\bibinfo{person}{{Samsung}}.}
  \bibinfo{year}{2022}\natexlab{}.
\newblock \bibinfo{booktitle}{\emph{{S}amsung {P}rivacy: {Y}ou're in
  {C}ontrol}}.
\newblock
\urldef\tempurl%
\url{https://www.youtube.com/watch?v=kZKK80urUZc}
\showURL{%
\tempurl}
\newblock
\shownote{Accessed on Nov. 07, 2022}.


\bibitem[Sheng et~al\mbox{.}(2007)]%
        {sheng2007}
\bibfield{author}{\bibinfo{person}{Steve Sheng}, \bibinfo{person}{Bryant
  Magnien}, \bibinfo{person}{Ponnurangam Kumaraguru},
  \bibinfo{person}{Alessandro Acquisti}, \bibinfo{person}{Lorrie~Faith Cranor},
  \bibinfo{person}{Jason Hong}, {and} \bibinfo{person}{Elizabeth Nunge}.}
  \bibinfo{year}{2007}\natexlab{}.
\newblock \showarticletitle{Anti-{P}hishing {P}hil: {T}he {D}esign and
  {E}valuation of a {G}game {T}hat {T}eaches {P}eople {N}ot to {F}all for
  {P}hish}. In \bibinfo{booktitle}{\emph{Proc. USENIX Security Symposium}}.
  \bibinfo{pages}{88--99}.
\newblock


\bibitem[Sohoraye et~al\mbox{.}(2015)]%
        {sohoraye_2015}
\bibfield{author}{\bibinfo{person}{Mrinal Sohoraye}, \bibinfo{person}{Vandanah
  Gooria}, \bibinfo{person}{Suniti Nundoo-Ghoorah}, {and}
  \bibinfo{person}{Premanand Koonjal}.} \bibinfo{year}{2015}\natexlab{}.
\newblock \showarticletitle{Do you {K}now {B}ig {B}rother is {W}atching you on
  {F}acebook? {A} {S}tudy of the {L}evel of {A}wareness of {P}rivacy and
  {S}ecurity {I}ssues {A}mong a {S}elected {S}ample of {F}acebook {U}sers in
  {M}auritius}. In \bibinfo{booktitle}{\emph{Proc. International Conference on
  Computing, Communication and Security (ICCCS)}}. \bibinfo{pages}{1--7}.
\newblock


\bibitem[Stransky et~al\mbox{.}(2021)]%
        {stransky2021l}
\bibfield{author}{\bibinfo{person}{Christian Stransky},
  \bibinfo{person}{Dominik Wermke}, \bibinfo{person}{Johanna Schrader},
  \bibinfo{person}{Nicolas Huaman}, \bibinfo{person}{Yasemin Acar},
  \bibinfo{person}{Anna~Lena Fehlhaber}, \bibinfo{person}{Miranda Wei},
  \bibinfo{person}{Blase Ur}, {and} \bibinfo{person}{Sascha Fahl}.}
  \bibinfo{year}{2021}\natexlab{}.
\newblock \showarticletitle{On the {L}imited {I}mpact of {V}isualizing
  {E}ncryption: {P}erceptions of {E2E} {M}essaging {S}ecurity}. In
  \bibinfo{booktitle}{\emph{Proc. Symposium on Usable Privacy and Security
  (SOUPS)}}. \bibinfo{pages}{437--454}.
\newblock


\bibitem[{The Police}(2023)]%
        {everybreath23}
\bibfield{author}{\bibinfo{person}{{The Police}}.}
  \bibinfo{year}{2023}\natexlab{}.
\newblock \bibinfo{booktitle}{\emph{The {P}olice - {E}very {B}reath {Y}ou
  {T}ake ({O}fficial {M}usic {V}ideo)}}.
\newblock
\urldef\tempurl%
\url{https://www.youtube.com/watch?v=OMOGaugKpzs}
\showURL{%
\tempurl}
\newblock
\shownote{Accessed on Oct. 18, 2023}.


\bibitem[Thompson and Irvine(2011)]%
        {thompson2011}
\bibfield{author}{\bibinfo{person}{Michael Thompson} {and}
  \bibinfo{person}{Cynthia Irvine}.} \bibinfo{year}{2011}\natexlab{}.
\newblock \showarticletitle{Active {L}earning with the {C}yber{C}iege {V}ideo
  {G}ame}. In \bibinfo{booktitle}{\emph{Proc. Workshop on Cyber Security
  Experimentation and Test (CSET)}}.
\newblock


\bibitem[Tsai et~al\mbox{.}(2011)]%
        {tsai2011}
\bibfield{author}{\bibinfo{person}{Janice~Y Tsai}, \bibinfo{person}{Serge
  Egelman}, \bibinfo{person}{Lorrie Cranor}, {and} \bibinfo{person}{Alessandro
  Acquisti}.} \bibinfo{year}{2011}\natexlab{}.
\newblock \showarticletitle{The {E}ffect of {O}nline {P}rivacy {I}nformation on
  {P}urchasing {B}ehavior: {A}n {E}xperimental {S}tudy}.
\newblock \bibinfo{journal}{\emph{Information Systems Research}}
  \bibinfo{volume}{22}, \bibinfo{number}{2} (\bibinfo{year}{2011}),
  \bibinfo{pages}{254--268}.
\newblock


\bibitem[Twitter(2022)]%
        {twitter_2022}
\bibfield{author}{\bibinfo{person}{Twitter}.} \bibinfo{year}{2022}\natexlab{}.
\newblock \bibinfo{booktitle}{\emph{Twitter {D}ata {D}ash: {L}evel {U}p {Y}our
  {P}rivacy {G}ame}}.
\newblock
\urldef\tempurl%
\url{https://twitterdatadash.com}
\showURL{%
\tempurl}
\newblock
\shownote{Accessed on Nov. 07, 2022}.


\bibitem[Ulanoff(2022)]%
        {ulanoff2022}
\bibfield{author}{\bibinfo{person}{Lance Ulanoff}.}
  \bibinfo{year}{2022}\natexlab{}.
\newblock \bibinfo{booktitle}{\emph{Apple's {D}ata {A}uction {P}rivacy {A}d is
  {O}nly {S}cary {B}ecause {I}t's {T}rue}}.
\newblock
\urldef\tempurl%
\url{https://www.techradar.com/news/apples-data-auction-privacy-ad-is-only-scary-because-its-true}
\showURL{%
\tempurl}
\newblock
\shownote{Accessed on Oct. 28, 2023}.


\bibitem[Ur et~al\mbox{.}(2012)]%
        {ur2012}
\bibfield{author}{\bibinfo{person}{Blase Ur}, \bibinfo{person}{Pedro~Giovanni
  Leon}, \bibinfo{person}{Lorrie~Faith Cranor}, \bibinfo{person}{Richard Shay},
  {and} \bibinfo{person}{Yang Wang}.} \bibinfo{year}{2012}\natexlab{}.
\newblock \showarticletitle{Smart, {U}seful, {S}cary, {C}reepy: {P}erceptions
  of {O}nline {B}ehavioral {A}dvertising}. In \bibinfo{booktitle}{\emph{Proc.
  Symposium on Usable Privacy and Security (SOUPS)}}. \bibinfo{pages}{1--15}.
\newblock


\bibitem[{WhatsApp}(2022)]%
        {whatsapp_2022}
\bibfield{author}{\bibinfo{person}{{WhatsApp}}.}
  \bibinfo{year}{2022}\natexlab{}.
\newblock \bibinfo{booktitle}{\emph{A {N}ew {E}ra of {P}ersonal {P}rivacy
  {W}ith {D}efault {E}nd-to-{E}ncryption}}.
\newblock
\urldef\tempurl%
\url{https://www.youtube.com/watch?v=zvI4cVGWJhM}
\showURL{%
\tempurl}
\newblock
\shownote{Accessed on Nov. 07, 2022}.


\bibitem[Winkie(2022)]%
        {winkie2022}
\bibfield{author}{\bibinfo{person}{Lance Winkie}.}
  \bibinfo{year}{2022}\natexlab{}.
\newblock \bibinfo{booktitle}{\emph{Data {T}he {D}og: {T}witter {T}urns its
  {P}rivacy {P}olicy {I}nto an {O}ld-{S}chool {V}ideo {G}ame}}.
\newblock
\urldef\tempurl%
\url{https://www.theguardian.com/technology/2022/may/14/twitter-data-dash}
\showURL{%
\tempurl}
\newblock
\shownote{Accessed on Oct. 28, 2023}.


\end{thebibliography}
\clearpage
\appendix
\section{Appendices} \label{sec:app}
In this section, we present the screening survey~(\autorefappendix{app:survey}) and interview script~(\autorefappendix{app:interview}). Text inside [square brackets] was not shown or read to survey and interview participants. Interview section headings were not read to interview participants. The nature of semi-structured interviews allowed the interviewer to slightly deviate from the interview script without affecting the essence of the interview.

\subsection{Screening Survey} \label{app:survey}
\textbf{Section 1 - General} \newline 
\textbf{Q.1:} What is your gender? \newline 
    \textbf{Answers:} \begin{inparaitem}[$\circ$]\item \quotes{Male} \item \quotes{Female} \item \quotes{Prefer not to answer}\item \quotes{Prefer to self describe (please specify)}\end{inparaitem} \newline 
\textbf{Q.2:} What is your age group? \newline 
    \textbf{Answers:} \begin{inparaitem}[$\circ$]\item \quotes{18 - 25 years} \item \quotes{26 - 35 years} \item \quotes{36 - 45 years}\item \quotes{46 - 55 years}\item \quotes{56 - 65 years}\item \quotes{65 -75 years}\item \quotes{76 years or more}\end{inparaitem} \newline 
\textbf{Q.3:} Which of the following best describes your highest achieved education level? \newline
    \textbf{Answers:} \begin{inparaitem}[$\circ$]\item \quotes{Some high school, no degree} \item \quotes{High school graduate} \item \quotes{Some college, no degree}\item \quotes{Associate's degree}\item \quotes{Bachelor's degree}\item \quotes{Graduate degree (Masters, Doctorate, etc.)}\item \quotes{Other (Please specify)}\end{inparaitem} \newline 
\textbf{Q.4:} Do you have (or are currently studying for) a degree, or work (or have worked) in one or more of the following areas: Computer Science, Information Systems, Information Technology, Computer Engineering, or a related field? \newline 
\textbf{Answers:} \begin{inparaitem}[$\circ$]\item \quotes{Yes} \item \quotes{No}\end{inparaitem} \newline 

\noindent \textbf{Section 2 - Phone} \newline 
\textbf{Q.5:} Do you use a smartphone? \newline
    \textbf{Answers:} \begin{inparaitem}[$\circ$]\item \quotes{Yes} \item \quotes{No}\end{inparaitem} \newline 
[If Q.5 answer is \quotes{No,} the participant skips the rest of this section and moves to Section 3] \newline 
\textbf{Q.6:} Approximately, how often do you use your smartphone? \newline
    \textbf{Answers:} \begin{inparaitem}[$\circ$]\item \quotes{At least once a day} \item \quotes{At least once a  week} \item \quotes{At least once a month} \item \quotes{Less than once a month}\item \quotes{I do not remember}\item \quotes{Other (Please specify)}\end{inparaitem} \newline 
\textbf{Q.7:} Please indicate your current, most used smartphone brand: \newline
    \textbf{Answers:} \begin{inparaitem}[$\circ$]\item \quotes{Apple smartphone} \item \quotes{Samsung smartphone} \item \quotes{Other Android smartphone} \item \quotes{Other (Please specify)}\end{inparaitem} \newline 
\textbf{Q.8:} Approximately, how long have you been using your current, most used smartphone device? \newline
    \textbf{Answers:} \begin{inparaitem}[$\circ$]\item \quotes{Less than a month} \item \quotes{At least for 3 months} \item \quotes{At least for 6 month} \item \quotes{At least for a year} \item \quotes{At least for 2 years} \item \quotes{At least for 3 years} \item \quotes{More than 3 years} \item \quotes{Other (Please specify)} \end{inparaitem} \newline 
\textbf{Q.9:} Have you switched mobile operating systems (OS), like from Apple to Android, in the last 5 years? \newline
    \textbf{Answers:} \begin{inparaitem}[$\circ$]\item \quotes{No, I have always used the same OS} \item \quotes{Yes, from Apple to Android} \item \quotes{Yes, from Android to Apple} \item \quotes{I bounced in between different OS - e.g., Apple to Android then back to Apple} \item \quotes{I owned two phones at the same time, one Apple and one Android.} \item \quotes{Other (Please specify)} \end{inparaitem} \newline 
    
\noindent \textbf{Section 3 - Gaming} \newline 
\textbf{Q.10:} Do you play or participate in any form of electronic, digital, and/or video gaming? \newline
    \textbf{Answers:} \begin{inparaitem}[$\circ$]\item \quotes{Yes} \item \quotes{No}\end{inparaitem} \newline 
[If Q.10 answer is \quotes{No}, the participant skips the rest of this section, and moves to Section 4] \newline 
\textbf{Q.11:} Approximately, how often do you play electronic games? \newline
    \textbf{Answers:} \begin{inparaitem}[$\circ$]\item \quotes{At least once a day} \item \quotes{At least once a week} \item \quotes{At least once a month} \item \quotes{Less than once a month}\item \quotes{I do not remember}\item \quotes{Other (Please specify)}\end{inparaitem} \newline 
\textbf{Q.12:} Please indicate all forms of electronic gaming you regularly participate in: \newline
    \textbf{Answers:} \begin{inparaitem}[$\circ$]\item \quotes{PlayStation games} \item \quotes{Nintendo Switch games} \item \quotes{PC games (e.g., League of Legends, SIM4, Minecraft, browser games like Tetris and other similar computer games)} \item \quotes{VR games (e.g. Beat Saber, Super Hot and other similar games)} \item \quotes{Other (Please specify)}\end{inparaitem} \newline

\noindent \textbf{Section 4 - Social Media} \newline 
\textbf{Q.13:} Do you use social media? \newline
    \textbf{Answers:} \begin{inparaitem}[$\circ$]\item \quotes{Yes} \item \quotes{No}\end{inparaitem} \newline 
[If Q.13 answer is \quotes{No,} the participant skips the rest of this section, and moves to Section 5] \newline
\textbf{Q.14:} Please indicate all social media platform you use: \newline
    \textbf{Answers:} \begin{inparaitem}[$\circ$]\item \quotes{Facebook} \item \quotes{Instagram} \item \quotes{Twitter} \item \quotes{LinkedIn} \item \quotes{Snapchat} \item \quotes{TikTok} \item \quotes{Reddit} \item \quotes{Other (Please specify)}\end{inparaitem} \newline 
[If Q.14 answer does not include \quotes{Twitter,} the participant skips the rest of this section, and moves to Section 5] \newline 
\textbf{Q.15:} Approximately, how often do you use Twitter? \newline
    \textbf{Answers:} \begin{inparaitem}[$\circ$]\item \quotes{At least once a day} \item \quotes{At least once a  week} \item \quotes{At least once a month} \item \quotes{Less than once a month}\item \quotes{I do not remember}\item \quotes{Other (Please specify)}\end{inparaitem}
\newline 

\noindent \textbf{Section 5 - Instant Messaging} \newline 
\textbf{Q.16:} Do you use instant messaging? \newline
    \textbf{Answers:} \begin{inparaitem}[$\circ$]\item \quotes{Yes} \item \quotes{No}\end{inparaitem} \newline 
[If Q.16 answer is \quotes{No,} the participant skips the rest of this section, and moves to Section 6]

\noindent\textbf{Q.17:} Please indicate all form of instant messaging service you use: \newline
    \textbf{Answers:} \begin{inparaitem}[$\circ$]\item \quotes{iMessage} \item \quotes{Facebook Messenger} \item \quotes{Telegram} \item \quotes{WhatsApp} \item \quotes{Discord} \item \quotes{WeChat} \item \quotes{Slack} \item \quotes{SMS text} \item \quotes{Other (Please specify)}\end{inparaitem} \newline 
[If Q.17 answer does not include \quotes{WhatsApp,} the participant skips Q.18 and moves to Section 6] \newline
\noindent\textbf{Q.18:} Approximately, how often do you use WhatsApp? \newline
    \textbf{Answers:} \begin{inparaitem}[$\circ$]\item \quotes{At least once a day} \item \quotes{At least once a week} \item \quotes{At least once a month} \item \quotes{Less than once a month}\item \quotes{I do not remember}\item \quotes{Other (Please specify)}\end{inparaitem} \newline 

\noindent \textbf{Section 6 - Internet Browsing} \newline 
\textbf{Q.19:} Do you use search engines and/or web browsers (e.g. Google Chrome, Safari)? \newline
    \textbf{Answers:} \begin{inparaitem}[$\circ$]\item \quotes{Yes} \item \quotes{No}\end{inparaitem} \newline 
[If Q.19 answer is \quotes{No,} the participant skips the rest of this section, and the survey is completed] \newline 
\textbf{Q.20:} Please select all search engines you use: \newline
    \textbf{Answers:} \begin{inparaitem}[$\circ$]\item \quotes{Google} \item \quotes{Bing} \item \quotes{Yahoo} \item \quotes{Baidu} \item \quotes{AOL} \item \quotes{Naver} \item \quotes{Ask.com} \item \quotes{DuckDuckGo} \item \quotes{Other (Please specify)}\end{inparaitem} \newline 
[If Q.20 answer does not include \quotes{DuckDuckGo,} the participant skips Q.21 and moves to Q.22] \newline 
\textbf{Q.21:} Approximately, how often do you use DuckDuckGo? \newline
    \textbf{Answers:} \begin{inparaitem}[$\circ$]\item \quotes{At least once a day} \item \quotes{At least once a week} \item \quotes{At least once a month} \item \quotes{Less than once a month}\item \quotes{I do not remember}\item \quotes{Other (Please specify)}\end{inparaitem} \newline 
\textbf{Q.22:} Please select all web browsers you use: \newline
    \textbf{Answers:} \begin{inparaitem}[$\circ$]\item \quotes{Google Chrome} \item \quotes{Microsoft Edge} \item \quotes{Mozilla Firefox} \item \quotes{Safari} \item \quotes{DuckDuckGo} \item \quotes{Vivaldi} \item \quotes{Brave} \item \quotes{Opera} \item \quotes{Other (Please specify)}\end{inparaitem} \newline 
[If Q.22 answer does not include \quotes{DuckDuckGo,} the participant skips Q.23, and the survey is completed] \newline 
\textbf{Q.23:} Approximately, how often do you use DuckDuckGo? \newline
    \textbf{Answers:} \begin{inparaitem}[$\circ$]\item \quotes{At least once a day} \item \quotes{At least once a week} \item \quotes{At least once a month} \item \quotes{Less than once a month}\item \quotes{I do not remember}\item \quotes{Other (Please specify)}\end{inparaitem} 
\subsection{Semi-structured Interview Script} \label{app:interview}
\textbf{Introduction}

Thank you for participating in our study. As specified in the consent form, we will be recording the interview to ensure that we do not miss parts of the conversation. Your name will not be associated with any data collected during this interview. Please leave your camera off during this process. The interview will last around 20 to 25 minutes. 

[Specific for Twitter] During the interview, you will be asked to share your screen for an online gaming activity, which will be screen recorded. We will not ask you to show any private information, such as photos, mail or any other related setting.

Before we start, do you have any questions regarding the consent form? Could I have your permission to start the recording?  

Thank you, I have started the recording. If there are short silences during the interview, please excuse them - I am also taking notes during this process. There are no right or wrong answers to what I am asking, let’s begin. 

[All [Prompt] questions are optional. A company name between square brackets [Company] means this question is specific for participants in that treatment group.] \newline 

\noindent \textbf{Section 1 - All Campaigns} \newline 
[Interviewer asks different variations of the question based on
participants' assigned treatment group and screening survey results. For Apple and Samsung groups we use the term \quotes{smartphone;} for WhatsApp we use \quotes{instant messaging;} for Twitter we use \quotes{social media;} and for DuckDuckGo we use \quotes{browser and search engine.} For all smartphone groups we use the term \quotes{device;} for apps groups we use \quotes{service.}]

\noindent \textbf{Q.1:} What is your most preferred (smartphone | instant messaging | social media | browser and search engine) (device | service)?
\begin{adjustwidth}{0.5cm}{}
\textbf{Q.1A:} Why is [Q.1 answer] your most preferred (smartphone | instant messaging | social media | browser and search engine)? 
\end{adjustwidth}
 
\noindent[For Q.2 to Q.4, ask the question that aligns with the participants' treatment group.]

\noindent\textbf{Q.2:} [Twitter/WhatsApp User] What is your main purpose of using (Twitter | WhatsApp)?
\begin{adjustwidth}{0.5cm}{}
\textbf{Q.2A:} [WhatsApp Non-User] Do you actively participate in instant message chats and group discussions?
\end{adjustwidth}

\noindent\textbf{Q.3:} [Apple/Samsung] Do you actively use smartphone apps and Internet browsing using your smartphone?
\begin{adjustwidth}{0.5cm}{}
\textbf{Q.3A:} To the best of your knowledge, when was the last time you updated your phone?
\end{adjustwidth}

\noindent\textbf{Q.4:} [DuckDuckGo] Do you actively browse the web and perform search queries in search engines?
\begin{adjustwidth}{0.5cm}{}
\noindent\textbf{Q.4A:} Do you use the same browser and search engine on your desktop and mobile? \newline 
[Prompt] Why/Why not?
\end{adjustwidth}

\noindent\textbf{Q.5:} What is your general impression of (Apple | Samsung | WhatsApp | Twitter | DuckDuckGo) as an organization?
\begin{adjustwidth}{0.5cm}{}
\textbf{Q.5A:} [WhatsApp] What about your impression of Meta? \newline
[Prompt] Is it the same with WhatsApp? 
\end{adjustwidth}
[Apple, Samsung, WhatsApp \& DuckDuckGo participants moves to Section 2. Twitter participants moves to Section 3.] \newline 

\noindent\textbf{Section 2 - Video Instruction (Apple, Samsung, WhatsApp \& DuckDuckGo)}

I am now going to share my screen, and show you an advertisement from (Apple | Samsung | WhatsApp | DuckDuckGo)

[Show video, then participants move to Section 4.] \newline

\noindent\textbf{Section 3 - Game Instruction (Twitter)}

Next, please open Twitter and log out of your own account. Once you have logged out please let me know, I will send you a username and password to log into an account we have created for the study. 

Now that you have logged in, I am going to send you a link to an online video game hosted by Twitter, called Data Dash. Once you have accessed the web page, please share your screen with me, and narrate your thoughts as we progress through each level. Please start from level 1 and we will have about 8 minutes for this activity. 

\noindent[Share DataDash Link, allow for 8 minutes of play time.]

[Time is up] Thank you, you can now stop screen sharing, Please return to Twitter and log out of the account.

\noindent[Participants move to Section 4.] \newline 

\noindent\textbf{Section 4 - Post Exposure}

\noindent\textbf{Q.6:} Have you (seen this video | heard about DataDash) before?
\begin{adjustwidth}{0.5cm}{}
\textbf{Q.6A:} [If Q.6 answer is \quotes{Yes}] Where did you (see this advertisement | hear about DataDash)? \newline 
\textbf{Q.6B:} [If Q.6 answer is \quotes{No} - Continue to next question.]
\end{adjustwidth}

\noindent\textbf{Q.7:} In your own words, could you describe the (ad | game) you had just (watched | played)?

\noindent\textbf{Q.8:} How do you feel about (the ad we just saw | the game you just played)?
\begin{adjustwidth}{0.5cm}{}
\textbf{Q.8A:} What stood out to you? \newline 
[Positive prompt] What did you like about the (ad | game)? \newline
[Negative prompt] What did you dislike about the (ad | game)? \newline 
\textbf{Q.8B:} [Twitter] Why did you (click | not click) on the setting button at the end of the level?
\end{adjustwidth}

\noindent\textbf{Q.9:} What do you think was the main purpose of the (video | game)? 

\noindent\textbf{Q.10:} Based on your impression of the (ad | game), what do you think (Apple - Auctioneer | Samsung - yellow umbrella | WhatsApp - Pigeon | Twitter - envelope avatar | DuckDuckGo - Vocalist) represents?

\noindent\textbf{Q.11:} On a scale of 1 - 5, with 1 being \quotes{Poor} and 5 being \quotes{Excellent,} how well do you think the (ad | game) communicated that message?
\begin{adjustwidth}{0.5cm}{}
[Prompt] Why did you give it that rating?
\end{adjustwidth}

\noindent [Q.12A is for the ad campaign participants, Q.12B for the game participants.] \newline
\textbf{Q.12A:} [Apple/Samsung/WhatsApp/DuckDuckGo] If you came across this advertisement online while browsing the internet, how long would you have stayed with the video? \newline 
\textbf{Q.12B:} [Twitter] If you came across this game while using Twitter, would you have played the game?
\begin{adjustwidth}{0.5cm}{}
    [Prompt] How long would you have played it? \newline
    [Prompt] Why/Why not?
\end{adjustwidth}

\noindent\textbf{Q.13:} Prior to watching this video/playing this game, were you aware of (Apple \& Samsung - smartphone features shown | WhatsApp - end-to-end encryption | Twitter - privacy settings | DuckDuckGo - web tracking)?
\begin{adjustwidth}{0.5cm}{}
    \textbf{Q.13A:} How would you describe (Apple \& Samsung - the features shown | WhatsApp - end-to-end encryption | Twitter - privacy settings | DuckDuckGo - web tracking)? \newline
    \textbf{Q.13B:} How did you come across it (or across this term for end-to-end encryption and web tracking)?
\end{adjustwidth}

\noindent\textbf{Q.14:} Having (seen this advertisement | played the game), what is your impression of (Apple | Samsung | WhatsApp | Twitter | DuckDuckGo)?

\noindent[Apple, Samsung, Twitter participants move to Section 5. WhatsApp \& DuckDuckGo participants move to Section 9.] \newline

\noindent\textbf{Section 5 - Actionable Steps [Apple/Samsung/Twitter]}

\noindent Relating back to the ad, 

\noindent \textbf{Q.15:} What actions do you think the advertisement was suggesting you could take? 

\noindent\textbf{Q.16:} What are the steps you would need to take to complete those actions?
\begin{adjustwidth}{0.5cm}{}
    [Prompt] How would you go about this?
\end{adjustwidth}

\noindent\textbf{Q.17:} On a scale of 1 \quotes{Not at all motivated} to 5 \quotes{Extremely motivated,} how would you rank your motivation to (perform the steps | go to the game-level setting) shown in the (advertisement | game)?

\noindent\textbf{Q.18:} On a scale of 1 \quotes{Not at all confident} to 5 \quotes{Fully confident,} how would you rank your confidence in being able to (perform the steps | go to the game-level setting) shown in the (advertisement | game)?

\noindent[Apple, Samsung to Section 6. Twitter to Section 7] \newline

\noindent\textbf{Section 6 - Activity Instruction (Apple/Samsung)} 

Let’s try a quick exercise. I am going to share my phone screen, next, you can give me verbal directions on where to go in order to find [Apple (1) Access App tracking permission page, (2) Turn on Protection Mail Activity; Samsung (1) Change Location permission for Weather App, (2) Turn On/Off overall microphone permission for Zoom.] During this process, feel free to try different approaches, ask to start from the beginning or request hints if needed. 

\noindent[Link phone to screen share into Zoom - 3 minutes max for both activity (1) and (2) combined, do not reveal time to participants]

Thank you. 

\noindent[Proceed to Section 8] \newline 

\noindent\textbf{Section 7 - Activity Instruction (Twitter)}

I am now going to share my screen and show you the Twitter Home page. Next, please give me verbal directions on how to reach ad preferences setting. 

\noindent[Watch for activity time, 3 minutes max]

Thank you

\noindent[Proceed to Section 8] \newline

\noindent\textbf{Section 8 - Activity Feedback (Apple/Samsung/Twitter)} 

\noindent\textbf{Q.19:} How do you feel about the exercise we just did?

\noindent\textbf{Q.20:} Having done the exercise, is there something new you have learned?
\begin{adjustwidth}{0.5cm}{}
    [Prompt] How do you feel about the (ad | game) now? \newline 
    [Prompt] Anything new you remembered about the (ad | game)? \newline
\end{adjustwidth} 

\noindent\textbf{Section 9 - User Motivation}

\noindent[For non-users]

\noindent\textbf{Q.21:} Did the (ad | game) motivate you to consider switching from your current (Apple \& Samsung - smartphone | WhatsApp - instant messaging | Twitter - social media | DuckDuckGo - browser and search engine) to the (ad | game)'s (smartphone | instant messaging | social media | browser and search engine)?
\begin{adjustwidth}{0.5cm}{}
    [Prompt] Why? \newline 
    [Prompt] Why not? 
\end{adjustwidth}

\noindent[For users]

\noindent\textbf{Q.22} Did the (ad | game) motivate you to continue using your (Apple \& Samsung - smartphone | WhatsApp - instant messaging | Twitter - social media | DuckDuckGo - browser and search engine) rather than switching to a different one?
\begin{adjustwidth}{0.5cm}{}
    [Prompt] Why? \newline 
    [Prompt] Why not?
\end{adjustwidth}

\noindent[Twitter skips Q.23, interview completed.]

Let's watch the video together one more time.

\noindent[Show advertisement again.]

\noindent\textbf{Q.23:} Having seen the advertisement again, did you notice anything new? 

\noindent[Proceed to End. WhatsApp DuckDuckGo group start again at S1, repeats entire process]

Thank you, that concludes the end of the interview. I will now stop the recording. Do you have any concluding remarks or questions regarding what we did today?

\clearpage
\onecolumn
\subsection{Ads Screens} \label{app:screen}

\begin{figure*}[!h]
\centering
\resizebox{\textwidth}{!}{%
\begin{minipage}[b]{\textwidth}
    \begin{minipage}[b]{0.49\textwidth}
        \subfloat[Data auction.\label{fig:apple-01}]{%
        \includegraphics[width=\textwidth]{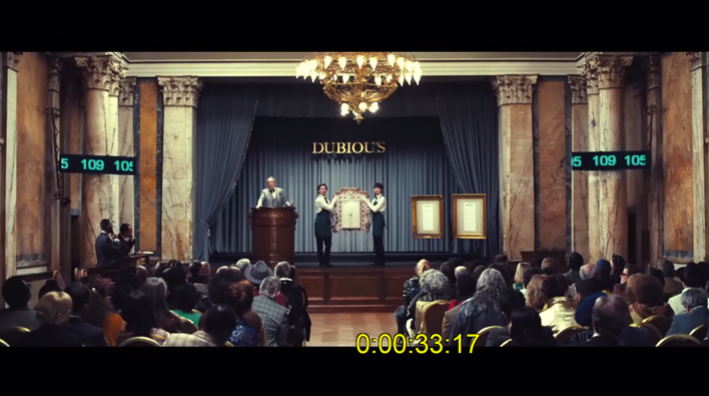}
        \Description{apple-01}
        }
    \end{minipage}
    \hfill
    \begin{minipage}[b]{0.49\textwidth}
        \subfloat[The protagonist uses the \quotes{Ask App Not To Track} feature.\label{fig:apple-02}]{%
        \includegraphics[width=\textwidth]{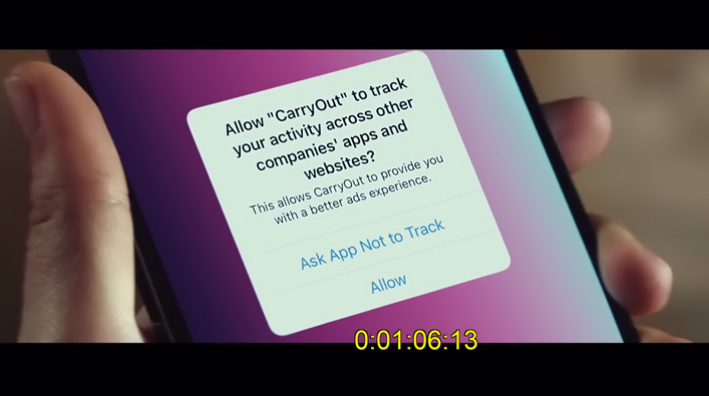}
        \Description{apple-02}
        }
    \end{minipage}

    \begin{minipage}[b]{0.49\textwidth}
        \subfloat[The protagonist uses the \quotes{Protect Mail Activity} feature.\label{fig:apple-03}]{%
        \includegraphics[width=\textwidth]{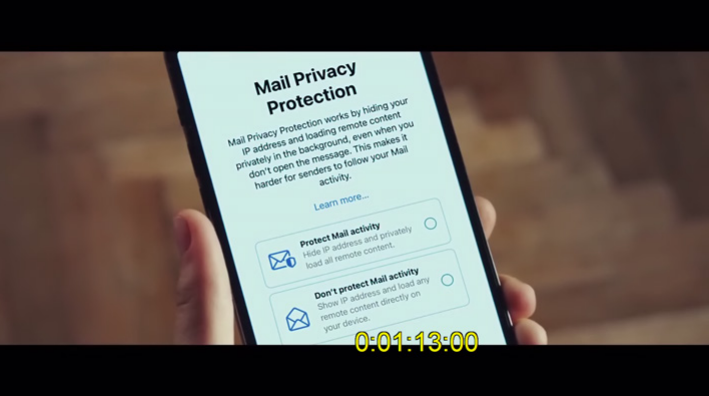}
        \Description{apple-03}
        }
    \end{minipage}
    \hfill
    \begin{minipage}[b]{0.49\textwidth}
        \subfloat[The auction participants disappeared into smoke after the protagonist used Apple's privacy features.\label{fig:apple-04}]{%
        \includegraphics[width=\textwidth]{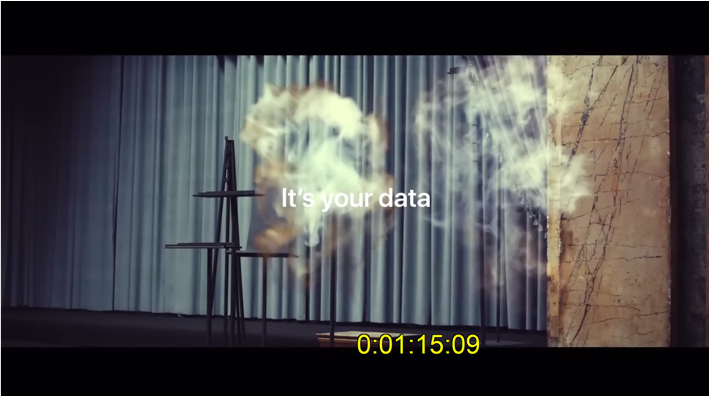}
        \Description{apple-04}
        }
    \end{minipage}
\end{minipage}    
}
\caption{Examples of Apple's ad screens.} \label{fig:apple_extended}
\end{figure*}

\begin{figure*}[!h]
\centering
\resizebox{\textwidth}{!}{%
\begin{minipage}[b]{\textwidth}
    \begin{minipage}[b]{0.49\textwidth}
        \subfloat[Samsung's privacy dashboard showing which apps accessing which permission (e.g., camera, microphone, and location).\label{fig:samsung-01}]{%
        \includegraphics[width=\textwidth]{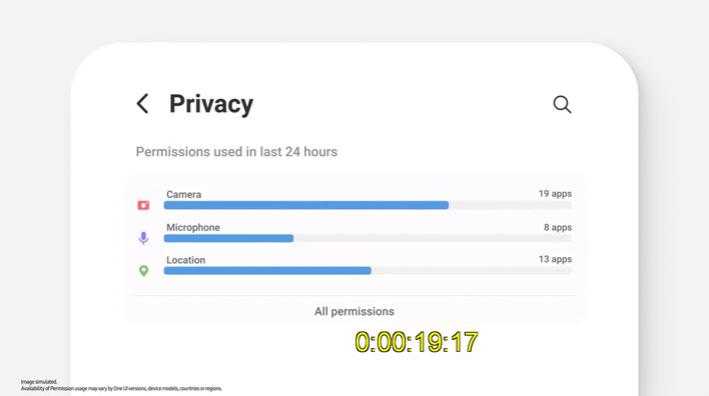}
        \Description{samsung-01}
        }
    \end{minipage}
    \hfill
    \begin{minipage}[b]{0.49\textwidth}
        \subfloat[The protagonist turned off the precise location for the Weather app, portrayed by a yellow umbrella.\label{fig:samsung-02}]{%
        \includegraphics[width=\textwidth]{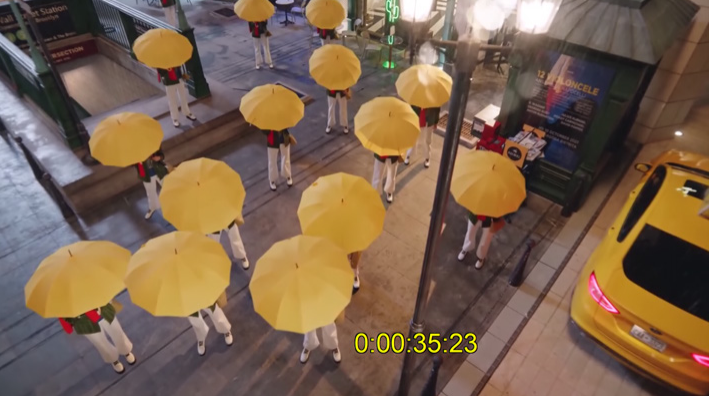}
        \Description{samsung-02}
        }
    \end{minipage}

    \begin{minipage}[b]{0.49\textwidth}
        \subfloat[The protagonist turned off the precise location for the Weather app.\label{fig:samsung-03}]{%
        \includegraphics[width=\textwidth]{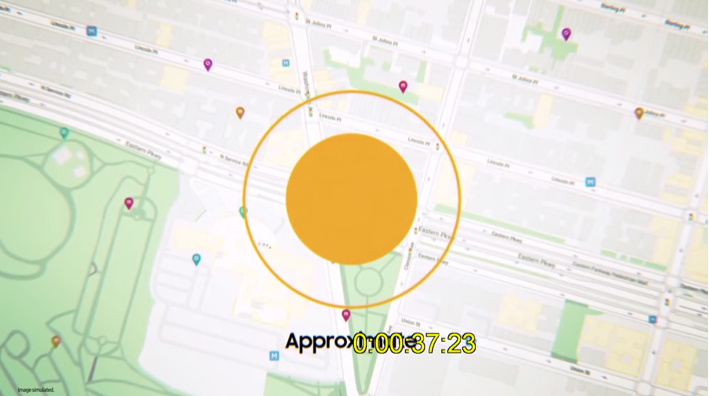}
        \Description{samsung-03}
        }
    \end{minipage}
    \hfill
    \begin{minipage}[b]{0.49\textwidth}
        \subfloat[The protagonist uses location access to control when to allow an app access their location. \label{fig:samsung-04}]{%
        \includegraphics[width=\textwidth]{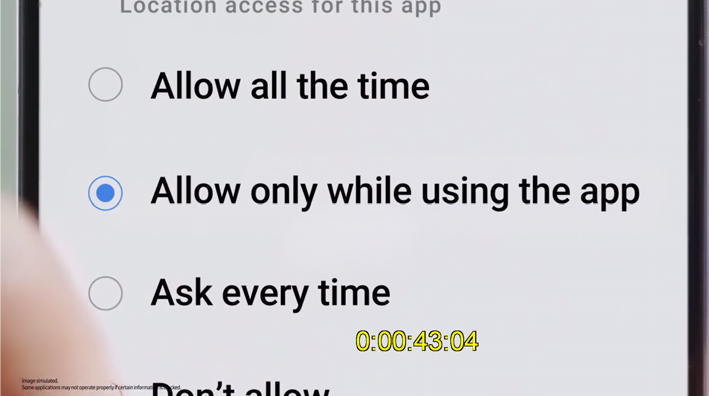}
        \Description{samsung-04}
        }
    \end{minipage}
\end{minipage}    
}
\caption{Examples of Samsung's ad screens.} \label{fig:samsung_extended}
\end{figure*}

\begin{figure*}[!h]
\centering
\resizebox{\textwidth}{!}{%
\begin{minipage}[b]{\textwidth}
    \begin{minipage}[b]{0.49\textwidth}
        \subfloat[The post office.\label{fig:whatsapp-01}]{%
        \includegraphics[width=\textwidth]{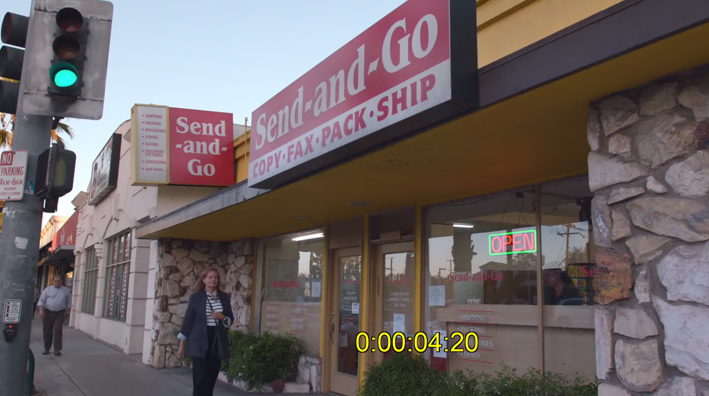}
        \Description{whatsapp-01}
        }
    \end{minipage}
    \hfill
    \begin{minipage}[b]{0.49\textwidth}
        \subfloat[The post office clerk uses pigeons as a message delivery method. \label{fig:whatsapp-02}]{%
        \includegraphics[width=\textwidth]{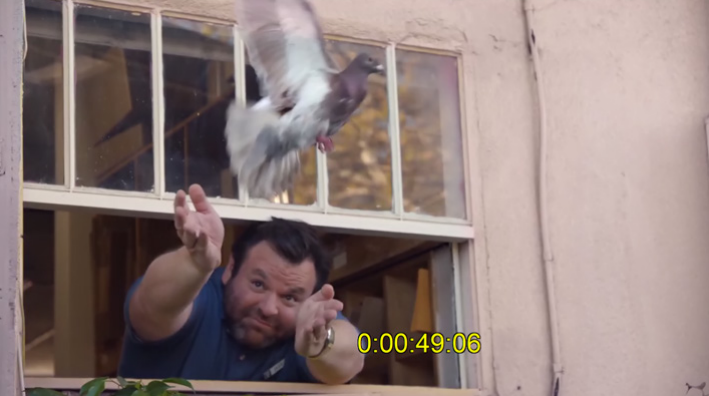}
        \Description{whatsapp-02}
        }
    \end{minipage}

    \begin{minipage}[b]{0.49\textwidth}
        \subfloat[The ad states that 5.5 billion texts
per day are sent without encryption.\label{fig:whatsapp-03}]{%
        \includegraphics[width=\textwidth]{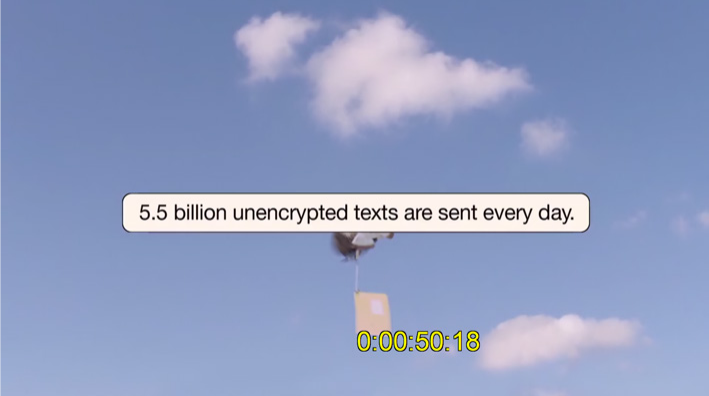}
        \Description{whatsapp-03}
        }
    \end{minipage}
    \hfill
    \begin{minipage}[b]{0.49\textwidth}
        \subfloat[The ad states that with WhatsApp, your messages (the encrypted messages) will not be one of them.\label{fig:whatsapp-04}]{%
        \includegraphics[width=\textwidth]{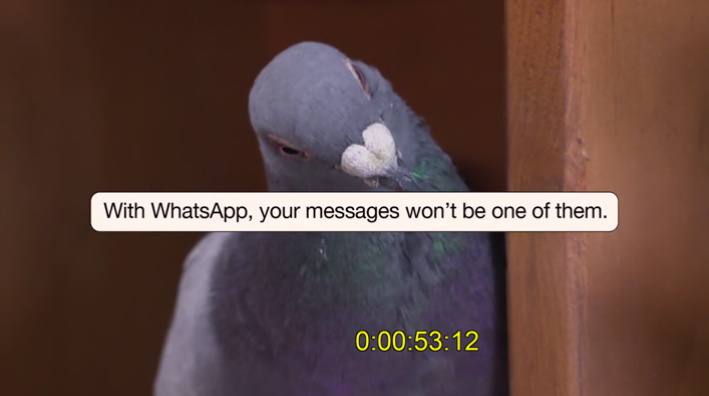}
        \Description{whatsapp-04}
        }
    \end{minipage}
\end{minipage}    
}
\caption{Examples of WhatsApp's ad screens.} \label{fig:whatsapp_extended}
\end{figure*}
\begin{figure*}[!h]
\centering
\resizebox{\textwidth}{!}{%
\begin{minipage}[b]{\textwidth}
    \begin{minipage}[b]{0.49\textwidth}
        \subfloat[The singer shoulder-surfs a person with their mobile browsing the Internet.\label{fig:ddg_01}]{%
        \includegraphics[width=\textwidth]{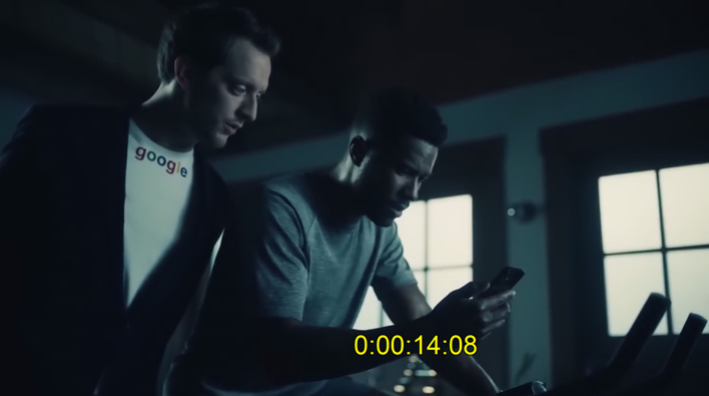}
        \Description{ddg-01}
        }
    \end{minipage}
    \hfill
    \begin{minipage}[b]{0.49\textwidth}
        \subfloat[A screen showing a search using DuckDuckGo.\label{fig:ddg_02}]{%
        \includegraphics[width=\textwidth]{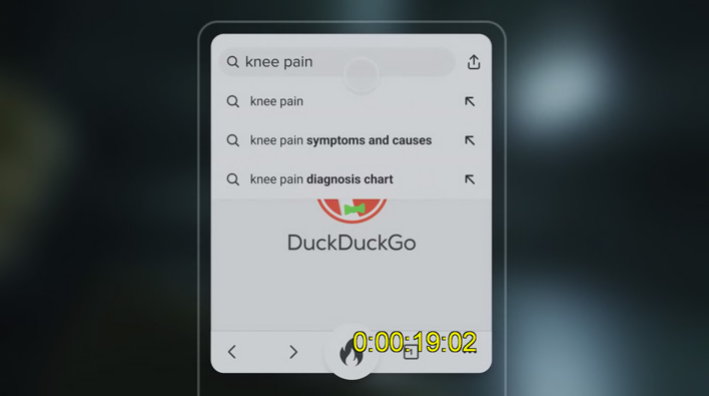}
        \Description{ddg-02}
        }
    \end{minipage}

    \begin{minipage}[b]{0.49\textwidth}
        \subfloat[A screen showing a medical web page using DuckDuckGo showing trackers blocked.\label{fig:ddg_03}]{%
        \includegraphics[width=\textwidth]{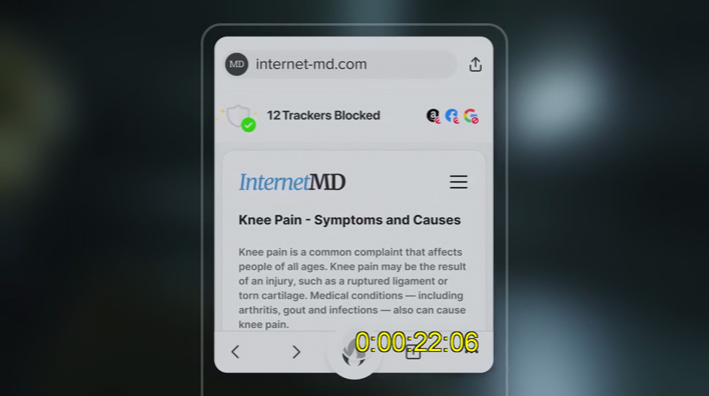}
        \Description{ddg-03}
        }
    \end{minipage}
    \hfill
    \begin{minipage}[b]{0.49\textwidth}
        \subfloat[After the person started using DuckDuckGo, the singer dragged away by an invisible force.\label{fig:ddg_04}]{%
        \includegraphics[width=\textwidth]{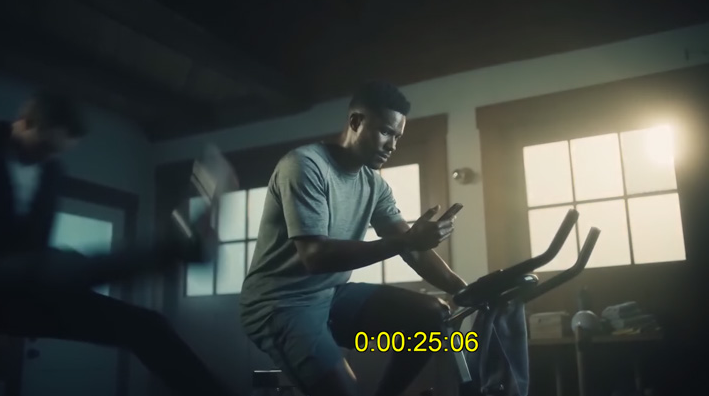}
        \Description{ddg-04}
        }
    \end{minipage}
\end{minipage}    
}
\caption{Examples of DuckDuckGo's ad screens.} \label{fig:ddg_extended}
\end{figure*}

\begin{figure*}[!h]
\centering
\resizebox{\textwidth}{!}{%
\begin{minipage}[b]{\textwidth}
    \begin{minipage}[b]{0.49\textwidth}
        \subfloat[During level 1 of the game.\label{fig:twitter-01}]{%
        \includegraphics[width=\textwidth]{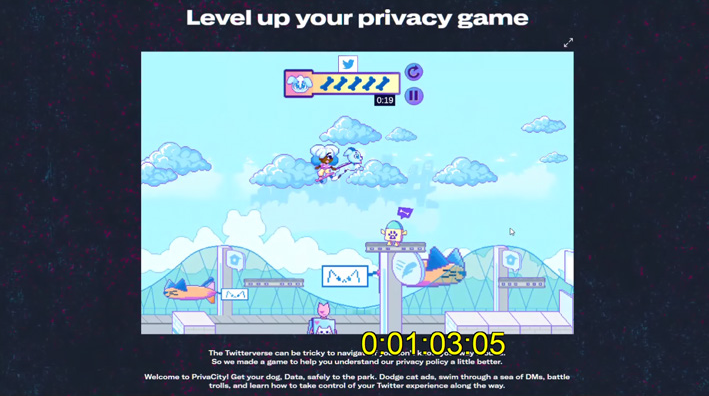}
        \Description{twitter-01}
        }
    \end{minipage}
    \hfill
    \begin{minipage}[b]{0.49\textwidth}
        \subfloat[End of level 1 of the game.\label{fig:twitter-02}]{%
        \includegraphics[width=\textwidth]{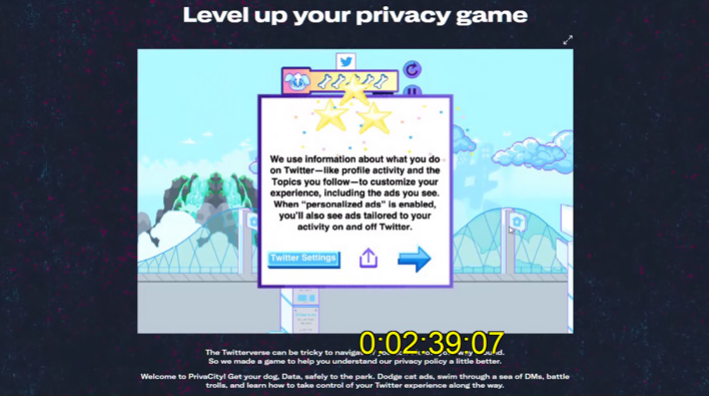}
        \Description{twitter-02}
        }
    \end{minipage}

    \begin{minipage}[b]{0.49\textwidth}
        \subfloat[Start of level 2 of the game.\label{fig:twitter-03}]{%
        \includegraphics[width=\textwidth]{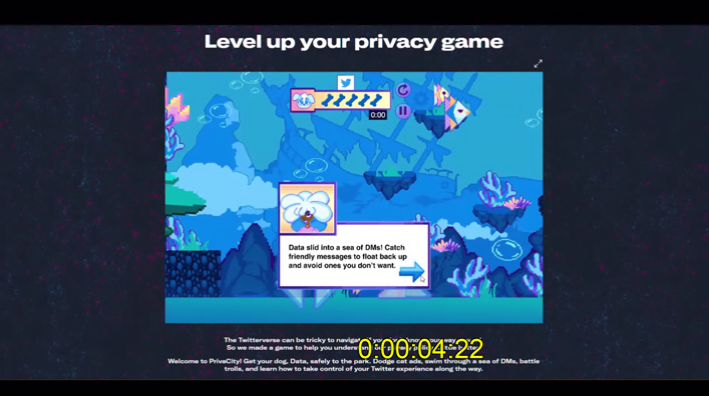}
        \Description{twitter-03}
        }
    \end{minipage}
    \hfill
    \begin{minipage}[b]{0.49\textwidth}
        \subfloat[During level 2 of the game\label{fig:twitter-04}]{%
        \includegraphics[width=\textwidth]{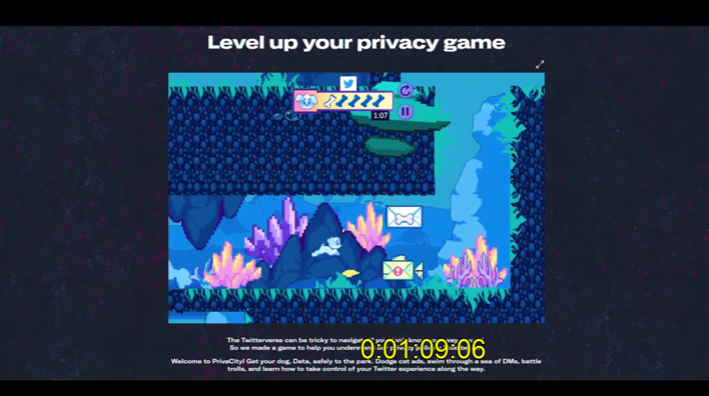}
        \Description{twitter-04}
        }
    \end{minipage}
\end{minipage}    
}
\caption{Examples of Twitter's game screens.} \label{fig:twitter_extended}
\end{figure*}


\clearpage
\onecolumn
\subsection{Participant Overview} \label{app:p-info}
\begin{table}[htbp!]
\footnotesize
    \centering
    \caption{Participant Overview. In the \quotes{User Type} column, U represents users, N represents non-users, W represents WhatsApp, and D represents DuckDuckGo. The \quotes{Tech} column indicates whether participants indicated in the screening survey that they had any experience learning or working in information technology or a related field.}
    \label{tab:demo}
        \begin{tabular}{l|l|l|l|l|l}
            \toprule
            \textbf{ID} & \textbf{Gender} & \textbf{Age} & \textbf{User Type} & \textbf{Tech} & \textbf{Recruitment}\\
            \midrule
            A1 & F & 46 - 55 & U & No & Prolific\\
            \hline
            A2 & F & 18 - 25 & N & No & Prolific\\
            \hline
            A3 & M & 46 - 55 & N & Yes & Prolific\\
            \hline
            A4 & M & 26 - 35 & N & No & Prolific\\
            \hline
            A5 & F & 26 - 35 & U & Yes & Snowball\\
            \hline
            A6 & M & 26 - 35 & U & Yes & Snowball\\
            \hline
            S1 & F & 36 - 45 & U & No & Prolific\\
            \hline
            S2 & M & 56 - 65 & N & No & Prolific\\
            \hline
            S3 & M & 46 - 55 & U & Yes & Prolific\\
            \hline
            S4 & M & 26 - 35 & U & No & Snowball\\
            \hline
            S5 & F & 18 - 25 & N & Yes & Snowball\\
            \hline
            S6 & F & 18 - 25 & N & Yes & Snowball\\
            \hline
            WD1 & M & 36 - 45 & W-U; D-N & No & Prolific\\
            \hline
            WD2 & M & 26 - 35 & W-N; D-U & No & Prolific\\
            \hline
            WD3 & F & 26 - 35 & W-U; D-N & No & Snowball\\
            \hline
            WD4 & M & 18 - 25 & W-U; D-U & Yes & Snowball\\
            \hline
            WD5 & M & 36 - 45 & W-N; D-U & Yes & Prolific\\
            \hline
            WD6 & M & 18 - 25 & W-N; D-N & No & Prolific\\
            \hline
            T1 & M & 26 - 35 & U & No & Prolific\\
            \hline
            T2 & F & 26 - 35 & U & Yes & Prolific\\
            \hline
            T3 & M & 46 - 55 & U & No & Prolific\\
            \hline
            T4 & F & 18 - 25 & U & No & Snowball\\
            \hline
            T5 & F & 18 - 25 & U & No & Snowball\\
            \hline
            T6 & F & 18 - 25 & U & Yes & Snowball\\
             \bottomrule
        \end{tabular}
\end{table}
\clearpage
\begin{landscape}
\subsection{Diagram for Our Qualitative Analysis Coding} \label{app:diagram}
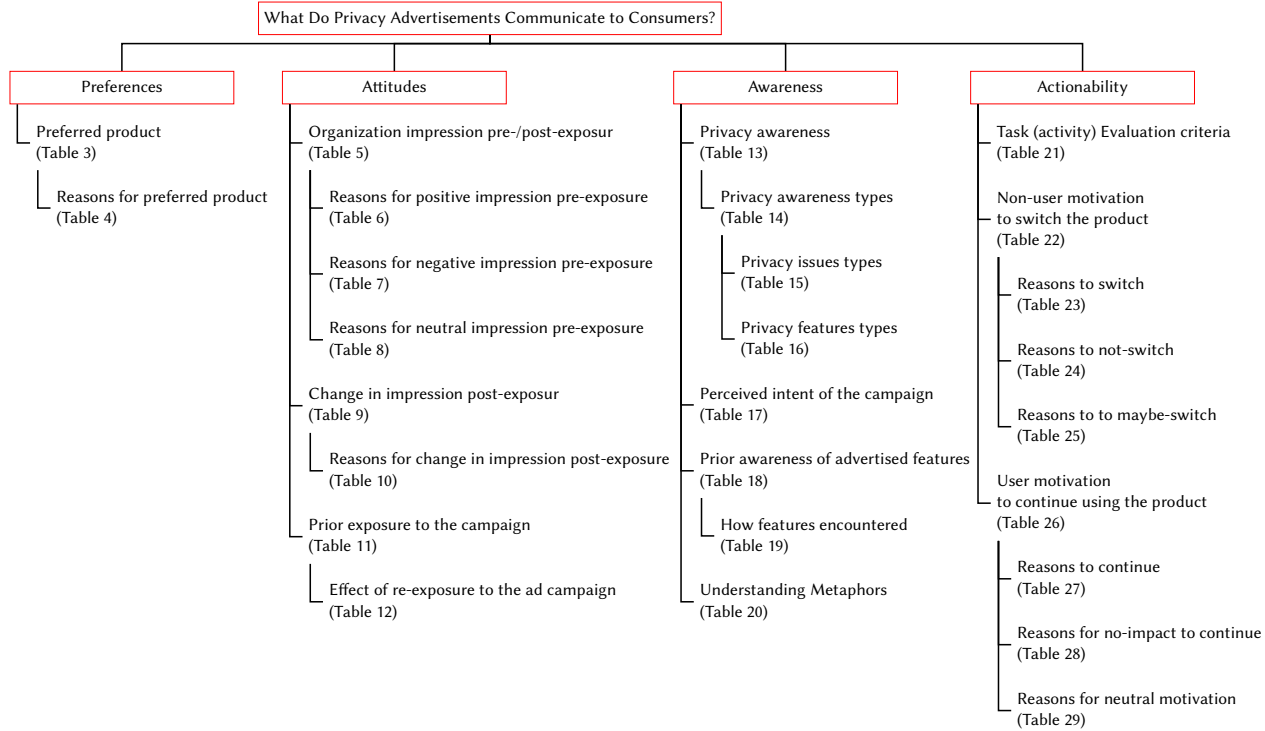
\begin{figure}[!h]
\centering
\caption{A diagram summarizing our qualitative analysis coding.}
\label{fig:diagram}

\resizebox{0.95\textwidth}{!}{%
\forestset{
	direction switch/.style={
		forked edges,
		for tree={
			align=left,
			edge+=thick, 
			font=\sffamily,
		},
		where level=1{minimum width=13em}{},
		where level<=1{draw=red}{},
		where level>=1{folder, grow'=0}{},
	},
}
	\newlength\gap
	\setlength\gap{10mm}

\begin{forest} 
		direction switch
		[What Do Privacy Advertisements Communicate to Consumers? 
			[Preferences
				[Preferred product\\(\autoref{tab:preferred})
                    [Reasons for preferred product\\(\autoref{tab:preferred_reason})]
				]
			] 
			[Attitudes
                [Organization impression pre-/post-exposur\\(\autoref{tab:attitude})
    				[Reasons for positive impression pre-exposure\\(\autoref{tab:attitude_reason_pos})]
                    [Reasons for negative impression pre-exposure\\(\autoref{tab:attitude_reason_neg})]
                    [Reasons for neutral impression pre-exposure\\(\autoref{tab:attitude_reason_neutral})]
                ]
                [Change in impression post-exposur\\(\autoref{tab:change})
    				[Reasons for change in impression post-exposure\\(\autoref{tab:change_reason})]
                ]
                [Prior exposure to the campaign\\(\autoref{tab:prior_exposure})
    				[Effect of re-exposure to the ad campaign\\(\autoref{tab:re-exposure})]
                ]
			] 
			[Awareness
				[Privacy awareness\\(\autoref{tab:awareness})
					[Privacy awareness types\\(\autoref{tab:priv_aware})
						[Privacy issues types\\(\autoref{tab:priv_issue_type})]
						[Privacy features types\\(\autoref{tab:priv_feature_type})]
					]
				]
				[Perceived intent of the campaign\\(\autoref{tab:intent})]
				[Prior awareness of advertised features\\(\autoref{tab:prior_awareness})
					[How features encountered\\(\autoref{tab:how_encountered})]
				]
                [Understanding Metaphors\\(\autoref{tab:metaphore})]
			] 
			[Actionability
                [Task (activity) Evaluation criteria\\ (\autoref{tab:actionability})]
				[Non-user motivation\\to switch the product\\(\autoref{tab:switch})
					[Reasons to switch\\(\autoref{tab:switch_reason})]
					[Reasons to not-switch\\(\autoref{tab:not_switch_reason})]
					[Reasons to to maybe-switch\\(\autoref{tab:maybe_switch_reason})]
				]
				[User motivation\\to continue using the product\\(\autoref{tab:continue})
					[Reasons to continue\\(\autoref{tab:continue_reason})]
                    [Reasons for no-impact to continue\\(\autoref{tab:not_continue_reason})]
                    [Reasons for neutral motivation\\(\autoref{tab:neutral_continue_reason})]
				]
			] 
		]
	\end{forest}
}
\end{figure}
\end{landscape}
\clearpage
\clearpage
\subsection{Codebook} \label{app:codebook}
In this section, we list our codebook divided into tables. Each table represents the codebook for a question or a set of related questions. All codes are in small letters only.~\autoref{fig:diagram} represents a diagram that draws the overall picture of the codes hierarchy.

\begin{table*}[!h]
\caption{The codebook for Q.1 about participants' preferred products. Note that a participant's preferred product can be different from the product of the treatment group the participant was assigned to.}
\label{tab:preferred}
\renewcommand{\arraystretch}{1.25}
\resizebox{0.85\textwidth}{!}{%
    \begin{tabular}{p{0.19\textwidth}|p{0.45\textwidth}|p{0.45\textwidth}}
    \toprule 
    \rowcolor{gray!50}
    \multicolumn{3}{l}{Q.1: Preferences $\rightarrow$ Preferred product} \\
    \midrule
    \textbf{Code (\# frequency) [non-exclusive]}   & \textbf{Code Definition}   & \textbf{Example}     \\
    \hline
    apple (6)               & The Apple smartphone              & \quotes{iPhone.} (A1); \quotes{Apple.} (A5)\\ 
    \hline 
    samsung (5)              & The Samsung smartphone            & \quotes{Samsung galaxy.} (S3); \quotes{Samsung.} (S4)\\
    \hline
    engine-google (4)        &  The Google search engine         &  \quotes{Google.} (WD3, DuckDuckGo)\\
    \hline
    instagram (3)            & The Instagram app                  &  \quotes{Instagram, I'd say.} (T1)\\
    \hline
    browser-google (3)       & The Google web browser            &  \quotes{I use Google and Safari on my laptops} (WD4, DuckDuckGo)\\
    \hline
    lg (2)                   & The LG smartphone                 & \quotes{Samsung and LG.} (A4)\\
    \hline
    facebook-messenger (2)   & The Facebook messenger app       &  \quotes{I use Facebook messenger usually to communicate with friends.} (WD2, WhatsApp) \\
    \hline
    twitter (2)              & The Twitter app                   &  \quotes{Probably Twitter.} (T2)\\
    \hline
    facebook (1)            & The Facebook app                  &  \quotes{Facebook is primarily what I use.} (T3)\\
    \hline
    tiktok (1)               & The TikTok app                    &  \quotes{Probably TikTok.} (T6) \\
    \hline
    discord (1)              & The Discord app                   &  \quotes{I use Discord the most.} (WD1, WhatsApp)\\
    \hline
    browser-firefox (1)      & The Firefox web browser           &  \quotes{My preferred browser is Firefox} (WD5, DuckDuckGo)\\
    \hline
    browser-edge (1)         & The Edge web browser              &  \quotes{I alternate between Microsoft Edge and Google Chrome.} (WD6, DuckDuckGo) \\
    \hline
    browser-safari (1)       & The Safari web browser            &  \quotes{I use Google and Safari on my laptops} (WD4, DuckDuckGo)\\
    \hline
    browser-ddg (1)         &  The DuckDuckGo web browser       &  \quotes{And I use DuckDuckGo and, mostly DuckDuckGo on my phone} (WD4, DuckDuckGo)\\
    \hline
   
    engine-ddg (1)          &  The DuckDuckGo search engine     &  \quotes{Sometimes I'll use DuckDuckGo or Bing} (WD2, DuckDuckGo)\\
    \hline
    engine-bing (1)          &  The Bing search engine           &  \quotes{Sometimes I'll use DuckDuckGo or Bing} (WD2, DuckDuckGo)\\
    \hline
    whatsapp (1)           &  The WhatsApp app                 &  \quotes{I prefer using Whatsapp.} (WD4, WhatsApp)\\
    \hline
    snapchat (1)             &  The Snapchat app                 &  \quotes{I generally prefer Snapchat as an instant messaging service.} (WD6, WhatsApp)\\
    \hline
    telegram (1)            &  The Telegram app                 &  \quotes{Telegram.} (WD5, WhatsApp)\\
    \bottomrule
    \end{tabular}
} 
\end{table*}

\begin{table*}[!h]
\caption{The codebook for Q.1A about the reasons for participants' preferred products. Note that a participant's preferred product can be different from the product of the treatment group the participant was assigned to.}
\label{tab:preferred_reason}
\renewcommand{\arraystretch}{1.25}
\resizebox{0.85\textwidth}{!}{%
    \begin{tabular}{p{0.19\textwidth}|p{0.45\textwidth}|p{0.45\textwidth}}
    \toprule 
    \rowcolor{gray!50}
    \multicolumn{3}{l}{Q.1A: Preferences $\rightarrow$ Reason for preferred product} \\
    \midrule
    \textbf{Code (\# frequency) [non-exclusive]}   & \textbf{Code Definition}   & \textbf{Example}     \\
    \hline 
    familiarity (8) & The participant is familiar with the product, or has been using it for long time & \quotes{because I am used to it.} (A4)  \\
    \hline
    privacy (7) & The product has good privacy & \quotes{the privacy features.} (WD5, DuckDuckGo) \\
    \hline
    sync (5) & The product is easy to synchronize with other products & \quotes{it is a lot easier to sync data.} (A5) \\
    \hline
    used-by-peers-family (5) & The product is used by peers and family & \quotes{most of the people I want to contact on that app} (WD3, WhatsApp)  \\
    \hline
    ease-use (3) & The product is easy to use &  \quotes{pretty simple and easy to use} (A3) \\
    \hline
    long-lasting (2) & The product lasts for long time  &  \quotes{The device has pretty long life} (S6) \\
    \hline
    good-product (2) & The product is good generally &  \quotes{So LGs always worked very well} (A3)\\
    \hline
    focused-function (2) & The product is focused on one function & \quotes{it's just always videos.} (T6)  \\
    \hline
    performance (2) & The product has good performance & \quotes{It seems to be faster.} (S2)  \\
    \hline
    skeptical (2) & The participant has skepticism about the campaign &  \quotes{end-to-end encryption. Even though I’m not 100\% sure about that anymore.} (WD4, WhatsApp)\\
    \hline
    open-env (1) & The product is not restricted to the company's app store & \quotes{I like that you are able to download apps outside of the Google play store} (A2) \\
    \hline
    security (1) & The product has good security & \quotes{your browsing, is, is private and secure.} (WD5, DuckDuckGo) \\
    \hline
    switched-bad-product (1) & The participant switched product because of previous bad experience with other products & \quotes{I had a Google phone before ... it was really slow, and it crashed a lot ... so I switched.} (A6) \\
    \hline
    fun (1) & The product is fun & \quotes{because it's very humorous} (T4) \\
    \hline
    developed-product (1) & The product is developed & \quotes{it's kind of the more developed search engine } (WD2, DuckDuckGo) \\
    \hline
    popularity (1) & The product is popular & \quotes{this is the most popular one that i've or It's the one that I sort of am most aware of.} (WD3, DuckDuckGo) \\
    \hline
    good-interface (1) & The product has good interface &  \quotes{Has a good interface overall.} (S6) \\
    \hline
    personalized (1) & The product is personalized & \quotes{it has channels to design toward my interest} (WD1, WhatsApp)  \\
    \hline
    combined-function (1) & The product combines multiple functions & \quotes{but kind of all rolled up into one with like stories and messages, but also posting.} (T5)\\
    \bottomrule
    \end{tabular}
} 
\end{table*}

\begin{table*}[!h]
\caption{The codebook for Q.5 \& Q.14 about participants' impressions about the advertising organization. The codes used for both pre-exposure (Q.5) and post-exposure (Q.14) to the campaign, except codes prefixed with \quotes{-plus} are applicable only for post-exposure.}
\label{tab:attitude}
\renewcommand{\arraystretch}{1.25}
\resizebox{0.85\textwidth}{!}{%
    \begin{tabular}{p{0.19\textwidth}|p{0.45\textwidth}|p{0.45\textwidth}}
    \toprule 
    \rowcolor{gray!50}
    \multicolumn{3}{l}{Q.5 \& Q.14: Attitudes $\rightarrow$ Organization impression pre-/post-exposure} \\
    \midrule
    \textbf{Code (\# frequency Q.5; Q14)\newline[exclusive]}   & \textbf{Code Definition}   & \textbf{Example}     \\
    \hline 
    positive (13; 15)   & The participant expressed positive impression only  & \quotes{I have positive feelings about it.} (WD5, DuckDuckGo) \\
    \hline
    mixed-pos (7; 5)  & The participant expressed mixed positives and negatives impression, with stronger positive. Stronger positive means either more positive points mentioned, or if equal positives and negative points mentioned, then positive came first  & \quotes{They seem better than the other tech companies. Relatively. They sell hardware rather than selling my data. I guess they probably sell my data, but they seem to do it less. So I like that.} (A6) \\
    \hline
     neutral  (4; 0)       &  The participant expressed neutral impression & \quotes{I don't really have an impression.} (T5) \\
    \hline 
    mixed-neg (3; 2)  &  The participant expressed mixed negative and positive impression, with stronger negative. Stronger negative means either more negative points mentioned, or if equal negative and positive points mentioned, then negative came first  & \quotes{I usually don’t trust the parent company Meta, but they seem to be doing all right.} (WD5, WhatsApp)\\
    \hline
    negative (3; 3)    & The participant expressed negative impression only  & \quotes{it's not great.} (T2) \\
    \hline
    pos-plus (n.a.; 4)        &  The participant expressed even more positive. This code is applicable only post-exposure to the campaign  & \quotes{Probably have a better impression on Apple.} (A5)\\
    \hline
    neg-plus (n.a.; 1)        &  The participant expressed even more negative. This code is applicable only post-exposure to the campaign  & \quotes{My opinion Twitter, is slightly worse} (T1) \\
    \bottomrule
    \end{tabular}
} 
\end{table*}

\begin{table*}[!h]
\caption{The codebook for Q.5 about the reasons for participants' positive impressions about the advertising organization pre-exposure to the campaign.}
\label{tab:attitude_reason_pos}
\renewcommand{\arraystretch}{1.25}
\resizebox{0.85\textwidth}{!}{%
    \begin{tabular}{p{0.19\textwidth}|p{0.45\textwidth}|p{0.45\textwidth}}
    \toprule 
    \rowcolor{gray!50}
    \multicolumn{3}{l}{Q.5: Attitudes $\rightarrow$ Reasons for positive impression pre-exposure} \\
    \midrule
    \textbf{Code (\# frequency) [non-exclusive]}   & \textbf{Code Definition}   & \textbf{Example}     \\
    \hline 
    good-product (12)        &  The product is generally good  & \quotes{It's pretty good.} (WD3, WhatsApp)  \\
    \hline 
    privacy-focused (6)    &  The product is privacy focused  &  \quotes{a search engine platform that doesn't steal your data and doesn’t record your info.} (WD6, DuckDuckGo) \\
    \hline 
    good-company (3)        & The company is generally good   & \quotes{Apple as an the organization I think they, they're pretty good.} (A3) \\
    \hline 
    good-competitor (2)    &  The company is a good competitor  &  \quotes{a well-respected competitors in the marketplace.} (S1) \\
    \hline 
    fun (2)                &  The product is fun   &  \quotes{it's still a fun place to be.} (T4)  \\
    \hline 
    reliable-product (1)    &  The product is reliable  & \quotes{I know whenever SnapChat is down or something, they'll always have people go straight to Twitter to talk about it} (T6) \\
    \hline 
    security-focused (1)    &  The product is security focused  & \quotes{Supposedly security forward.} (WD4, DuckDuckGo) \\
    \hline 
    better-than-other-product (1) &  The product is better than other products  &  \quotes{probably better than Facebook Messenger} (WD4, WhatsApp) \\
    \hline 
    efficient-product (1)   &  The product is efficient &  \quotes{pretty efficient app} (WD6, WhatsApp) \\
    \hline
    good-product-feature (1)    & The product has good features (specific)   &  \quotes{There is some good features on them ... for example, Android has this feature, where, if you message people on either a messenger or text, and they can appear as text bubble on top of your apps.} (S6) \\
    \hline 
    innovative-company (1)  &  Innovative company  &  \quotes{I think of it as a pretty innovative company} (A1) \\
    \hline 
    sell-hardware (1)      & The company sells hardware not data   &  \quotes{They sell hardware rather than selling my data.} (A6)  \\
    \hline 
    needed-product (1)     &  The product is a necessity   &  \quotes{I definitely see the need.} (T3) \\
    \bottomrule
    \end{tabular}
} 
\end{table*}

\begin{table*}[!h]
\caption{The codebook for Q.5 about the reasons for participants' negative impressions about the advertising organization pre-exposure to the campaign.}
\label{tab:attitude_reason_neg}
\renewcommand{\arraystretch}{1.25}
\resizebox{0.85\textwidth}{!}{%
    \begin{tabular}{p{0.19\textwidth}|p{0.45\textwidth}|p{0.45\textwidth}}
    \toprule 
    \rowcolor{gray!50}
    \multicolumn{3}{l}{Q.5: Attitudes $\rightarrow$ Reasons for negative impression pre-exposure} \\
    \midrule
    \textbf{Code (\# frequency) [non-exclusive]}   & \textbf{Code Definition}   & \textbf{Example}     \\
     \hline 
     skeptical (3) & The participant is skeptical about the campaign & \quotes{I don't know how truthfully accurate that is} (WD6, DuckDuckGo)  \\
   \hline 
     false-news (2)         & The product spreads false news & \quotes{being on Twitter requires a little bit more research, and not believing the first thing I see, because news outlets} (T4)\\
    \hline 
    new-leadership (2)      & The company has new leadership that is bad  & \quotes{Currently under the new leadership of Elon Musk, it's a mess} (T4) \\
   \hline
   lack-moderation (2)      & The product lacks moderation         & \quotes{I just think there needs to be some sort of like more sort of like regulation on it.} (T1) \\
   \hline 
    high-price (1)          & The product has a high price    & \quotes{an iMac will cost on average, about twice as much as a, like given, as a given windows machine} (A4) \\
    \hline 
   noisy (1)               & The product is noisy or disrupting  & \quotes{it can be a little noisy if you follow it too much} (T3) \\
   \hline 
    marketing-strategy (1)   & The company has a bad marketing strategy & \quotes{their marketing doesn't quite click with me ... maximize what they can get out of a smaller user base rather than like necessarily expanding and like making their products more accessible to a wider based users.} (A4) \\
    \hline 
    sell-data (1)          & The company sells their customers' data  &  \quotes{I guess they probably sell my data, but they seem to do it less.} (A6) \\
    \hline 
    lack-customer-care (1) &  The company doesn't care about their customers & \quotes{they don't really care about their individual customers so much.} (S3) \\
    \hline 
    
   political-game (1)      & The product is used for political games & \quotes{it can be used they as the platform can be used, for like a political game} (T1)\\
   \hline 
    censoring (1) & The product applies censorship & \quotes{there's more censoring in Twitter than there has been before.} (T4) \\
    \hline
    spam (1)   & The product has spam  & \quotes{I do get spam on it that's always kind of annoying} (WD1, WhatsApp) \\
    \hline 
    lots-ads (1) & The product has lots of ads  & \quotes{it's just so lame with ads.} (WD2, WhatsApp) \\
    \bottomrule
    \end{tabular}
} 
\end{table*}

\begin{table*}[!h]
\caption{The codebook for Q.5 about the reasons for participants' neutral impressions about the advertising organization pre-exposure to the campaign.}
\label{tab:attitude_reason_neutral}
\renewcommand{\arraystretch}{1.25}
\resizebox{0.85\textwidth}{!}{%
    \begin{tabular}{p{0.19\textwidth}|p{0.45\textwidth}|p{0.45\textwidth}}
    \toprule 
    \rowcolor{gray!50}
    \multicolumn{3}{l}{Q.5: Attitudes $\rightarrow$ Reasons for neutral impression pre-exposure} \\
    \midrule
    \textbf{Code (\# frequency) [non-exclusive]}   & \textbf{Code Definition}   & \textbf{Example}     \\
    \hline 
    not-user (3)    & The participant is not a user of the product & \quotes{I actually don't have any impression of it ... I don't know anything about it.} (WD1, DuckDuckGo) \\
    \hline 
    new-leadership (1) & The participant mentioned the new leadership of the product or company & \quotes{I don't really have an impression ... I know that Elon Musk just bought them so he's quite a character, so I'm not super sure how I feel about them now.} (T5) \\
    
    \bottomrule
    \end{tabular}
} 
\end{table*}

\begin{table*}[!h]
\caption{The codebook for Q.14 about the change in participants' impressions about the advertising organization post-exposure to the campaign.}
\label{tab:change}
\renewcommand{\arraystretch}{1.25}
\resizebox{0.85\textwidth}{!}{%
    \begin{tabular}{p{0.19\textwidth}|p{0.45\textwidth}|p{0.45\textwidth}}
    \toprule 
    \rowcolor{gray!50}
    \multicolumn{3}{l}{Q.14: Attitudes $\rightarrow$ change in impression post-exposure} \\
    \midrule
    \textbf{Code (\# frequency)\newline[exclusive]}   & \textbf{Code Definition}   & \textbf{Example}     \\
    \hline 
    more-pos (13)   & The participant impression changed towards more positive & \quotes{I guess it's more positive} (A6)\\
    \hline 
    same (12)       & The participant impression did not change. This either expressed explicitly, or implicitly by giving an impression similar to the pre-exposure impression & \quotes{I'd say it hasn't changed.} (T3)          \\
    \hline
    more-neg (5)    & The participant impression changed towards more negative & \quotes{Slightly more negative} (WD4, WhatsApp)   \\
    
    \bottomrule
    \end{tabular}
} 
\end{table*}

\begin{table*}[!h]
\caption{The codebook for Q.14 about the reasons for the change in participants' impressions about the advertising organization post-exposure to the campaign. Note that the frequency of each code here represents the aggregate number where this code occurred in any combination of impressions change between pre- and post-exposure to the campaign.}
\label{tab:change_reason}
\renewcommand{\arraystretch}{1.25}
\resizebox{0.85\textwidth}{!}{%
    \begin{tabular}{p{0.19\textwidth}|p{0.45\textwidth}|p{0.45\textwidth}}
    \toprule 
    \rowcolor{gray!50}
    \multicolumn{3}{l}{Q. 14: Attitudes $\rightarrow$ Reasons for change in impression post-exposure} \\
    \midrule
    \textbf{Code (\# frequency) [non-exclusive]}   & \textbf{Code Definition}   & \textbf{Example}     \\
    \hline 
    informed-abt-feature (8) & The participant is informed about the product's advertised features  & \quotes{The fact that it mentions that it uses end-to-end encryption.} (WD5, WhatsApp)  \\
    \hline 
    skeptical (6)   & The participant has skepticism about the campaign  & \quotes{I don't really trust advertisements to really indicative of how corporate that you know the, the true nature of corporations does it?} (S1)  \\
    \hline 
    priv-as-selling-point (4) & The company is using privacy as a selling point  &  \quotes{because it shows that they are using data privacy as part of their company brand.} (A6) \\
    \hline 
    no-reason [exclusive] (4)   & The participant did not provide a reason     &  \quotes{It's unchanged.} (A1)\\
    \hline 
     good-company (2) & The company is good generally  & \quotes{I still think overall pretty good company.} (A3)   \\
    \hline 
    already-aware (2)   & The participant already aware about the advertised features  &  \quotes{I'm already slightly familiar with the feature in general ... I don't feel like it's changed my image of Apple so much} (A4)  \\
    \hline 
    priv-feature (2) & The product has privacy features  & \quotes{I like the concept I like that it’s a messaging app that helps with end-to-end encryption} (WD2, WhatsApp)  \\
    \hline 
    motivated-use-feature (2) & The participant is motivated to use the advertised feature   & \quotes{I'd probably gonna look it up afterwards.} (WD3, DuckDuckGo) \\
    \hline 
    address-priv-concerns (2) & The company is addressing consumers privacy concerns  & \quotes{definitely addressing the most pressing concerns that most smartphone users are raising right now.} (S5)  \\
    \hline 
    not-informative (2) & The campaign is not informative & \quotes{As far as the lesson the game was trying to teach. I wasn't really picking up on that.} (T3) \\
    \hline 
    useful-app (1) & The app is useful  & \quotes{I still think Twitter is a useful tool.} (T3) \\
    \hline 
    feel-control (1) & The participant feels in control of their privacy & \quotes{that I have the power to ask the app not to track.} (A5)  \\
    \hline 
    ambitious-ad (1) & The advertisement is ambitious  & \quotes{I guess it's ambitious}   \\
    \hline 
    exploiting-priv (1) & The company is exploiting their customers' data  &  \quotes{That they have probably no issue selling our data and like exploiting our privacy.} (T5) \\
    \hline 
    not-serious (1) &  The campaign was not serious & \quotes{this game didn't really do a good job of taking it like very seriously} (T1)  \\
    \hline 
    not-real-examples (1) & The company does not use real examples in the campaign  & \quotes{I think if they did it again have like more real life implications, and it was more tied to real examples of what can happen with the privacy and data} (T1)  \\
    \bottomrule
    \end{tabular}
} 
\end{table*}

\begin{table*}[!h]
\caption{The codebook for Q.6 about participants' prior-exposure to the campaign.}
\label{tab:prior_exposure}
\renewcommand{\arraystretch}{1.25}
\resizebox{0.85\textwidth}{!}{%
    \begin{tabular}{p{0.19\textwidth}|p{0.45\textwidth}|p{0.45\textwidth}}
    \toprule 
    \rowcolor{gray!50}
    \multicolumn{3}{l}{Q.6: Attitudes $\rightarrow$ Prior exposure to the campaign} \\
    \midrule
    \textbf{Code (\# frequency)\newline[exclusive]}   & \textbf{Code Definition}   & \textbf{Example}     \\
    \hline 
    no (25) & The participant had not seen, or heard about, the campaign before participating in the study & \quotes{No, I haven't} (WD1, WhatsApp)\\
    \hline 
    yes (4) & The participant had seen, or heard about, the campaign before participating in the study & \quotes{Yes ... On television.} (WD5, WhatsApp) \\
    \hline 
    not-sure (1) & The participant is not sure of they had seen, or heard about, the campaign before & \quotes{I will say it looks familiar, but I'm not 100\% sure} (WD6, DuckDuckGo) \\
    \bottomrule
    \end{tabular}
} 
\end{table*}

\begin{table*}[!h]
\caption{The codebook for Q.23 about the effect of re-exposure to the ad campaign on participants.}
\label{tab:re-exposure}
\renewcommand{\arraystretch}{1.25}
\resizebox{0.85\textwidth}{!}{%
    \begin{tabular}{p{0.19\textwidth}|p{0.45\textwidth}|p{0.45\textwidth}}
    \toprule 
    \rowcolor{gray!50}
    \multicolumn{3}{l}{Q. 23: Attitudes $\rightarrow$ Effect of re-exposure to the ad campaign} \\
    \midrule
    \textbf{Code (\# frequency) [non-exclusive]}   & \textbf{Code Definition}   & \textbf{Example}     \\
    \hline 
    more-details (11)    & The participant saw more details in the ad. This can be mentioned explicitly, or implicitly by mentioning further details & \quotes{I feel like I saw more of the advertisement} (WD6, DuckDuckGo) \\
    \hline 
     more-features (4) & The participant saw more features in the ad & \quotes{I realized that there was the Protect mail feature that was at the end that I did not notice before.} (A5)\\
    \hline 
    more-understanding (2) & The participant has more understanding of the ad & \quotes{I got a chance to read the mail protect button this time, and I see I- I think I have a sense of what it does now} (A6) \\
    \hline 
    more-questions (2) & The participant has more questions of the ad & \quotes{So does that mean the auctioneer is Apple? and by the consumer saying, Ask app not to track. Is that just us preventing Apple legally from being able to sell that data?} (A5)\\
    \hline 
    more-attention (1)  &  The participant paid more attention to the ad & \quotes{I paid more attention to what they said about the, the messages they said} (WD4, WhatsApp) \\
    \hline 
    more-curious (1)    & The participant became more curious about the ad, company, or product & \quotes{It's again it's made me feel a bit more curious about DuckDuckGo} (WD3, DuckDuckGo)\\
    \hline 
    consider-switch (1) & The participant (non-user) considers switching to the advertised product in the ad & \quotes{kind of convinced me more and more to get iPhone somehow} (A2) \\
    \hline 
    misleading (1)  & The participant finds the ad misleading & \quotes{that's kinda misleading to me.} (A3)\\
    \hline 
    oversimplified (1)  & The participant finds the ad oversimplified the process of using the feature & \quotes{it's representation of the process is slightly oversimplified.} (A4) \\
    \hline 
    misunderstood (1)   & The participant realized they misunderstood the ad the first time & \quotes{I misunderstood the ad the first time around} (S2)\\
    \hline 
    enjoyable (1)   & The participant enjoyed watching the ad again  & \quotes{it's a really enjoyable ad to watch or ad to watch to be honest with you.} (S3)\\
    \hline 
    same (16)   & The participant did not notice any additional things after watching the ad again & \quotes{You know, kind of the same.} (A3)\\
    \hline 
    uncomfortable (1)  & The participant felt uncomfortable after watching the ad again & \quotes{coming back and seeing it again makes me feel like I know the apps are stealing my data, but I'm not sure as a user I'm comfortable, seeing that much level for detail} (S5)\\
    \hline 
    boring (1)  & The participant would feel boring if they keep watching it over and over & \quotes{if I would want to watch this advertisement like a 100 times. I'm sure I’ll get sick of it.} (WD1, WhatsApp)\\
    \hline 
    want-more-features (1)  & The participant wants more features & \quotes{would like to have the ability to like kind of save some of the cookies or cache for ... if there's some kind of function where it will, you can save some cookies} (WD2, DuckDuckGo)\\
    \hline 
    fear (1)   & The participant has fear & \quotes{That it is like physically watching you. That, that was a bit scary. That's what I've noticed.} (WD3, DuckDuckGo)\\
    \hline 
    better (1)  & The participant feels better about the ad & \quotes{Honestly, I feel better about it.} (WD6, DuckDuckGo)\\
    \hline 
    funny (1)  & The participant found funny things about the ad & \quotes{He said he said, it's not creepy. It's commerce which I thought was kind of funny.} (A1) \\
    \hline 
    convincing (1)  & The participant finds the ad convincing & \quotes{But it's very convincing.} (A2)\\
    \hline 
    interesting (1) & The participant finds the ad interesting & \quotes{So that was kind of interesting to me.} (WD3, DuckDuckGo)\\
    \hline 
    skeptical (1)  & The participant has skepticism about the campaign & \quotes{I just feel like that it's because it's Facebook. I don't really trust that they could do it very well.} (WD2, WhatsApp)\\
    \bottomrule
    \end{tabular}
} 
\end{table*}

\begin{table*}[!h]
\caption{The codebook for Q.7, Q.8, \& Q.9 about participants' privacy awareness.}
\label{tab:awareness}
\renewcommand{\arraystretch}{1.25}
\resizebox{0.85\textwidth}{!}{%
    \begin{tabular}{p{0.19\textwidth}|p{0.45\textwidth}|p{0.45\textwidth}}
    \toprule 
    \rowcolor{gray!50}
    \multicolumn{3}{l}{Q.7, Q.8, and Q.9: Awareness $\rightarrow$ Privacy awareness} \\
    \midrule
    \textbf{Code (\# frequency)\newline[exclusive]}   & \textbf{Code Definition}   & \textbf{Example}     \\
    \hline 
    yes (28) &  The participant mentioned privacy and/or security aspects in their response &  \quotes{to sell iPhones to present them as the most secure option for protecting your identity and your data from unauthorized use.} (A1) \\
    \hline 
    no (2) &   The participant did not mention privacy and/or security aspects in their response &  \quotes{I'm not super sure but it was fun.} (T5) \\
    \bottomrule
    \end{tabular}
} 
\end{table*}

\begin{table*}[!h]
\caption{The codebook for Q.7, Q.8, \& Q.9 about the types of participants' privacy awareness.}
\label{tab:priv_aware}
\renewcommand{\arraystretch}{1.25}
\resizebox{0.85\textwidth}{!}{%
    \begin{tabular}{p{0.19\textwidth}|p{0.45\textwidth}|p{0.45\textwidth}}
    \toprule 
    \rowcolor{gray!50}
    \multicolumn{3}{l}{Q.7, Q.8, and Q.9: Awareness $\rightarrow$ Privacy awareness type} \\
    \midrule
    \textbf{Code (\# frequency) [non-exclusive]}   & \textbf{Code Definition}   & \textbf{Example}     \\
    \hline 
    aware-issues (26) & The participant mentioned specific privacy issues  &  \quotes{whether or not an app was tracking where you were.} (S2). See~\autoref{tab:priv_issue_type} for further examples. \\
    \hline 
    aware-features (25) & The participant mentioned specific privacy features & \quotes{asking apps not to track her} (A1). See~\autoref{tab:priv_feature_type} for further examples.\\
    \hline 
    n.a. (2)        &  Not applicable as the participant is not privacy aware (did not mention any privacy or security aspects in their Q.7, Q.8, and Q.9 responses -- see previous table) & n.a. \\
    \bottomrule
    \end{tabular}
} 
\end{table*}

\begin{table*}[!h]
\caption{The codebook for Q.7, Q.8, \& Q.9 about the types of privacy issues participants are aware of.}
\label{tab:priv_issue_type}
\renewcommand{\arraystretch}{1.25}
\resizebox{0.85\textwidth}{!}{%
    \begin{tabular}{p{0.19\textwidth}|p{0.45\textwidth}|p{0.45\textwidth}}
    \toprule 
    \rowcolor{gray!50}
    \multicolumn{3}{l}{Q.7, Q.8, and Q.9: Awareness $\rightarrow$ Privacy issues types} \\
    \midrule
    \textbf{Code (\# frequency) [non-exclusive]}   & \textbf{Code Definition}   & \textbf{Example}     \\
    \hline 
    mentioned-privacy (14) & The participant mentioned the word \quotes{privacy} explicitly & \quotes{it’s supposed to be more privacy} (WD5, DuckDuckGo)\\
    \hline 
    tracking (11) & Tracking issue & \quotes{relating to phone security in terms of whether or not an app was tracking where you were.} (S2)\\
    \hline 
    control (7) & Privacy controls issue & \quotes{to help control your privacy like how much information like what information is being shared.} (S1)\\
    \hline
     personal-data (7) & Personal data issue & \quotes{different people bidding on, like someone's personal information.} (A2)\\
    \hline 
    msg-safety (6) & Security of an instant message, such as secure vs. insecure messages, encrypted vs. unencrypted messages & \quotes{comparing unencrypted SMS messaging to old fashioned carrier pigeon} (WD1, WhatsApp)\\
    \hline
     location-data (6) & Location data issue & \quotes{her personal information like location, history} (A4)\\
    \hline 
     n.a. (4) & Not applicable, as the participant did not mention any privacy issue in their answers for Q.7, Q.8, and Q.9 & n.a. \\
    \hline
     browsing-data (3) & Browsing data issue & \quotes{Google keeps track and monitors your search history} (WD1, DuckDuckGo)\\
    \hline 
     concerns (3) & Privacy concerns issue & \quotes{It gave me an impression that it can be getting dangerous to be on Twitter with all the different elements of privacy concerns} (T1)\\
    \hline
    data-sharing (3) & Data sharing issue & \quotes{to help control your privacy like how much information like what information is being shared.} (S1)\\
    \hline 
    data-stealing (3) & Data stealing issue & \quotes{calling out Google ... like, yeah, they, they can steal your data off and it should be something that you’re - you should be worried about.} (WD3, DuckDuckGo) \\
    \hline 
    permission (3) & Privacy permissions issue & \quotes{only use location that when the app is actively in use, or all the time, or ask for permission every time} (S3)\\
    \hline 
    data-selling (3) & Data selling issue & \quotes{a woman who was looking around and discovered that her data was being sold at auction.} (A6)\\
    \hline 
    privacy-invasion (3) & Privacy invasion issue & \quotes{her data was being sold at auction. It was very invasive} (A6)\\
    \hline 
    access (2) &  Controlling access to private data issue & \quotes{privacy control of allowing users to choose when and where they can be accessed for private information} (S5)\\
    \hline 
    camera-data (2) & Camera data issue & \quotes{Specifically the option to see which apps are using which features such as camera} (S6)\\
    \hline 
    contacts-data (2) & Contact data issue & \quotes{her personal information like location, history, contacts, browsing history, email information being auctioned off} (A4) \\
    \hline
    data-collection (2) & Data collection issue & \quotes{I was not surprised that Google collects your personal information.} (WD3, DuckDuckGo) \\
    \hline 
    data-usage (2) & Data use issue & \quotes{I guess how Twitter uses our data.} (T2) \\
    \hline 
    email-data (2) & Email data issue & \quotes{asking apps not to track her or protecting her email data.} (A1) \\
    \hline 
     awareness (1) & Privacy awareness issue & \quotes{a game designed to help you understand how Twitter uses data and their privacy policy} (T1)\\
    \hline 
    camera-tape (1) & Camera tape to protect one's privacy & \quotes{I have a, a coworker ... and he had a piece of tape over the camera.} (S3) \\
    \hline 
    encryption (1) & Data encryption issue & \quotes{it's a little bold, you know, and not believable. I mean. And the encryption thing is one thing you know.} (WD1, DuckDuckGo)\\
    \hline 
    hackers (1) & Hackers issue & \quotes{The hackers are very smart and they know how to get into your, your devices} (A3)\\
    \hline 
    history-data (1) & General history data issue & \quotes{her personal information like location, history, contacts, browsing history, email information being auctioned off} (A4)\\
    \hline 
    identifying-data (1) & Identifiable data issue & \quotes{also some other things identifying you} (S2)\\
    \hline 
    identity-protection (1) & Identity protection issue & \quotes{to present them as the most secure option for protecting your identity and your data}\\
    \hline 
    lurking (1) & Lurking issue & \quotes{the lurking guy behind people singing} (WD2, DuckDuckGo) \\
    \hline 
    microphone-data (1) & Microphone data issue & \quotes{see which apps are using which features such as camera, microphone} (S6)\\
    \hline 
    minimize-data-sharing (1) & Minimizing data sharing issue & \quotes{would like to use a device that does that to the minimal degree.} (S6)\\
    \hline  
    personalization (1) & Personalization issue & \quotes{because they can yield more personalized search results is what they claim they do.} (WD4, DuckDuckGo)\\
    \hline 
    priv-msg (1) & Private messaging issue & \quotes{sending private messages using a messenger pigeon} (WD5, WhatsApp)\\
    \hline 
    priv-policy (1) & Privacy policy issue &  \quotes{help you understand how Twitter uses data and their privacy policy} (T1)\\
    \hline 
    priv-protective (1) & Privacy protective behavior issue & \quotes{being more protective of, you know, user privacy} (S1)\\
    \hline 
    snooping (1) & Snooping issue & \quotes{the search snooping by search engine companies.} (WD4, DuckDuckGo)\\
    \hline 
    spoofing (1) & Spoofing issue & \quotes{I know that your texts and all that can get like spoofed} (WD2, WhatsApp) \\
    \hline 
    unauthorized-access (1) & Unauthorized access issue & \quotes{they know how to get into your, your devices} (A3)\\
    \hline 
    unauthorized-use (1) & Unauthorized use issue & \quotes{protecting your identity and your data from unauthorized use} (A1)\\
    \bottomrule
    \end{tabular}
} 
\end{table*}

\begin{table*}[!h]
\caption{The codebook for Q.7, Q.8, \& Q.9 about the types of privacy features participants are aware of.}
\label{tab:priv_feature_type}
\renewcommand{\arraystretch}{1.25}
\resizebox{0.85\textwidth}{!}{%
    \begin{tabular}{p{0.19\textwidth}|p{0.45\textwidth}|p{0.45\textwidth}}
    \toprule 
    \rowcolor{gray!50}
    \multicolumn{3}{l}{Q.7, Q.8, and Q.9: Awareness $\rightarrow$ Privacy features types} \\
    \midrule
    \textbf{Code (\# frequency) [non-exclusive]}   & \textbf{Code Definition}   & \textbf{Example}     \\
    \hline 
    priv-features (10) & Privacy and/or security features generally (not specific to a company) & \quotes{And then she used an Apple security feature and made it stop.} (A6)\\
    \hline
    ddg-priv (6) & DDG's privacy features generally & \quotes{we're a search engine that you know. Let you have more control over your data, and we won't pry into it.} (WD1, DuckDuckGo)\\
    \hline
    permission-control (6) & Samsung's permission control general feature & \quotes{Samsung is giving users the power to control who's using your data.} (S4)\\
    \hline
    encryption (5) & WahtsApp's encryption feature  & \quotes{saying that whatsapp was you know, has encryption} (WD2, WhatsApp)\\
    \hline
    n.a. (5) & Not applicable, as the participant did not mention any privacy feature in their answers for Q.7, Q.8, and Q.9 & n.a. \\
    \hline
    location-control (4) & Samsung's location permission control feature & \quotes{Samsung gives you. Specifically the option to see which apps are using which features such as camera, microphone, location,} (S6)\\
    \hline
    ask-app (4) & Apple's ask app not to track feature & \quotes{selecting options on her iPhone, asking apps not to track her} (A1)\\
    \hline 
    camera-control (2) & Samsung's camera permission control feature & \quotes{Samsung gives you. Specifically the option to see which apps are using which features such as camera, microphone, location,} (S6)\\
    \hline
    no-tracking (2) & DDG's no tracking as a feature & \quotes{Google keeps track and monitors your search history, and DuckDuckGo doesn't.} (WD1, DuckDuckGo)\\
    \hline
    priv-control (2) & Privacy controls feature (not specific to a company) & \quotes{the game was to show that I have more control over my data than I realize.} (T4)\\
    \hline
    rcs-encrypt-type (1) & RCS encryption type as a feature & \quotes{think it’s called RCS right, the rich clear ... whatever, whatever the end-to-end encrypted Google Message thing is} (WD4, WhatsApp)\\
    \hline
    ddg-app (1) & DDG's app as a feature & \quotes{introduce people to DuckDuckGo the app, as, as opposed to just a search, as opposed to a just a search engine} (WD5, DuckDuckGo)\\
    \hline 
    protect-mail (1) & Apple's protect mail feature & \quotes{asking apps not to track her or protecting her email data.} (A1)\\
    \hline
    direct-msg (1) &  Twitter's direct message as a feature & \quotes{when it said that you have access to direct messaging or something, letting you know that people can send you things} (T6)\\
    \hline
    msg-deletion (1) & Twitter's message deletion as a feature & \quotes{and then if you delete them it only deletes on your end.} (T6)\\
    \hline
    mic-control (1) & Samsung's microphone permission control feature & \quotes{Samsung gives you, specifically the option to see which apps are using which features such as camera, microphone, location} (S6)\\
    \bottomrule
    \end{tabular}
} 
\end{table*}

\begin{table*}[!h]
\caption{The codebook for Q.9 about participants' perceived intent of the campaign.}
\label{tab:intent}
\renewcommand{\arraystretch}{1.25}
\resizebox{0.85\textwidth}{!}{%
    \begin{tabular}{p{0.19\textwidth}|p{0.45\textwidth}|p{0.45\textwidth}}
    \toprule 
    \rowcolor{gray!50}
    \multicolumn{3}{l}{Q.9: Awareness $\rightarrow$ Perceived intent of the campaign} \\
    \midrule
    \textbf{Code (\# frequency) [non-exclusive]}   & \textbf{Code Definition}   & \textbf{Example}     \\
    \hline 
    advert-priv-as-feature (24) & To advertise privacy as a feature in the product & \quotes{the main purpose was like, there's other stuff that will actually keep your stuff private} (WD2, DuckDuckGo)\\
    \hline 
    better-thn-competitors (15) & To stand out from their competitors, or to encourage viewers to switch product & \quotes{it comes from the idea of yeah, you know, I know you guys using Google and all that stuff. But you know. Think about it.} (WD1, DuckDuckGo)\\
    \hline 
    educate (13) & To raise awareness or educate viewers about some sort of privacy information & \quotes{I think it was to educate people about how Twitter uses our data} (T2)\\
    \hline 
    entertainment (2)  & The campaign is for entertainment and fun  & \quotes{I would think a diversion, or just something to do?} (T3)\\
    \bottomrule
    \end{tabular}
} 
\end{table*}

\begin{table*}[!h]
\caption{The codebook for Q.13 about participants' prior awareness of the the advertised features.}
\label{tab:prior_awareness}
\renewcommand{\arraystretch}{1.25}
\resizebox{0.85\textwidth}{!}{%
    \begin{tabular}{p{0.19\textwidth}|p{0.45\textwidth}|p{0.45\textwidth}}
    \toprule 
    \rowcolor{gray!50}
    \multicolumn{3}{l}{Q.13: Awareness $\rightarrow$ Prior awareness of the advertised features} \\
    \midrule
    \textbf{Code (\# frequency)\newline[exclusive]}   & \textbf{Code Definition}   & \textbf{Example}     \\
    \hline 
   yes (18) & The participant has prior awareness of the advertised feature & \quotes{Yes, I was.} (A5) \\
   \hline
   no (7) & The participant does not have prior awareness of the advertised feature at all, or not the way it was shown & \quotes{No, as I said, I haven't used a Samsung device in a few years.} (S6)\\
   \hline
   somewhat (5) & The participant has heard of the advertised feature, but not fully aware of what it is, or is aware of similar features in other systems, but not exactly as shown in campaign & \quotes{No, not in it's full - not, not in like that} (WD3, WhatsApp)\\
    \bottomrule
    \end{tabular}
} 
\end{table*}

\begin{table*}[!h]
\caption{The codebook for Q.13B about how participants came across the advertised features.}
\label{tab:how_encountered}
\renewcommand{\arraystretch}{1.25}
\resizebox{0.85\textwidth}{!}{%
    \begin{tabular}{p{0.19\textwidth}|p{0.45\textwidth}|p{0.45\textwidth}}
    \toprule 
    \rowcolor{gray!50}
    \multicolumn{3}{l}{Q. 13b: Awareness $\rightarrow$ How features encountered} \\
    \midrule
    \textbf{Code (\# frequency) [non-exclusive]}   & \textbf{Code Definition}   & \textbf{Example}     \\
    \hline 
    install-prompt (5) & In a prompt during app installation & \quotes{it was probably when I was installing apps.} (A1) \\
    \hline 
    settings (3) & In the settings during app use or unspecified, but not during installation & \quotes{I came across a tracking setting by going through my own settings} (A2) \\
    \hline
    usage-prompt (3) & During app use in general or unspecified but not during installation & \quotes{it seems like every other time I log into Twitter I get a pop up saying you can come and review it.} (T2)  \\
    \hline
    post-update-prompt (3) & In a prompt after OS or app update & \quotes{when they push their updates they have a license for why they have the updates and web tracking is one of the subjects that come up.} (WD5, DuckDuckGo) \\
    \hline
     news (3) & From news  & \quotes{Oh, probably news media.} (WD4, WhatsApp) \\
    \hline
    general-knowledge (3) & From general knowledge & \quotes{that's pretty general knowledge, even if you're not up on privacy and security stuff.} (T6) \\
    \hline
    word-of-mouth (3) & Through someone  & \quotes{But I think someone had to tell me about turning off the personalized ads thing} (T6) \\
    \hline 
    seen-in-passing (2) & The participant encountered the feature while using the app or in a sidebar & \quotes{I see them in the sidebar} (T5) \\
    \hline
    work (2) & At work & \quotes{at work.} (WD1, DuckDuckGo) \\
    \hline 
     use-other-service (2) & While using other services such as VPN or PGP & \quotes{I used VPN Services in the past, and also PGP.} (WD5, WhatsApp) \\
    \hline
    work-policy (1) & In work communication about policy documents &  \quotes{It was work related to our company shift about careful communications and that's probably the first time I'd heard about, or even that's when I probably would start to pay attention to.} (T3) \\
    \hline
    terms-update-prompt (1) & In a prompt when the company updated their terms & \quotes{I definitely got a notification when they update the terms.} (T4) \\
    \hline
    email-notif (1) & In an email when the company updated their terms & \quotes{sometimes, if they change the terms fully, they send an email.} (T4) \\
    \hline 
    dont-remember (1) & The participant cannot remember where they encountered the feature & \quotes{ I don't remember where I first heard it.} (WD2, DuckDuckGo)\\
    \hline
    ads (1) & In ads & \quotes{I came across that term in a lot of advertising I see on Youtube for VPN} (WD6, WhatsApp) \\
    \hline
    self-education (1) & Through self-education & \quotes{I've grown up with the Internet and learn more about search engines and how different browsers stores your data and record your data.} (WD6, DuckDuckGo)\\
    \bottomrule
    \end{tabular}
} 
\end{table*}

\begin{table*}[!h]
\caption{The codebook for Q.10 about participants' perception of the metaphors shown in the campaigns.}
\label{tab:metaphore}
\renewcommand{\arraystretch}{1.25}
\resizebox{0.85\textwidth}{!}{%
    \begin{tabular}{p{0.19\textwidth}|p{0.45\textwidth}|p{0.45\textwidth}}
    \toprule 
    \rowcolor{gray!50}
    \multicolumn{3}{l}{Q. 10: Awareness $\rightarrow$ Understanding metaphor} \\
    \midrule
    \textbf{Code (\# frequency) [non-exclusive]}   & \textbf{Code Definition}   & \textbf{Example}     \\
    \hline 
    voc-google (4) & DDG's vocalist represents Google & \quotes{the lurking guy behind people singing, Oh, you know, representing Google} (WD2, DuckDuckGo)\\
    \hline
    auct-other-parties (3) & Apple's auctioneer represents other parties such as other smartphone companies, ISP, other app providers, etc. & \quotes{The auctioneer would represent ... your Internet provider or ... other cell phone companies.} (A1) \\
    \hline 
    auct-data-sale (3) & Apple's auctioneer represents data sale of the consumer's data by apps or websites to other parties such as advertisers & \quotes{ad sales auctions.} (A6)\\
    \hline 
    umbr-warning (2) & Samsung's umbrella represents warning or caution & \quotes{a mark of security warning?} (S5)\\
    \hline 
    env-data-collection (2) & Twitter's envelope represents data collection & \quotes{Maybe data, like pieces of data.} (T5)\\
    \hline 
    umbr-dont-know (2) & Samsung's umbrella represents don't know & \quotes{I don't know.} (S4) \\
    \hline 
     pigeon-unsafe-msg (2) & WhatsApp's pigeon represents unsafe messages & \quotes{sending messages in the open.} (WD5, WhatsApp)\\
    \hline 
     pigeon-sms (2) & WhatsApp's pigeon represents SMS & \quotes{old fashioned SMS style communication.} (WD1, WhatsApp)\\
    \hline 
    pigeon-old-tech (2) & WhatsApp's pigeon represents old technologies & \quotes{The pigeon represents antiqu- antiquated ... essentially older technology that has been outdated} (WD6, WhatsApp) \\
    \hline 
    umbr-protection (1) & Samsung's umbrella represents protection of the consumer's data such as location from outsider's access  & \quotes{shielding like your information, like, I guess protection} (S1)\\
    \hline 
    umbr-location (1) & Samsung's umbrella represents sharing location data & \quotes{they are talking about sharing location data} (S3)\\
    \hline 
    env-tweet (1) & Twitter's envelope represents tweets & \quotes{I just understood the envelope was like a tweet in general} (T2)\\
    \hline 
    env-cat (1) & Twitter's envelope represents a cat & \quotes{I kind of thought it was a cat} (T3)\\
    \hline 
    env-spam (1) & Twitter's envelope represents spam & \quotes{Spam messaging} (T4) \\
    \hline 
    env-opening-dm (1) & Twitter's envelope represents direct messages & \quotes{it was probably related to opening a message like a DM} (T6)\\
    \hline 
    pigeon-platform (1) & WhatsApp's pigeon represents platforms & \quotes{The pigeon is more like, yeah, the, the platform that you send messages to and from people} (WD3, WhatsApp) \\
    \hline 
    pigeon-network (1) & WhatsApp's pigeon represents network & \quotes{the network?} (WD4, WhatsApp)\\
    \hline  
    voc-search-engine (1) & DDG's vocalist represents the search engine companies & \quotes{maybe the Search Engine Company} (WD4, DuckDuckGo)\\
    \hline 
    voc-cheesy (1) & DDG's vocalist represents something cheesy & \quotes{So the song, okay, yeah ... it was, was I mean, it's little cheesy}\\
    \hline 
    voc-data-buyer (1) & DDG's vocalist represents data buyers & \quotes{the search engine sells its data to} (WD4, DuckDuckGo) \\
    \bottomrule
    \end{tabular}
} 
\end{table*}

\begin{table*}[!h]
\caption{The codebook for Section 6 \& 7 about the actionability task evaluation criteria.}
\label{tab:actionability}
\renewcommand{\arraystretch}{1.25}
\resizebox{0.85\textwidth}{!}{%
    \begin{tabular}{p{0.19\textwidth}|p{0.45\textwidth}|p{0.45\textwidth}}
    \toprule 
    \rowcolor{gray!50}
    \multicolumn{3}{l}{Q. 10: Actionability $\rightarrow$ Task (activity) evaluation criteria} \\
    \midrule
    \textbf{Code (\# frequency) [non-exclusive]}   & \textbf{Code Definition}   & \textbf{Example}     \\
    \hline
    know-end (20) & The participant knows that they have completed the activity & n.a. -- task \\
    \hline 
    hint (8) &	The participant asked for help & n.a. -- task  \\
    \hline 
     unsure-end (6) & The participant asked interviewer for reassurance that they completed the activity & n.a. -- task \\
    \hline 
     google (3) & The participant used Google search & n.a. -- task \\ 
     \hline 
    clarify (2) &	The participant asked for clarification. This is not considered as intervention from the interviewer & n.a. -- task \\
    \hline 
    use-own-device (2) & The participant looked at their own device during the activity & n.a. -- task \\
    \hline 
    skipped (2) &	The participant did not complete the tasks - time out & n.a. -- task \\
    \bottomrule
    \end{tabular}
} 
\end{table*}
\begin{table*}[!h]
\caption{The codebook for Q.21 about participants' (non-users of the product) motivation to switch to the advertised product post-exposure to the campaigns.}
\label{tab:switch}
\renewcommand{\arraystretch}{1.25}
\resizebox{0.85\textwidth}{!}{%
    \begin{tabular}{p{0.19\textwidth}|p{0.45\textwidth}|p{0.45\textwidth}}
    \toprule 
    \rowcolor{gray!50}
    \multicolumn{3}{l}{Q. 10: Actionability $\rightarrow$ Non-user motivation to switch product} \\
    \midrule
    \textbf{Code (\# frequency)\newline[exclusive]}   & \textbf{Code Definition}   & \textbf{Example}     \\
    \hline 
    no  (5) & The participant felt that the campaign did not motivate them to switch their current product to the advertised one & \quotes{I would still say no} (S5)\\
    \hline 
    maybe (5) & The participant felt that the campaign motivated them to some extent, to switch their current product to the advertised one, but not fully motivated & \quotes{Like not necessarily, but kind of convinced me a tiny bit. Just a little bit.} (A2)\\
    \hline 
    yes (2) & The participant felt that the campaign motivated them to switch their current product to the advertised one & \quotes{Yeah, I think I would. Yes} (S2)\\
    \bottomrule
    \end{tabular}
} 
\end{table*}

\begin{table*}[!h]
\caption{The codebook for Q.21 about reasons for participants' (non-users of the product) motivations to switch to the advertised products post-exposure to the campaigns.}
\label{tab:switch_reason}
\renewcommand{\arraystretch}{1.25}
\resizebox{0.85\textwidth}{!}{%
    \begin{tabular}{p{0.19\textwidth}|p{0.45\textwidth}|p{0.45\textwidth}}
    \toprule 
    \rowcolor{gray!50}
    \multicolumn{3}{l}{Q. 21: Actionability $\rightarrow$ Reasons to switch product} \\
    \midrule
    \textbf{Code (\# frequency) [non-exclusive]}   & \textbf{Code Definition}   & \textbf{Example}     \\
    \hline
    simpler (1) & The advertised product seems simpler & \quotes{Yeah, I think I would. Yes ... It seemed like it was a simpler process based on the ad} (S2)\\
    \hline
    increase-usage (1) & The participant is motivated to increase using the advertised product and eventually shift to it & \quotes{it did ... it has me thinking about potentially using that to communicate with some of the people I currently talk to on Snapchat.} (WD6, WhatsApp)\\
    \bottomrule
    \end{tabular}
} 
\end{table*}

\begin{table*}[!h]
\caption{The codebook for Q.21 about the reasons for participants (non-users of the product) motivations to not-switch to the advertised products post-exposure to the campaigns.}
\label{tab:not_switch_reason}
\renewcommand{\arraystretch}{1.25}
\resizebox{0.85\textwidth}{!}{%
    \begin{tabular}{p{0.19\textwidth}|p{0.45\textwidth}|p{0.45\textwidth}}
    \toprule 
    \rowcolor{gray!50}
    \multicolumn{3}{l}{Q. 21: Actionability $\rightarrow$ Reasons to not-switch product} \\
    \midrule
    \textbf{Code (\# frequency) [non-exclusive]}   & \textbf{Code Definition}   & \textbf{Example}     \\
    \hline 
    already-have-features (2) & The participant's current product already has similar feature & \quotes{mainly showed features that iPhones already really have.} (S6)\\
    \hline 
    switching-cost (2) &  Cost of switching product & \quotes{would need to spend a fair deal of money like kind of getting on another contract as well because Samsung and Apple use like different chargers and peripherals.} (A4)\\
    \hline
    familiarity (2) &  The participant is familiar with their current product & \quotes{I'd still want to stick with an operating system that I'm more comfortable with and more familiar with.} (A4) \\
    \hline 
    same-company (1) & The advertised product is owned by same company of the current used product (e.g., Meta) & \quotes{they’re from the same company, and I really only use like Facebook messenger for just to contact friends with [too] personal stuff ... and it's owned by the same company} (WD2, WhatsApp)\\
    \hline
    data-sync (1) &  Moving personal data (synchronization) issues & \quotes{there's more exit cost for me as an iOS user to switch to an Android operating system, like moving all the contacts, or like photos, or like even memos.} (S5)\\
    \bottomrule
    \end{tabular}
} 
\end{table*}

\begin{table*}[!h]
\caption{The codebook for Q.21 about the reasons for participants' (non-users of the product) motivations to maybe-switch to the advertised products post-exposure to the campaigns.}
\label{tab:maybe_switch_reason}
\renewcommand{\arraystretch}{1.25}
\resizebox{0.85\textwidth}{!}{%
    \begin{tabular}{p{0.19\textwidth}|p{0.45\textwidth}|p{0.45\textwidth}}
    \toprule 
    \rowcolor{gray!50}
    \multicolumn{3}{l}{Q. 21: Actionability $\rightarrow$ Reasons to maybe-switch product} \\
    \midrule
    \textbf{Code (\# frequency) [non-exclusive]}   & \textbf{Code Definition}   & \textbf{Example}     \\
    \hline 
    encouraged-explore (3) & The campaign encouraged the participant to explore and try the advertised product  & \quotes{at least like encouraged me to look into what, what it is} (WD3, DuckDuckGo) \\
    \hline 
     familiarity (1) &  The participant is familiar with their current product & \quotes{maybe think about it ... because Google's already, you know it in our lives.} (WD1, DuckDuckGo)\\
    \hline 
    nice-features (1) & The participant liked the features shown & \quotes{Like not necessarily, but kind of convinced me a tiny bit. Just a little bit ... because those features are kind of like neat and pretty nice} (A2)\\
    \bottomrule
    \end{tabular}
} 
\end{table*}

\begin{table*}[!h]
\caption{The codebook for Q.22 about participants' (users of the product) motivations to continue using the advertised products post-exposure to the campaigns.}
\label{tab:continue}
\renewcommand{\arraystretch}{1.25}
\resizebox{0.85\textwidth}{!}{%
    \begin{tabular}{p{0.19\textwidth}|p{0.45\textwidth}|p{0.45\textwidth}}
    \toprule 
    \rowcolor{gray!50}
    \multicolumn{3}{l}{Q. 21: Actionability $\rightarrow$ User motivation to continue using the product} \\
    \midrule
    \textbf{Code (\# frequency)\newline[exclusive]}   & \textbf{Code Definition}   & \textbf{Example}     \\
    \hline
    yes (8) & The participant felt that the campaign motivated them to continue using their current product & \quotes{Yeah ... that makes me more motivated to stay with Samsung} (S3)\\
    \hline 
    no-impact (7) & The participant felt that the campaign has no impact in their motivation to continue using their current product. Participants may still be motivated to continue using the product but for reasons other than the campaign & \quotes{No, not really. There's other factors.} (A1)\\
    \hline 
    neutral (3) &  The participant provided neutral impression (neither motivated or not) to continue using their current product. This can be explicitly mentioned or implicitly by not taking any position in their answer & \quotes{neutral. I already use WhatsApp} (WD4, WhatsApp)\\	
    \bottomrule
    \end{tabular}
} 
\end{table*}

\begin{table*}[!h]
\caption{The codebook for Q.22 about the reasons for participants' (users of the product) motivations to continue using the advertised products post-exposure to the campaigns.}
\label{tab:continue_reason}
\renewcommand{\arraystretch}{1.25}
\resizebox{0.85\textwidth}{!}{%
    \begin{tabular}{p{0.19\textwidth}|p{0.45\textwidth}|p{0.45\textwidth}}
    \toprule 
    \rowcolor{gray!50}
    \multicolumn{3}{l}{Q. 21: Actionability $\rightarrow$ Reasons for user motivation to continue} \\
    \midrule
    \textbf{Code (\# frequency) [non-exclusive]}   & \textbf{Code Definition}   & \textbf{Example}     \\
    \hline
    already-user (4) & They are already users and would continue using the product  & \quotes{I already use WhatsApp, I would not consider switching to another service.} (WD4, WhatsApp) \\
    \hline 
    privacy-focus (2) &  The company is privacy focused  & \quotes{Yeah, I mean, especially with more um, privacy focused} (WD2, DuckDuckGo)\\
    \hline 
    unique-value (1) &  It provides a unique value & \quotes{at least from what I've seen from other brands, I haven't seen that.} (A5) \\
    \hline 
    control (1) &  It provides them with more control over their data & \quotes{By suggesting that you, as a consumer, you have control over what other things, of what other apps are tracking.} (A5) \\
    \hline 
    familiarity (1) & The participant is familiar with their current product & \quotes{I'm quite familiar with how Samsung works and the permissions and things. So that's definitely a benefit.} (S4)\\
    \hline 
    increased-usage (1) & The participant feels the campaign motivated them to increase using their product & \quotes{I would probably want to use it even more} (WD2, DuckDuckGo)\\
    \hline 
    new-feature (1) &  To try new features & \quotes{I didn't know DuckDuckGo was an app. Now that I know I like to try it.} (WD5, DuckDuckGo)\\
    \bottomrule 
    \end{tabular}
} 
\end{table*}

\begin{table*}[!h]
\caption{The codebook for Q.22 about the reasons for participants' (users of the product) no-impact on motivations to continue using the advertised products post-exposure to the campaigns.}
\label{tab:not_continue_reason}
\renewcommand{\arraystretch}{1.25}
\resizebox{0.85\textwidth}{!}{%
    \begin{tabular}{p{0.19\textwidth}|p{0.45\textwidth}|p{0.45\textwidth}}
    \toprule 
    \rowcolor{gray!50}
    \multicolumn{3}{l}{Q. 21: Actionability $\rightarrow$ Reasons for no-impact to continue} \\
    \midrule
    \textbf{Code (\# frequency) [non-exclusive]}   & \textbf{Code Definition}   & \textbf{Example}     \\
    \hline
    not-serious (2) & The campaign is not serious & \quotes{I would see that as ... more of a diversion or something to pass the time, it wouldn't be something I would go to search. I, I wouldn't. I wouldn't learn that way} (T3)\\
    \hline 
    data-sync (1) & Moving personal data (synchronization) issues & \quotes{because it would mess everything up. You know. I've got my iCloud calendars and all that I mean all that would have to be redone.} (A1)\\
    \hline 
    convenience (1) & It is convenient and they need to be motivated by something to switch & \quotes{I would have to really be motivated by something to try to to switch from Apple, so.} (A1)\\
    \hline 
    more-privacy-conscious (1) & The campaign just made the participant more privacy conscious, but no impact in motivation to continue using the product & \quotes{definitely made me a little more conscious} (T5) \\
    \hline 
    others-use (1) & The participant will continue using the product because others in their social circle use it, not because of the campaign & \quotes{Oh, it doesn't affect me ... I use it because some people use it.} (WD1, WhatsApp)\\
    \hline
    need-more-info (1) & The participant needs to research more about the the advertised product or feature & \quotes{I need to research, more in depth research} (T1)\\
    \hline
    nothing (1) &  Nothing in the campaign motivated them to change &  \quotes{nothing in the AD that you know help, that help me change my perception} (S1)  \\
    \bottomrule
    \end{tabular}
} 
\end{table*}
\begin{table*}[!h]
\caption{The codebook for Q.22 about the reasons for participants' (users of the product) neutral motivations to continue using the advertised products post-exposure to the campaign.}
\label{tab:neutral_continue_reason}
\renewcommand{\arraystretch}{1.25}
\resizebox{0.85\textwidth}{!}{%
    \begin{tabular}{p{0.19\textwidth}|p{0.45\textwidth}|p{0.45\textwidth}}
    \toprule 
    \rowcolor{gray!50}
    \multicolumn{3}{l}{Q. 21: Actionability $\rightarrow$ Reasons for neutral motivation} \\
    \midrule
    \textbf{Code (\# frequency) [non-exclusive]}   & \textbf{Code Definition}   & \textbf{Example}     \\
    \hline 
    user-friendly (1) & The campaign reminded the participant how user-friendly the product is, but the participant has neutral impression about motivating them to continue using the product & \quotes{It just, I guess you reminded me of how user friendly Twitter is} (T6) \\ 
    \hline 
    privacy-focus (1) &  The company is privacy focused, but the participant has neutral impression about motivating them to continue using the product  & \quotes{neutral ... I know that among browsers and among search engines DuckDuckGo is probably one of the best in terms of not tracking users.} (WD4, DuckDuckGo) \\
    \hline 
    already-user (1) & They are already users, but the participant has neutral impression about motivating them to continue using the product & \quotes{neutral. I already use WhatsApp} (WD4, WhatsApp) \\
    \bottomrule
    \end{tabular}
} 
\end{table*}

\end{document}